\documentclass[lettersize,journal]{IEEEtran}
\usepackage{amsmath,amsfonts}
\usepackage{algorithmic}
\usepackage{algorithm}
\usepackage{array,amssymb}
\usepackage{textcomp}
\usepackage{stfloats}
\usepackage{url}
\usepackage{verbatim,bm}
\usepackage{graphicx}
\usepackage{booktabs,xcolor,tabularx}
\usepackage{longtable}
\usepackage{cite}
\usepackage{subfigure}
\usepackage{epsfig}
\usepackage{epstopdf}
\usepackage{makecell}
\usepackage{multirow} 
\usepackage{geometry}

 \usepackage{enumitem}
\begin{document}

\title{Reconfigurable Antennas for Next-generation Mobile Communication Networks: A Comprehensive Survey and Tutorial}
\author{Yizhe Zhao, \IEEEmembership{Member,IEEE}, Long Zhang, \IEEEmembership{Graduate Student Member ,IEEE}, Halvin Yang,\IEEEmembership{ Member,IEEE}, Kun Yang,\IEEEmembership{ Fellow,IEEE}, Rui Zhang,\IEEEmembership{ Fellow,IEEE}, Lingyang Song,\IEEEmembership{ Fellow,IEEE}, Yuanwei Liu,\IEEEmembership{ Fellow,IEEE}
\thanks{This work was supported in part by the Natural Science Foundation of China (NSFC)  (Grant No.62571091, No.62531008, No.62431002, No.62132004, No.62331022), in part by Jiangsu Major Project on Fundamental Research (Grant No. BK20243059), in part by Gusu Innovation Project (Grant No. ZXL2024360) and High-Tech District of Suzhou City (Grant No. RC2025001), in part by National University of Singapore under Research Grants A-8003646-00-00 and A-8003676-00-00, and the Guangdong Major Project of Basic and Applied Basic Research (Grant No. 2023B0303000001), in part by Beijing Outstanding Young Scientist Program JWZ020240102001 and Key Project of Beijing Nat-ural Science Foundation under grant L253004. (\textit{Yizhe Zhao and Long Zhang contributed equally to this work.})}
\thanks{Yizhe Zhao and Long Zhang are with the School of Information and Communication Engineering, University of Electronic Science and Technology of China, Chengdu, China, e-mail:  yzzhao@uestc.edu.cn, l.zhang@std.uestc.edu.cn.}
\thanks{Halvin Yang is  with the Department of Electrical and Electronic Engineering, Imperial College London, SW7 2AZ London, U.K., e-mail: halvin.yang@imperial.ac.uk.}
\thanks{Kun Yang is with the State Key Laboratory of Novel Software Technology, Nanjing University, Nanjing, 210008, China, Institute of Intelligent Networks and Communications (NINE), Collaborative Innovation Center of Novel Software Technology and Industrialization, and School of Intelligent Software and Engineering, Nanjing University (Suzhou Campus), Suzhou, 215163, China, email: kunyang@nju.edu.cn.}

\thanks{Rui Zhang is with the Department of Electrical and Computer Engineering, National University of Singapore, Singapore 117583, e-mail: elezhang@nus.edu.sg.}


\thanks{ Lingyang Song is with the State Key Laboratory of Photonics and Communications, School of Electronics, Peking University, Beijing, 100871, China, e-mail: lingyang.song@pku.edu.cn.}

\thanks{Yuanwei Liu is with the Department of Electrical and Electronic Engineering, the University of Hong Kong, Hong Kong, and also with Faculty of Applied Sciences, Macao Polytechnic University, e-mail: yuanwei@hku.hk.}
}
\maketitle


\begin{abstract}
The transition to next-generation mobile communication networks, particularly 6G, demands advanced technologies to meet the requirements for ultra-reliable, low-latency communication, massive connectivity, and intelligent applications. Reconfigurable antennas (RAs) play a crucial role in achieving these objectives by enabling dynamic adjustments to the radio frequency (RF) characteristics of antennas, such as gain, radiation pattern, impedance, and polarization. Unlike traditional fixed-position antennas, RAs can alter both their radiation patterns and positions, offering flexibility in response to varying communication environments. This paper presents a comprehensive survey and tutorial on RAs, with a focus on fluid antennas (FAs), movable antennas (MAs), pinching antennas (PAs), and reconfigurable holographic antennas (RHAs), examining their potential in next-generation mobile networks. We explore the channel modelling and estimation, performance analysis, resource allocation strategies, and their synergy with other emerging wireless technologies for each type of RA. Finally, we provide a comparative analysis of different RAs and discuss the open challenges and future research directions, offering insights and guidance for future investigations  in the exciting research area.

\end{abstract}


%
\IEEEpeerreviewmaketitle

\section{Introduction}

\subsection{The Need for Advanced Antenna Technologies for Next-Generation Mobile Communications}
The advent of six-generation (6G) wireless communication systems is set to revolutionize the way we connect and interact, aiming for a hyper-connected and intelligent world \cite{9390169}. As per the expectations, 6G is envisioned to achieve a peak data rate in the range of 100 Gbps-1 Tbps, a significant leap from five-generation's (5G) 20 Gbps. The latency is targeted to drop to the 0.1-1 millisecond level, and the connection density is expected to reach over a hundred devices per cubic meter \cite{10054381}. These ambitious goals, along with the need for enhanced mobility support and precise positioning, bring unprecedented challenges to antenna technologies. Traditional antenna systems are ill-equipped to meet the complex and dynamic demands of such advanced communication scenarios \cite{10041914}:
\begin{itemize}
	\item Ultra-High Data Rates. 6G anticipates data rates far surpassing 5G, crucial for applications like immersive extended reality (XR), real-time high-definition video streaming, and large-scale industrial Internet of Things (IIoT) data transfer \cite{10158439}. Antennas must handle wider bandwidths and transmit/receive signals with high spectral efficiency to achieve this.
	\item Ultra-Low Latency. For applications such as autonomous transportation and remote surgery, 6G antennas need to enable rapid signal processing and transmission, minimizing end-to-end latency.
	\item Massive Connectivity. The proliferation of IoT devices means 6G networks must support an extremely high device density \cite{9509294}. Antenna systems should distinguish between devices, manage interference, and ensure reliable communication in congested environments. Moreover, the hardware size of antennas should also be reduced to align with the increasingly miniaturized devices in such a dense connection scenario.
	\item Enhanced Mobility Support. With more high-speed transportation like autonomous vehicles, 6G antennas must maintain stable, high-quality communication under high-mobility conditions, adapting quickly to changing Doppler shifts and multipath propagation \cite{9540839}.
\end{itemize}
As the interface between the electrical signals in the communication system and the electromagnetic waves in the air, antennas directly determine the efficiency of signal transmission and reception. Their ability to adapt to different frequencies, control radiation patterns, and manage interference is crucial for achieving the high data rates, low latency, and massive connectivity that 6G demands. By developing advanced antenna technologies \cite{10379539,10753482,10906511,10232975}, we can optimize the use of the radio frequency spectrum, improve the quality of communication in diverse scenarios, and ultimately enable the full potential of 6G to be realized.
\subsection{Development and Limitations of Traditional MIMO Technology}
Multiple-Input Multiple-Output (MIMO) technology, which deploys multiple antennas at both the transmitting and receiving ends, enables parallel signal transmission through spatial dimensions. It has become one of the core technologies in 5G and 6G communication systems \cite{9665444}. This technology can significantly enhance channel capacity, link reliability, and spectral efficiency without increasing bandwidth or transmit power. Theoretical studies indicate that as the number of antennas at the transmitting and receiving ends approaches infinity, massive MIMO systems can approach the Shannon capacity limit while effectively suppressing fading effects and improving energy efficiency. In practical applications, large-scale MIMO arrays at base stations utilize beamforming techniques to achieve precise signal coverage for multiple users, effectively reducing interference and enhancing system throughput \cite{7389996}.

However, MIMO technology faces multiple challenges during its evolution towards 6G \cite{9896861}. Firstly, traditional MIMO systems rely on fixed antenna array configurations, making them difficult to adapt to dynamically changing wireless environments. {If fixed position antenna (FPA) systems are deployed at the BS, their static radiation patterns make it difficult to track rapid channel variations under drastic condition changes, such as high-mobility scenarios or rich multipath environments. Likewise, when fixed antennas are deployed at the UE, miniaturization forces antenna elements to be densely packed, which exacerbates mutual coupling and efficiency degradation and increases the difficulty of array design.} Additionally, the spectral efficiency improvement of MIMO systems depends on rich multipath environments. However, in high-frequency communications (such as millimeter-wave and terahertz bands), severe path loss and directional transmission characteristics of the channels reduce the number of multipath components, weakening the performance advantages of MIMO technology \cite{10422712}.

These technical bottlenecks have prompted academia and industry to explore more flexible antenna solutions \cite{10858129}. Reconfigurable antenna technology offers a new approach to address the limitations of MIMO systems by dynamically adjusting the radiation characteristics of antennas, such as frequency, radiation pattern, and polarization. By real-time sensing of channel conditions and adaptively optimizing antenna parameters, reconfigurable antennas can effectively alleviate spatial resource constraints and enhance the robustness of the system in complex scenarios. Thus, they have become a key technology to support the requirements of ultra-high speed, low latency, and high reliability in 6G communication.
\subsection{Reconfigurable Antennas: A Viable Solution}
Reconfigurable antennas (RAs) offer a promising solution to overcome traditional antenna limitations and meet 6G demands. These antennas can dynamically adjust operating parameters like radiation pattern, frequency, polarization, and impedance based on the communication environment or user needs. One of the primary advantages of reconfigurable antennas is their adaptability. They can direct signal energy towards the intended receiver, avoid interference, and enhance the signal-to-interference-and-noise ratio (SINR) by modifying the radiation pattern \cite{10753482}. For example, in a crowded urban area, a reconfigurable antenna can detect interference and adjust its pattern to minimize interference and maximize signal strength for the desired user. Reconfigurable antennas also enhance spectral efficiency. They can operate across different frequency bands and match impedance to the transmission medium, enabling more efficient spectrum use, which is vital for 6G. In addition, they can reduce power consumption. Instead of fixed-power omnidirectional transmission, they can adjust power and transmission direction according to the receiver's distance and location, reducing unnecessary power use and extending mobile device battery life. 

It is worth noting that intelligent reflecting surfaces (IRS) (also known as reconfigurable intelligent surface (RIS)) also provide substantial spatial and spectral control by manipulating the wireless propagation environment \cite{11007277,9982493}. {RAs and IRS both enable electromagnetic reconfigurability, but they operate at different parts of the link. RAs reconfigure the {transceiver-side effective aperture}, directly shaping the radiation/receiving patterns and providing local spatial degree of freedom (DoF) gains. In contrast, IRS reconfigures the {propagation environment} as a largely passive reflector/scatterer, creating a cascaded Tx--IRS--Rx channel whose performance depends on the two-hop path losses and cascaded-channel state information (CSI) acquisition. Thus, RAs are typically favored for transceiver-side DoF enhancement under mobility and local overhead constraints, whereas IRS is mainly used for coverage extension and blockage mitigation; the two are complementary and can be jointly deployed.}

According to the recent researches, reconfigurable antennas have significantly evolved with their main categories summarized as follows:
\begin{itemize}
	\item \textbf{Fluid Antenna (FA)}. Fluid antenna  \cite{9264694} uses fluids like liquid metals or conductive electrolytes as radiating elements. By controlling fluid shape, position, or volume, the antenna's electrical properties can be adjusted. A more generative concept of FA also includes any other antennas having fluidic radiation patterns, such as the pixel-based antenna. This flexibility makes them suitable for compact and adaptable applications such as wearables or small-form-factor sensors. However, fluid stability under different environmental conditions can affect performance, and precise control mechanisms are needed for consistent reconfiguration.
	\item \textbf{Movable Antenna (MA)}. Movable antenna  \cite{10318061} or generally known as six-dimensional MA (6DMA) \cite{11142311} changes position and/or orientation through mechanical means like motors, actuators. This adaptability is useful in scenarios with significant transmitter or receiver position changes. In mobile devices, they can be adjusted to face base stations for better signal quality. However, mechanical components add complexity, size, weight, and require additional power for operation.
	\item \textbf{Pinching Antenna (PA)}. Pinching antenna \cite{10945421} relies on mechanical deformation, such as pinching or squeezing, to alter electrical properties. They typically consist of a flexible or elastic material with conductive elements. When pinched, the shape and conductivity of these elements change, modifying the antenna's radiation pattern and operating frequency.  While offering a potentially low-cost reconfiguration method, their mechanical nature may limit response speed and reconfiguration range, and they may be prone to wear and tear.
	\item \textbf{Reconfigurable Holographic Antenna (RHA)}. Reconfigurable holographic antenna  \cite{10163760} is based on holography principles, using structures like metasurfaces or diffractive optical elements to control radiation patterns. By changing the phase and amplitude of incident electromagnetic waves, they can generate complex and reconfigurable radiation patterns, enabling advanced beam-shaping and interference mitigation. This makes them well-suited for high-capacity 6G networks, where precise radiation pattern control is crucial for efficient spectrum use. However, their design and fabrication are complex, requiring advanced techniques, and a high-precision control system to manipulate incident waves.
	\item \textbf{Other Reconfigurable Antennas}. In addition to the aforementioned categories of RAs, other designs have been developed that are capable of dynamically tuning the operating frequency, reconfiguring the radiation pattern, and altering the polarization state through the integration of active components that modify the antenna structure \cite{11039551}, such as dielectric resonator antenna \cite{9103974}, tapered slot antenna \cite{10652229} and Fabry-P\'{e}rot antenna \cite{9093995} \textit{et al.}. Such antennas can autonomously adapt to environmental variations, thereby obviating the need for multiple fixed-function antennas and reducing both implementation cost and design complexity. By leveraging these reconfigurable capabilities, wireless systems can be endowed with enhanced flexibility, improved resource utilization, and superior overall performance.
\end{itemize}

\begin{table*}[t]
		\centering
		\caption{Unified taxonomy for representative RA paradigms.}
		\label{tab:RA_taxonomy}
		\setlength{\tabcolsep}{6pt}
		\renewcommand{\arraystretch}{1.2}

	\begin{tabular}{|l|p{2.6cm}|p{2.9cm}|p{2.4cm}|p{2.4cm}|p{2.6cm}|}
			\hline
			\textbf{RA type} 
			& \textbf{Reconfiguration domain} 
			& \textbf{Control mechanism} 
			& \textbf{Spatial DoF} 
			& \textbf{Timescale / latency} 
			& \textbf{Hardware complexity} \\
			\hline
			\textbf{FA} 
			& Discrete/continuous position 
			& Radio frequency (RF) switches / microelectromechanical system (MEMS)
			& Discrete spatial diversity / radiation pattern selection 
			& Fast (typically $\mu$s--ms, implementation-dependent) 
			& Low--medium (switch network + control) \\
			\hline
			\textbf{MA} 
			& Continuous geometry (three-dimensional (3D) translation and possibly 3D rotation) 
			& Actuation/motion control with position feedback 
			& Continuous position/pose DoF
			& Slower (typically ms--s, limited by mechanical actuation) 
			& Medium--high (actuators + sensing + calibration; higher overhead) \\
			\hline
			\textbf{PA} 
			& Continuous/discrete antenna location along waveguide
			& Passive pinching / manual insertion 
			& one-dimensional (1D) position DoF + pinch-number
			& Slower (ms-s)
			& Low (simple waveguide structure) \\
			\hline
			\textbf{RHA} 
			& Continuous (or quasi-continuous) aperture field distribution 
			& Surface excitation/modulation subject to EM constraints 
			& Near-continuous field shaping; strong near-field focusing potential 
			& Fast ($\mu$s) 
			& Low \\
			\hline
		\end{tabular}
	
\end{table*}

{
Table~\ref{tab:RA_taxonomy} provides a unified conceptual framework to connect FA, MA, PA, and RHA through five common dimensions, namely reconfiguration domain, control mechanism, effective spatial DoF, reconfiguration timescale/latency, and hardware/implementation complexity. This taxonomy serves two purposes in this survey. First, it enables a coherent organization: each RA section is reviewed following the same template, starting from its reconfiguration domain/mechanism and hardware constraints, and then covering channel modeling/estimation, performance characteristics, resource allocation, and integration with other wireless technologies. Second, it supports fair cross-paradigm comparison by explicitly linking the achievable spatial DoF gains to the associated overhead and practical constraints (e.g., control, calibration, and CSI acquisition), thereby avoiding a fragmented "technology-by-technology" narrative.}

{
	Beyond the conceptual taxonomy, the practical value of RA technologies also depends on how configurations are controlled and signaled in real networks. In typical cellular deployments, RA adaptation is mainly RAN-centric: a centralized controller at the BS (e.g., a gNB) selects the RA configuration based on measurement reports and scheduling objectives, while the core network primarily provides higher-layer policy/management support and is usually not involved in the fast control loop. In practice, configuration updates are triggered by link/beam-quality changes or periodic scheduling updates, and are executed via control signaling to the RF/front-end controller. Moreover, ``per-user'' adaptation is usually realized as scheduling-timescale selection from a finite configuration set (or group-based updates), rather than continuous physical reconfiguration for each individual user. These control mechanisms also may introduce attack surfaces, which motivates integrity-protected signaling and robust measurement validation against jamming, spoofing/poisoning, and misconfiguration.}

\begin{figure*}[t]
	\centering
	\includegraphics[width=0.6\linewidth]{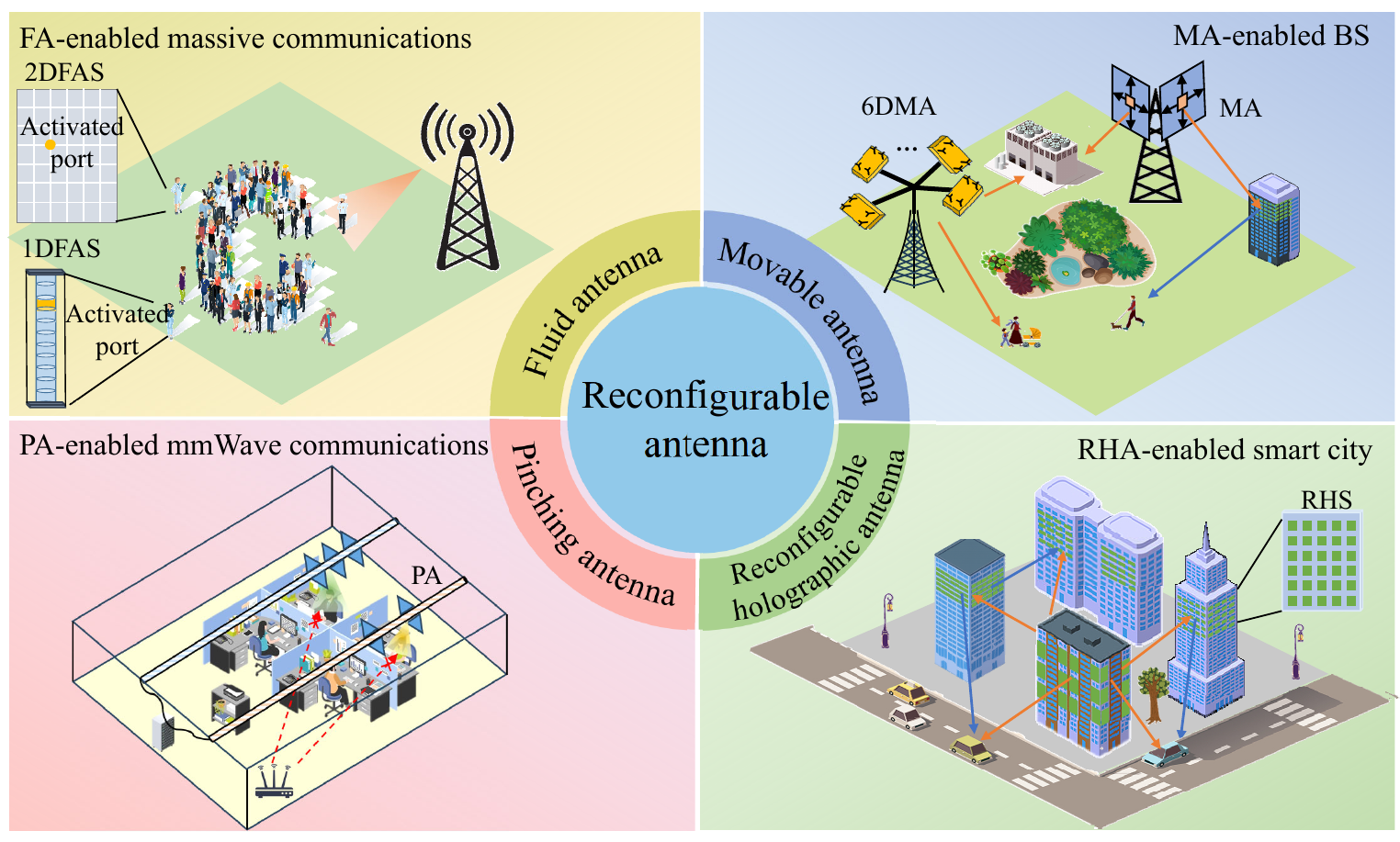}
	\caption{Application scenarios of RA-enabled wireless networks. }
	\label{scenario}
\end{figure*}

By leveraging their inherent reconfigurability, RAs offer significant opportunities to enhance the adaptability, robustness, and efficiency of future wireless networks. Fig. \ref{scenario} illustrates representative applications of the four major RA types.
\begin{itemize}
	\item \textbf{FA enabled massive communications:} FA can be deployed at either the transmitter or receiver, and dynamically adjusts the active port within a confined region to provide spatial diversity without additional antennas or complex processing. This reconfigurability suppresses interference, improves reliability, and enhances network capacity. With proper scheduling and activation, FA-assisted systems can support massive user connectivity, making them a promising solution for 6G networks.
	\item \textbf{MA enabled base station (BS):}  MAs and 6DMAs are typically deployed at the BS, where movable elements add spatial degrees of freedom by adjusting position and orientation. This enables the BS to adapt to channel variations, enhance link robustness, and improve user experience.
	\item \textbf{PA enabled millimeter-wave (mmWave) communications:} PAs have a low-complexity structure and can be flexibly installed on facades, rooftops, or indoor walls. By adjusting their position, PAs establish short-range LoS channels that bypass blockages and mitigate path loss, making them highly effective for mmWave and terahertz (THz) systems.
	\item \textbf{RHA enabled smart city:} RHAs, composed of densely packed low-cost elements, can function as either active transceivers or passive reflectors. Their versatility supports BS extension, relaying, and advanced services such as localization, vehicular sensing, and real-time tracking in urban environments, which contributes the development of future smart cities.
\end{itemize} 

\subsection{Related Surveys}
{To better position our contributions and clarify the novelty relative to prior literature, we first review representative recent tutorials and surveys on individual RA paradigms, and then highlight the limitations that motivate a unified cross-paradigm survey.} In \cite{10753482}, the authors present a comprehensive tutorial on fluid antenna systems (FAS), covering channel modelling, signal processing and estimation techniques, information-theoretic insights, novel multiple access schemes, and hardware designs. A more focused study on fluid antenna multiple access (FAMA) is presented in \cite{10568953}, where the authors detail its core mechanisms and explore its integration with emerging technologies such as IRS, MIMO systems, and artificial intelligent (AI)-based optimisation techniques. The work in \cite{hong2025surveyfa} offers a state-of-the-art survey on FAS, spanning fundamental principles to advanced applications, including quality-of-service (QoS) provisioning, power allocation, and content placement strategies. In \cite{10920759}, the authors explore the role of deep reinforcement learning (DRL) in resource allocation for FAS-enabled systems, demonstrating how DRL can effectively leverage the spatial flexibility of fluid antennas to improve signal to noise ratio (SNR), network capacity, and QoS in dynamic environments. MA have also attracted attention, as reflected in \cite{10906511}, where the authors provide a tutorial on MA-based wireless networks, including historical development, channel modelling, antenna architectures, movement optimisation, and practical prototypes. Extending this concept, \cite{11142311} introduces 6DMA, which enables 3D translation and 3D rotation to fully exploit spatial degrees of freedom. This paper also discusses corresponding channel models, hardware limitations, and potential for integration into future wireless systems. With regard to reconfigurable holographic antennas, \cite{an_tutorial_2023} establishes a solid foundation by proposing EM-compliant channel models and efficient estimation techniques for spatially continuous holographic MIMO (HMIMO) systems. The study in \cite{10130647} focuses on performance metrics such as spatial degrees of freedom, ergodic capacity, and holographic beamforming under various communication scenarios. In \cite{10130638}, the authors address practical implementation challenges- such as mutual coupling, EM field sampling, and electromagnetic information theory (EMIT)- and discuss HMIMO’s synergy with wireless power transfer (WPT), satellite communications, and integrated sensing and communication (ISAC). Finally, \cite{10232975} provides a broad perspective on enabling technologies and open research directions for HMIMO, reinforcing its potential as a key enabler for 6G networks.

\begin{figure}[t]
	\centering
	\includegraphics[width=1\linewidth]{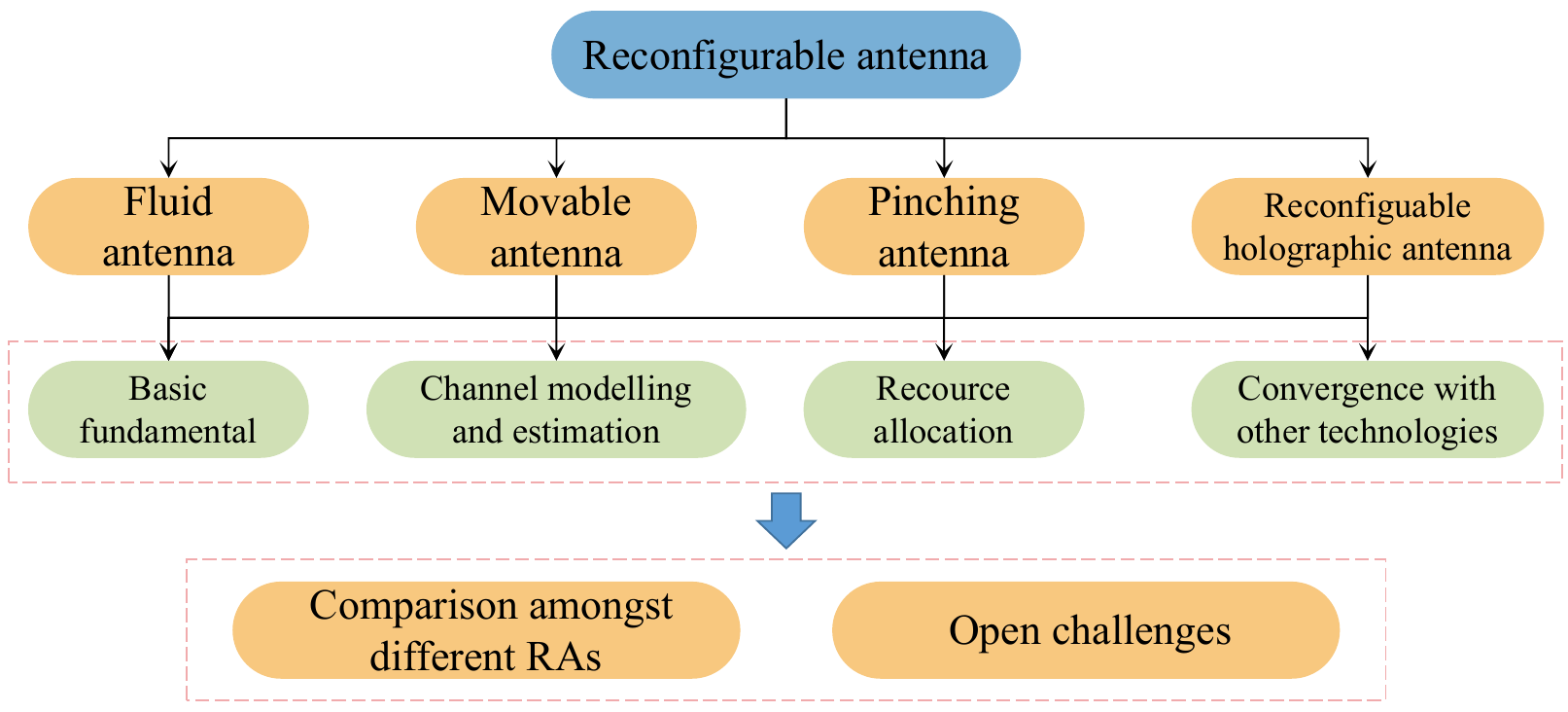}
	\caption{The organization of the tutorial.}
	\label{fig:organization}
\end{figure}

\subsection{Scope and Contribution of Our Survey}
{Although the above surveys have thoroughly investigated the applications and advantages of specific RA paradigms, a notable gap remains: there is still no comprehensive and systematic review that jointly covers the major RA families under a unified framework and enables fair cross-paradigm comparison.} Given the substantial technical potential and fertile application prospects of RAs, this paper aims to provide a comprehensive tutorial {and critical survey} on the fundamentals and recent advancements of RA technologies for next-generation mobile communications. {Unlike existing tutorials/surveys that predominantly focus on a single paradigm (e.g., FA, MA, or RHA) and thus rely on paradigm-specific assumptions and evaluation metrics, our goal is to establish a unified, system-oriented perspective that supports consistent comparison and yields actionable design insights.} Specifically, we introduce the core concepts of various RA types, covering their operating principles, channel modelling and estimation techniques, performance analysis, resource allocation strategies, and integration with other wireless technologies. Finally, we discuss open challenges {from a cross-paradigm viewpoint, with particular emphasis on practical constraints such as reconfiguration overhead, control/calibration complexity, CSI acquisition, and deployment scalability} faced by RA-enabled systems in next-generation mobile communications. Overall, this survey serves as a comprehensive reference {that goes beyond summarizing the literature by offering a unifying framework, a decision-relevant comparison, and system-level guidelines} for researchers and practitioners interested in the theory, implementation, and integration of reconfigurable antennas in 6G and beyond. The major contributions of this survey are summarized as follows.

\begin{itemize}
	\item We focus on four representative RA types- FA, MA, PA and RHA. For each RA type, we detail their operating fundamental principles and hardware design, offering a coherent foundation for understanding diverse RA technologies {and highlighting the limitations of existing single-technology tutorials in capturing cross-paradigm commonalities and trade-offs}.
	\item This survey comprehensively reviews the channel modelling, channel estimation techniques and performance characteristics associated with different RA types. {Importantly, we organize these results under a unified set of modelling/estimation considerations (e.g., near-/far-field applicability, mobility and coherence-time constraints, training/feedback overhead, and hardware non-idealities),} Furthermore, we highlight recent advancements in RA-enabled resource allocation and optimisation process {with emphasis on overhead-aware and implementable designs}.
	\item We explore how different RA types can be effectively integrated with other cutting-edge technologies, such as IRS, ISAC, semantic communication, over-the-air computing (AirComp), simultaneous wireless information and power transfer (SWIPT) and physical-layer security, enabling new capabilities in 6G and beyond {and revealing transferable design principles that are not apparent from isolated single-paradigm studies}.
	\item We conduct a comprehensive comparison of the four RA types across multiple criteria, including reconfiguration domain, reconfiguration mechanism,  reconfiguration latency, energy consumption, hardware cost, deployment role, and application scenarios. {This survey further develops a unifying comparative framework that explicitly links achievable spatial DoF/beamforming flexibility to the associated overhead (reconfiguration, training/CSI, and control/calibration), thereby strengthening the novelty of jointly analyzing FA, MA, PA, and RHA.} This comparative framework helps researchers evaluate trade-offs and choose suitable solutions for specific deployment contexts.
\end{itemize}

In this paper, we mainly consider four representative reconfigurable antennas {(FA, MA, PA, and RHA) since they cover distinct reconfiguration domains and hardware mechanisms and collectively represent the major design space of emerging RA-enabled systems}, while other reconfigurable antennas are not taken into account {and are left for future extensions}. The rest of this survey is organized as follows. Sections \ref{s2} to \ref{s5} present the four representative types of reconfigurable antennas-FA, MA, PA and RHA-in the context of next-generation mobile communications. For each RA type, we discuss the fundamental operating principles, channel modelling and estimation techniques, resource allocation strategies, and integration with other wireless technologies {under a consistent set of comparison metrics and assumptions}. Section \ref{s6} outlines key research challenges and potential future directions in the design and deployment of RA-enabled systems {with an emphasis on cross-paradigm scalability and practical implementation constraints}. The overall structure of this survey is schematically illustrated in Fig. \ref{fig:organization}. {The key
mathematical notations used in this paper are summarized in Table~\ref{tab:notations}.}

\begin{table*}[t]
	\centering
	\caption{Mathematical Notations.}
	\label{tab:notations}
	\small
	\setlength{\tabcolsep}{6pt}
	\renewcommand{\arraystretch}{1.15}
{	\begin{tabularx}{\textwidth}{|c|X|c|X|}
		\hline
		\textbf{Symbol} & \textbf{Description} & \textbf{Symbol} & \textbf{Description} \\
		\hline
		$\lambda$ & Carrier wavelength 
		& $M$ & Total number of pinching antennas \\
		\hline
		$N$ & Port number of FAS 
		& $x_p$ & Position of a pinching antenna along the waveguide \\
		\hline
		$W$ & Scaling constant of FA/FAS 
		& $c_0$ & Baseband signal fed to the waveguide input \\
		\hline
		$\mathbf{h}$ & Channel coefficient vector 
		& $\kappa$ & Mode coupling coefficient \\
		\hline
		$\mathbf{H}$ & Channel coefficient matrix 
		& $L$ & Coupling length of a pinching antenna \\
		\hline
		$\mathbf{J}$ &Spatial covariance matrix 
		& $\beta_g$ & Propagation constant of the waveguide mode \\
		\hline
		$K$ &  Rician $K$-factor 
		& $\beta_p$ & Propagation constant of the pinching-antenna mode \\
		\hline
		$L_p$ & Number of scattering paths 
		& $\beta_0$ & Wavenumber, $\beta_0=2\pi/\lambda$ \\
		\hline
		$\mathcal{C}_t,\mathcal{C}_r$ &   2D spatial regions of Tx/Rx MAs 
		& $s_{\mathrm{rad}}$ & Radiated signal from a pinching antenna \\
		\hline
		$\mathbf{t},\mathbf{r}$ & Coordinate vectors of Tx/Rx MAs 
		& $\eta$ & Combined channel gain and radiation-pattern factor \\
		\hline
		$M_t,M_r$ & Number of MAs at the transmitter and receiver 
		& $r$ & Distance between a pinching antenna and the user \\
		\hline
		$A$ & Length of the antenna moving region per dimension 
		& $x_m$ & Transmit signal at the $m$-th element \\
		\hline
		$L_t,L_r$ & Number of propagation paths at Tx/Rx 
		& $\Phi_m$ & Beamforming coefficient at the $m$-th element \\
		\hline
		$\mathbf{g}(\mathbf{t})$ & Field response vector at the transmitter 
		& $\Psi_{q\rightarrow m}$ &  EM response from the $q$-th feed to the $m$-th element \\
		\hline
		$\mathbf{f}(\mathbf{r})$ & Field response vector at the receiver 
		& $s_q$ & Source signal at the $q$-th feed \\
		\hline
	 	$\mathbf{\Sigma}$ &Path response matrix of the scattering environment 
		& $C(\mathbf{\Delta})$ & Spatial autocorrelation function \\
		\hline
		$\mathbf{v}_t,\mathbf{v}_r$ & Scatterer coordinates in near-field modeling 
	& $H_a(\cdot)$ & Complex channel gain at a wavenumber grid \\
		\hline
	$\mathbf{T}$ & Coordinate transform matrix between Tx and Rx 
		& $\mathbf{a}_r,\mathbf{a}_t$ & Receive/transmit steering vectors \\
		\hline
			$\mathcal{E}_r,\mathcal{E}_t$ & Sets of receiver/transmitter wavenumber grids 
		& $\mathbf{Y}_L$ & Uplink received pilot signal matrix \\
		\hline
	\end{tabularx}
}
\end{table*}

\section{Fluid antenna for next-generation mobile communications}\label{s2}
In this section, we focus on fluid antenna systems, including the basic principle, channel modelling and estimation as well as the resource allocation. A detailed discussion and investigation on fluid antenna is provided. 

\subsection{Basic Principle of FA}
\subsubsection{Fundamental Principle}

\begin{figure}
	\centering
	\subfigure[Single-user scenario]{\includegraphics[width=0.49\linewidth]{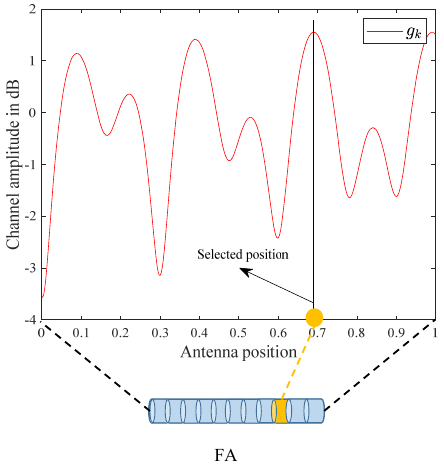}}
	\subfigure[Multi-user scenario]{\includegraphics[width=0.49\linewidth]{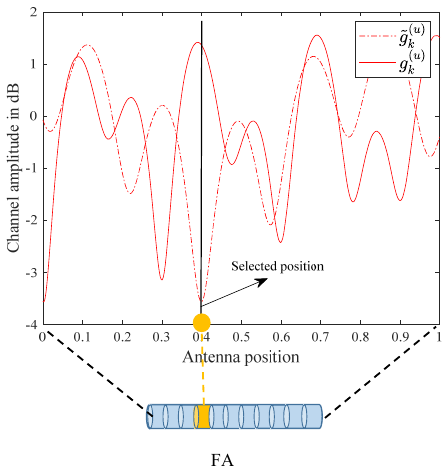}}
	\caption{ The port selection criterion based on the channel amplitude maximization for single-user scenario (a) and the SINR maximization for multi-user scenario (b). }
	\label{select}
\end{figure}
Fluid antennas  was first introduced by Wong \textit{et al.} in \cite{9264694} to denote that any software-controllable fluidic, conductive or dielectric radiating structure can change their shape or position to reconfigure the operating frequency, radiation pattern and other characteristics. It can leverage full spatial diversity within a predefined space by controlling the positions of radiating elements, which can provide additional degree of freedoms and achieve significant performance gains. Unlike MIMO or antenna selection techniques that require multiple fixed-antennas and RF chains to get more diversity gains, fluid antennas can achieve this with fewer antennas and RF chains, reducing hardware costs and power consumption. Moreover, the received signal strength can be enhanced and the interference can be alleviated through the position adjustment of FA without any signal processing. For instance, the received signal can be maximized by selecting the port with the strongest channel in single-user scenario. For multi-user scenarios, the interference of other users can be mitigated by selecting ports where the interference naturally fades due to channel variations. The port selection criteria for both single-user and multi-user scenarios are illustrated in Fig. \ref{select}. Note that the port selection strategy is not confined to the aforementioned cases. It can be adjusted according to specific scenarios to satisfy diverse performance objectives.

\begin{figure}
	\centering
	\subfigure[]{\includegraphics[width=0.25\linewidth]{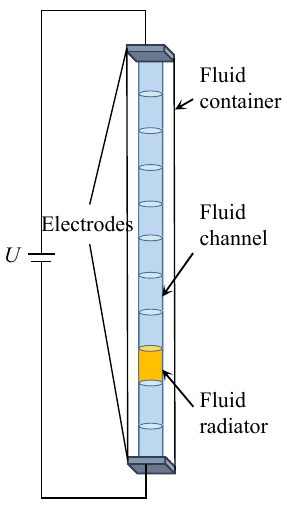}}
	\subfigure[]{\includegraphics[width=0.45\linewidth]{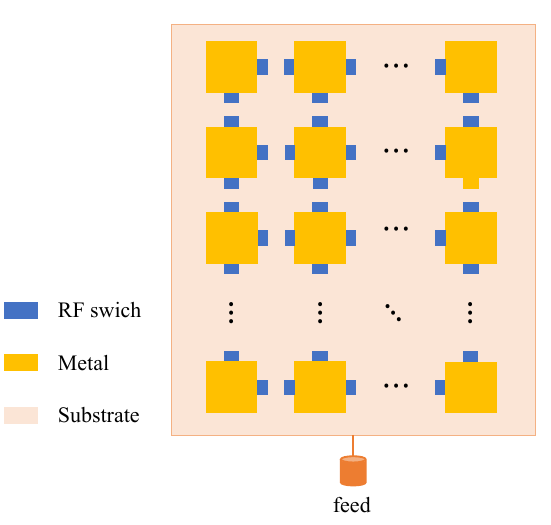}}
	\caption{The hardware design of FA: (a) liquid based FA; and (b) reconfigurable pixel based FA.}
	\label{hardware}
\end{figure}

\subsubsection{Hardware Design}
\begin{itemize}

 	\item \textbf{Liquid based fluid antenna:} The above advancements are enabled by recent breakthroughs in flexible material for antennas, such as liquid-based antennas and reconfigurable pixel-based antennas \cite{10146274}. The liquid based fluid antennas adopt metallic or non-metallic liquids as radiating elements. Most of them should be non-toxic and non-flammable. Gallium-based alloys exhibit superior thermal and electrical conductivity with low viscosity and non-toxic properties, which attract intensive research within the field of liquid antennas \cite{s20010177}. However, before FAS was proposed, most of studies about liquid antenna are focus on the reconfigurability in frequency, radiation pattern and polarization. In recent years,  flexible liquid based FAS began to be studied in \cite{9977471}, where the liquid antenna is in a dielectric holder and controlled by a micro-pump or a dedicated microelectromechanical system (MEMS). Fig. \ref{hardware} (a) shows the hardware design of liquid based FA. However, the viscosity of liquid material employed in FA can affect the flow speed and response time, limiting their use in fast-changing communication environments. Therefore, the switching speed of the antenna position need to be millisecond or sub-millisecond to unlock the full potential of FAS. 

	\item \textbf{Reconfigurable pixel based fluid antenna:} Reconfigurable pixel antenna is composed of a grid of metallic patches interconnected by RF-switches, enabling dynamic reconfiguration of its radiation characteristics through selective activation/deactivation of individual patches \cite{10035928}. The hardware design of reconfigurable pixel antenna is depicted in Fig. \ref{hardware} (b). Different RF-switches activation states correspond to different radiation patterns of pixel antenna. Although there are at most $2^L$ radiation patterns when the number of RF-switches is $L$, only a subset of these configurations is practically viable since many of patterns do not function effectively as antenna. With the assistance of RF switches, the response time of pixel-based FAS can be ignored, making it suitable for various modern communication system demands.
	\item \textbf{Meta fluid antenna:}{ A meta-fluid antenna enables “fluid-like” reconfiguration of the effective aperture without mechanical motion or pumped liquid metal. It typically combines a single-RF-chain feed with a one-dimensional (2D) programmable surface of slot-based meta-atoms. By electronically switching each meta-atom between radiating and non-radiating (and sometimes phase-shifted) states, it can rapidly generate many low-correlation spatial configurations, even at mmWave \cite{liu2025metafass}. However, it demands complex, tightly synchronized control and real-time optimisation across hundreds of switches. Moreover, mmWave operation imposes stringent hardware requirements-parasitics, thermal management, and tight fabrication tolerances can readily degrade performance.}

\end{itemize}

\subsection{Channel Modelling and Estimation}

Since FAS concentrate multi-ports over a tiny space, the channels between different ports have a strong correlation. Modelling the spatial correlation at FAS is vital to characterize and investigate the theoretical performance. Moreover, different channel models represent different implementation contexts, indicating the different directions of channel estimation. In this section, we introduce the channel modelling, channel estimation and performance analysis of FAS.   
\subsubsection{Channel Modelling}
\begin{itemize}
	\item \textbf{Simplified channel model:} Due to the small space  of FAS, the channel modelling of FAS mainly focuses on the small-scale fading. We firstly consider a point-to-point communication system where the transmitter is equipped with a fixed-position antenna and the receiver is equipped with a 1D FA.   Assuming that the FA has $N$ predefined locations (also known as ports) and has a length of $W\lambda$, where $\lambda$ is the wavelength and $W$ is a scaling constant, the channel coefficients between transmitter and receiver can be denoted as $\mathbf{h}=[h_1,\cdots,h_N]^T$, where $h_k\sim\mathcal{CN}(0,1)$ is the channel gain between the transmitter and $k$-th receiver's port. Since the ports are located much closer, the entries of $\mathbf{h}$ are strongly correlated. In \cite{9131873}, the channel gain is modeled as 
	\begin{align}
		h_k&=(\sqrt{1-\mu_k}x_k+\mu_kx_0)\notag\\
		&+j(\sqrt{1-\mu_k}y_k+\mu_ky_0),\text{for}\: k=1,\cdots,N
	\end{align}
	where $x_0,x_1,\cdots,x_N$ and $y_0,y_1,\cdots,y_N$ are all Gaussian distributed variables with zero mean and variance of $\frac{1}{2}$, and $\left\lbrace \mu_k \right\rbrace$ are correlation parameters. However, the correlation of different port is based on a reference port, which means that the spatial correlation between any two ports cannot be observed without the reference port. This will present the performance analysis of FAS in a more favorable light than it deserves. In order to depict the spatial correlation between any two ports, \cite{EL} used a common correlation parameter, which is determined by equating the correlation coefficient between any two antenna ports to the average correlation coefficient measured across an actual fluid antenna over a linear spatial space. 
	\item \textbf{Fully correlated channel model and its approximation:} Nevertheless, the above channel models may still not effectively capture the spatial correlation between FAS ports as the correlation matrix dose not exactly follow Jake’s model, resulting an overestimated performance of FAS. \cite{10103838} proposed a more accurate  correlation model that closely follows Jake’s model, named fully correlated channel model. Assuming two dimensional isotropic scattering over an linear FAS port, the covariance between $n$-th port and $m$-th port can be given by
	$J_{n,m}=\text{Cov}[h_n,h_m]=J_0(2\pi\frac{|n-m|}{N-1}W)$,
	which constitutes the covariance matrix $\mathbf{J}=[J_{n,m}]\in \mathbb{R}^{N\times N}$. Performing eigenvalue decomposition on $\mathbf{J}$, we can obtain $\mathbf{J}=\mathbf{U}\mathbf{\Lambda}\mathbf{U}^H$, where  $\mathbf{U}$ and $\mathbf{\Lambda}$ are a unitary matrix composed of the eigenvectors $\left\lbrace\mathbf{u}_n\right\rbrace $ of $\mathbf{J}$ and a diagonal matrix consisting of the eigenvalues $\left\lbrace \lambda_n \right\rbrace $ of $\mathbf{J}$, respectively. Without  loss of generality, the eigenvalues of $\mathbf{J}$ is assumed to be in descending order, i.e. $\lambda_1 \ge \cdots \ge \lambda_N$. Therefore,  the channel gain at $k$-th port can be rewritten as 
	\begin{align}\label{JK2}
		h_k=\sum_{m=1}^{N}u_{k,m}\sqrt{\lambda_m}z_m
	\end{align}
	where $u_{k,m}$ is the $(k,m)$-th entry of $\mathbf{U}$, $z_m$ is complex Gaussian distributed variable with zero mean and variance of 1.  {However, although \eqref{JK2} is convenient for simulation, it is often difficult to use for performance analysis. In particular, under this correlated model, the cumulative distribution function (CDF) of the FAS channel typically involves $N$-fold integrals, which makes analytical performance evaluation mathematically intractable \cite{10103838}. To obtain a tractable yet accurate alternative, \cite{10623405} proposes a block-correlation channel model. The correlation coefficient is assumed approximately constant within each block, while different blocks are treated as independent. This block-correlation model closely matches  \eqref{JK2} correlation behavior while significantly improving analytical tractability. }
	  
	\item \textbf{Finite Scattering Channel Model:} The above channel modelling focuses on the Rayleigh distributed channel, while the line-of-sight (LOS) component is not considered. In \cite{9953084}, a general channel model that includes a LOS path and several non-LOS paths is proposed, which can depict the microwave and  mmWave channels. The finite scattering channel model for 1D FAS can be expressed as 
	\begin{align}\label{FSCM}
		h_k&=\sqrt{\frac{K}{K+1}}e^{j\alpha}e^{-j\frac{2\pi(k-1)W}{N-1}\cos(\phi_0)\sin(\theta_0)}\notag\\
		&+\sum_{l=1}^{L_p}a_le^{-j\frac{2\pi(k-1)W}{N-1}\cos(\phi_l)\sin(\theta_l)}
	\end{align}
	where $K$ is the Rice factor, $\alpha$ is the random phase of the specular component, and $a_l$ is the random complex coefficient of the $l$-th scattered path and satisfies $\text{E}[\sum_{l}|a_l|^2]=\frac{1}{K+1}$. $\left\lbrace \theta_m\right\rbrace $ and  $\left\lbrace \phi_m\right\rbrace$ denote the azimuth and elevation angle of arrivals (AoAs) of corresponding paths. $L_p$ is the total number of scattering paths.  
	\item \textbf{Channel Model for 2D FAS:} In the previous content, we introduced 1D FAS channel models. In this part, we assume that the receiver is equipped with a planar FA, which has a size of $W=W_1\lambda\times W_2\lambda$ and $N=N_1\times N_2$ ports. Different from 1D FAS channel following Jake's model, the spatial correlation of 2D FAS channel  follows  3D Clarke’s model \cite{1622630}. For clarity, we will refer to the 2D indices of the ports as the new port indices. For instance, the $(k_1,k_2)$-th port can be mapped to a new index $n_{(k_1,k_2)}=(k_2-1)N_1+k_1$.
	Therefore, the spatial correlation between different ports can be given by
	\begin{align}
		&J_{n_{(k_1,k_2)},n_{(\tilde{k}_1,\tilde{k}_2)}}=\notag\\
		&j_0\left( 2\pi\sqrt{\left( \frac{|k_1-\tilde{k}_1|W_1}{N_1-1}\right)^2+\left( \frac{|k_2-\tilde{k}_2|W_2}{N_2-1}\right)^2 }\right), 
	\end{align}
	where $j_0(\cdot)$ is the zero-order spherical Bessel function or the
	sinc function. Similar to Jake's model, directly analyzing performance with this channel model is quite challenging. The block-diagonal approximation is also effective for this channel model, providing a more tractable approach for evaluating the performance of FAS. 
	
\end{itemize}

{
In addition to analytical modeling, measurement-driven studies have begun to validate FA channel characteristics in practice. For instance, \cite{11196946} reports a broadband THz measurement campaign for a physical two-dimensional FA operating at 300~GHz, where the channel is sampled over a dense $32\times32$ grid of candidate port positions. Based on the measured data, the spatial covariance is extracted and used to parameterize a spatially correlated 2D FA channel model. The same work also uses the measured channels to benchmark practical port/position selection strategies, providing empirical support for covariance-based FA modeling and design.
}

\subsubsection{Channel Estimation}

Accurate channel estimation is indispensable for harnessing the full potential of FAS. Conventional schemes that estimate CSI over all ports are impractical, as the large number of ports leads to prohibitive signaling overhead. Owing to the inherent spatial correlation and sparsity of FAS channels, complete CSI can instead be reconstructed from a limited subset of observed ports. In the following, we provide an overview of advanced estimation strategies specifically developed for FAS.
\begin{itemize}
	\item \textbf{Linear minimum mean-squared error (LMMSE) method:} 
	In \cite{10278813}, a skip-enabled LMMSE channel estimation (SeCE) scheme is proposed, where CSI is estimated only for a subset of FA ports, while the remaining ports are approximated by the nearest estimated neighbor based on spatial correlation. Specifically, LMMSE is used for selected ports based on pilot observations, while skipped ports inherit the CSI of the nearest estimated port. This scheme reduces signaling overhead and achieves a good balance between accuracy and efficiency, and has also been extended to cellular \cite{9992289} and full-duplex networks \cite{10184308}, though its performance is limited by both estimation and approximation errors.
	
	\item \textbf{Least square regression method:} In millimeter-wave (mmWave) FAS, a least square regression method for channel estimation is proposed in \cite{10233765}. The channel model is based on the finite scattering channel model of 1D FAS, as given in \eqref{FSCM}. In this model, if $L_p$ and the AoAs are given and $\sqrt{\frac{K}{K+1}}e^{j\alpha}$ is seemed as one variable, there are $L_p+1$ parameters unknown variables, i.e. $\left\lbrace a_l \right\rbrace $. Therefore, it is required to estimate at least $L_p+1$ ports. Assuming that there are $N_s\ge L_p+1 $  observed ports, the parameters can be estimated via least square (LS) criterion. 	Consequently, the channel can be reconstructed using the estimated parameters. However, this method assumes that the scattering path $L_p$ and AoAs are known, which is impractical.
	
	\item \textbf{L3SCR method:} The authors of \cite{10375559} propose a low-sample size sparse channel reconstruction (L3SCR) method base on the fact that mmWave channels have sparsity. This method works in three steps. In the first step, the channels of a few preset locations are estimated using LS estimation method. In the second step, the sparse parameters, including AoAs, angles of departure (AoDs) and path gains, are extracted using a compressed-sensing method. Finally, the mmWave channel can be reconstructed based on the planar-wave geometric model and estimated sparse parameters. This approach can reconstruct the full CSI  while reducing hardware switching and pilot overhead greatly.
	\item \textbf{Nyquist sampling and maximum likelihood (ML) method:} Previous methods required an exhaustive search to obtain the minimum number of estimated channels and the minimum distance between these channels for full CSI reconstruction. This will consume a lot of time and computing resources. The authors in \cite{10751774} develop a novel electromagnetic-compliant channel model for FAS and apply Nyquist sampling theory and maximum likelihood estimation (MLE) to investigate how accurately a channel can be reconstructed from a limited number of sampled locations. A key insight is that half-wavelength sampling-typically considered sufficient-is inadequate for perfect reconstruction in practical FAS scenarios due to spatial limitations and spectral leakage.
	
	\item \textbf{Successive Bayesian Reconstructor (S-BAR) method:} 
	While the aforementioned methods can approach the performance of perfect CSI with proper parameter tuning, they rely on assumptions such as slow channel variation or angular sparsity. When these conditions are violated, model mismatch leads to significant performance loss. To address this, \cite{10571302,10807122} propose the stochastic Bayesian active regression (S-BAR) framework, which models the FAS channel as a Gaussian process and employs kernel-based regression to capture spatial correlations. The estimation involves an offline stage for designing sampling patterns and regression weights using an empirical kernel, followed by an online stage for real-time reconstruction. S-BAR achieves robust performance under both matched and mismatched models with few pilots, though its reliance on an empirical covariance kernel may limit practicality.
	
	\item \textbf{Sparse Bayesian Learning (SBL) method:} In \cite{10755170}, two SBL algorithms for FAS channel estimation are proposed: a standard SBL using EM for sparse channel representation, and an improved version that exploits FAS switching to maximize mutual information gain. The latter reduces complexity by limiting iterations and pre-selecting ports, while maintaining better estimation accuracy. However, both methods depend on spatial sparsity, and their performance degrades when the channel is not sparse.
	
	\item \textbf{Learning based method:} In \cite{10495003}, an asymmetric graph masked autoencoder (AGMAE) is proposed to predict unobserved CSI from partial observations, employing transformer-based encoding and graph attention network (GAT)-based decoding to capture nonlinear mappings and spatial correlations. AGMAE achieves high extrapolation accuracy with low complexity and good generalization, but requires large-scale training and may be sensitive to dynamic environments. More recently, \cite{tang2025} introduced a diffusion model for two-dimensional FAS, enabling efficient CSI recovery from limited port observations via denoising diffusion with skipped sampling. In addition, \cite{11308117} proposed FAS-LLM, a large language model (LLM)-based architecture for orthogonal time frequency space (OTFS)-enabled satellite downlinks with FAS, which performs delay--Doppler-aware channel compression and time-series forecasting to predict future channel states.
\end{itemize}

\subsubsection{Performance Analysis}

In FAS, there are multiple preset locations to be chosen for receiving signals. The port selection strategies are mainly based on the maximum channel amplitude, or the maximum SNR, or other criterion. Therefore, the performance of FAS need to be reanalyzed according different the port selection strategies. In this section, we review the literature about the performance analysis of FAS. 

The authors of \cite{9131873} derive key performance metrics such as ergodic capacity, level crossing rate (LCR), and average fade duration (AFD) for an $N$-port FAS. In \cite{10309171}, a continuous FAS is induced and analyzed, where the antenna position follows a continuous process. The LCR and AFD are derived as closed-form expressions and the outage performance is characterized by utilizing the LCR expression. The outage probability performance of FAS  is analyzed in \cite{10375698}, where multiple antenna ports can be simultaneously activated and combined using maximum ratio combining (MRC). Under correlated Nakagami-$m$ fading channels, the outage performance of FAS is analyzed by exploiting copula theory in \cite{10253941}  and an asymptotic matching method in \cite{10308603}, respectively. The proposed performance analysis approaches significantly reduce computational effort and memory usage compared to exact analytical solutions, while maintaining favorable accuracy.  The outage probability of FAS-RIS systems is derived based on the block diagonal matrix approximation model in \cite{10858773}. 

Due to spatial diversity, FAS can innovate multiple access techniques without requiring CSI at the transmitter or successive interference cancellation at the receiver.  The inter-user interference can be significantly minimized by strategically selecting the optimal locations across the spatial domain. This innovative technique is known as FAMA. According to the FA port switching time, FAMA can be divided into two types:  fast FAMA \cite{9650760,9953084} and slow FAMA \cite{10066316,10436574,10960401,10078147,10318083}. In slow FAMA, the receiver’s antenna position changes whenever the channel changes. In contrast, fast FAMA allows the receiver to adjust its position on a symbol-by-symbol basis. The outage performance of fast FAMA and slow FAMA is analyzed in \cite{9650760} and \cite{10066316,10960401,10436574}, respectively. Specifically,  the outage probability, its upper bound and multiplexing gain of fast FAMA and slow FAMA are obtained in \cite{9650760} and \cite{10066316}, respectively. In order to make the performance analysis more tractable, the authors of \cite{10960401} provide precise and accessible closed-form approximate expressions for calculating outage probability (OP) and multiplexing gain metrics by using Gamma distribution.  However, the performance analysis is based on simplified spatial correlation models, which overestimate the performance of FAMA. In \cite{10436574}, the authors improves analytical accuracy by incorporating the approximate fully correlated channel model, which reflects real-world spatial correlations between ports.  In addition, the variations of FAMA contains opportunistic FAMA \cite{10078147} and compact ultra massive antenna array (CUMA) \cite{10318083}. Opportunistic FAMA  integrates opportunistic scheduling and FAMA to reduce the FA port number required for achieving the needed interference immunity of FAMA. The outage probability and multiplexing gain of opportunistic FAMA are obtained in \cite{10078147}. CUMA \cite{10318083} allows each UE to activate a large number of closely packed antenna ports-rather than switching among them-to receive signals in a way that constructively combines the desired signal while randomly summing interference. Each UE uses only two RF chains, and no sophisticated signal processing is required.  {However, most of the above studies analyze the theoretical performance under the assumption  of near
instant reconfiguration, perfect channel knowledge,
static or slowly varying propagation environments,
and ideal material properties, which may overestimate the FAS performance \cite{11297439}. In \cite{11184548},  an OTFS assisted FAS for satellite IoT is proposed. The outage probability and ergodic capacity are analyzed under a general channel model that accounts for aggregate interference, spatial correlation, and
imperfect-CSI port selection.    
}

\subsection{Resource Allocation}

Resource allocation is a consistent research topic to achieve desired performance objective by intelligently distribute the limited communication resources. Compared with the traditional resource allocation, the resource allocation of FAS includes an additional option of port selection, which makes it more flexible. In this section, we review the resource allocation of FAS. 
\subsubsection{Beamforming Design}
Beamforming design can enhance the performance of FAS by aligning the beam towards the target user.  The study in \cite{10303274} presents a comprehensive information-theoretic analysis of the MIMO-FAS. The rate of MIMO-FAS is maximized via joint port selection, beamforming, and power allocation. Considering only statistical CSI available, \cite{10328751} develops an alternating optimisation framework for transmit covariance and antenna positions, significantly improving ergodic achievable rate.  In  \cite{10767351}, a near-field communication FAS is studied. The energy efficiency is maximized by jointly optimizing the BS transmit beamforming matrix and the FA positions. In \cite{10794752}, the capacity region of FAS-assisted multiple access channels is studied, where joint optimisation of transmit covariance and antenna positions is employed. In \cite{10473750}, a fluid antenna-enabled multiuser MISO (MU-MISO) framework is proposed, where FA positions at the base station are dynamically optimized with transmit beamforming to maximize downlink sum-rate. The non-convex problem is addressed via an fractional programming (FP)-based algorithm, and a low-complexity zero-forcing (ZF)-based alternative is also developed. In \cite{hu2025}, a two-timescale uplink transmission framework is proposed for MIMO-FAS under Rician fading with imperfect CSI. The authors maximize the minimum user rate by optimizing the fluid antenna positions at the BS based on statistical CSI, while the transmit beamforming is adapted to instantaneous CSI using a low-complexity MRC detector. The authors of \cite{liao2025} investigate the downlink multi-user MIMO-FAS networks, where the beamforming matrices and FA positions are jointly optimized to maximize the weight sum-rate. The IRS and FAS  can significantly enhance channel conditions by jointly adjusting the positions of fluid antennas and phase shifts of IRS \cite{10992286}. In \cite{10992286}, the transmit power is minimized while satisfying SINR-based QoS constraints by jointly optimizing the transmit precoding, IRS phase shifts, and the positions of fluid antennas. The authors in \cite{zhang2025} presents a two-timescale design framework for FAS and IRS-assisted multi-user MISO downlink systems. Linear precoding is designed based on instantaneous CSI, whereas FAS port selection and IRS phase shifts are determined from long-term statistical CSI.

\subsubsection{Power Allocation}
An appropriate power allocation scheme can further improve the performance of FAS \cite{10354059,10208068,10750660}. In \cite{10354059}, energy efficiency (EE) optimisation for downlink slow FAMA is studied by formulating joint transmit power and port selection as a mean-field game (MFG). Under a FAS assisted point-to-point communication system, the EE is maximized by optimizing the power allocation, while satisfying the delay-outage probability constraint \cite{10208068}.  They model the system using queuing theory and the optimal transmit power is obtained by using a Lagrangian optimisation approach. In \cite{10750660}, a FAS-unmanned aerial vehicle (UAV)-non-orthogonal multiple access (NOMA) framework is proposed to enhance downlink sum-rate in 6G networks. The joint optimisation of UAV altitude, power allocation, and port selection is performed to maximize the sum-rate of the system.
\subsubsection{Port Selection}
In \cite{10677535}, the channel capacity of MIMO-FAS is maximized via joint port selection at the transmitter and receiver. A convex relaxation based on a capacity upper bound is formulated, and two low-complexity algorithms are proposed, both achieving significant gains over conventional MIMO and random port selection. In \cite{10388242}, a low-complexity algorithm is introduced for power minimization in a fluid antenna-enabled multiuser uplink, where BS antenna positions are optimized under users’ rate constraints. The problem is reformulated as a non-convex eigenvalue optimisation, and a projected gradient descent algorithm with closed-form gradient computation is developed, yielding superior performance to FPA and random placement.

\subsection{Convergence with Other Wireless Technologies}
FAS operates independently from other cutting-edge technologies, yet it possesses the remarkable ability to synergize with them, creating powerful outcomes that enhance overall effectiveness. This section is dedicated to the discussion of FAS convergence with other wireless technologies, such as  ISAC, SWIPT, AI \textit{et al.}. 
\subsubsection{ISAC}
A typical FAS-assisted ISAC system features a BS equipped with multiple FAs as transmitting antennas and FPAs as receiving antennas, serving multiple single antenna users while simultaneously sensing a target. The authors of \cite{11016053} study the FAS-assisted ISAC system, where the sensing SNR is maximized by jointly optimizing the dual-functional beamforming and positions of FAs under perfect CSI and imperfect CSI. In \cite{10772590}, a low-complexity algorithm is proposed for sum-rate maximization in FAS-assisted ISAC systems by jointly optimizing beamforming and antenna positions. The method integrates block successive upper bound minimization (BSUM) with proximal distance algorithm (PDA) for beamforming and extrapolated projected gradient (EPG) for antenna positioning. In \cite{10705114}, the authors investigate how fluid FAS can enhance the trade-off between sensing and communication in ISAC networks. The authors propose a novel joint optimisation of transmit beamforming and FAS port selection to minimize total transmit power while meeting both communication and sensing SNR requirements.  In \cite{10477314},  a novel learning-based framework is proposed  to enhance the performance of FAS-ISAC systems in multiuser MIMO downlink system. Specifically, the authors jointly optimize port selection and precoding to maximize the sum-rate of downlink users under sensing constraints by using deep reinforcement learning (DRL) approach. For the case of partial CSI, a masked autoencoder (MAE) is introduced to extrapolate the full CSI. For multi-target scenarios, the authors of \cite{yang2025} propose a block coordinate descent (BCD) framework with DRL based approach for intelligent port selection, enabling real-time decision making. The authors of \cite{10707252} study the FAS-ISAC system where both transmitter and receiver are equipped with fluid antennas. The transmitting beamforming and antenna locations are jointly optimized to maximize the downlink communication rate. In \cite{10949741}, a FAS assisted MIMO ISAC system is investigated, where the radar sensing signal-clutter-noise ratio (SCNR) is maximized by optimizing the transmit precoding matrix and the antenna position. In \cite{zhangtian2025}, the fundamental limits of FAS-ISAC under unsourced random access are analyzed, where a BS with fluid antennas performs joint user activity detection and signal estimation. Using finite blocklength information theory and random matrix tools, the trade-off among rate, sensing accuracy, and error probability is derived, showing clear performance gains. In \cite{ghadi2024}, a two-user FAS-ISAC system with NOMA is studied, and closed-form outage probability, ergodic communication rates, and sensing rate are obtained using Gauss–Laguerre quadrature and CRB analysis.

\subsubsection{AI}
In FAS assisted MIMO systems, the optimisation for port selection is usually high dimensional and NP-hard due to the near-continuous antenna positions.  With the rapid advances in AI, the challenges of FAS can be alleviated by using AI. In \cite{11004433}, a two-stage graph neural network (GNN) is proposed to address the problems of maximizing sum-rate and energy efficiency in multiuser MISO-FAS systems. The first stage focuses on inferring antenna positions, while the second stage is designed for optimizing beamforming vectors. In \cite{10960693}, deep unfolding neural networks are applied to optimize beamforming and antenna positions in FAS-enabled vehicular systems for weighted sum-rate maximization. By unfolding BSUM iterations into learnable layers, the method accelerates convergence and reduces complexity, outperforming conventional BSUM and fixed-position designs under vehicular channels. In \cite{zhangyl2025}, Port-Large Language Models (LLMs) is proposed as the first LLM-based method for port prediction in mobile FAS. By fine-tuning GPT-2 with Low-Rank Adaptation, it predicts future CSI from historical data and selects the optimal port to maintain channel stability under mobility. In \cite{siyun2025}, a fair resource allocation framework is proposed for UAV-based semantic communications with FAS, maximizing the minimum user rate by jointly optimizing trajectory, beamforming, semantic compression, user association, and port selection. The max-min problem is solved via alternating optimisation with Dinkelbach’s method, ant colony search, and iterative port updates.

\subsubsection{Physical Layer Security}
Deploying FAS in secure and covert communications systems can offer  great benifits. It can not only reduce the probability of detection by potential wardens but improve the secrecy rate for legitimate users by adjusting the antenna positions.  In \cite{10912516}, a FAS-enabled secure and covert communication system is studied, where antenna positions and beamforming are jointly optimized to maximize secrecy rate under covertness constraints. A penalty-based method is used for beamforming and a majorization minimization (MM) algorithm for antenna positioning. In \cite{ghadi2025}, the covertness outage is derived in FAS assisted covert communication system, where the receiver and the warden are equipped with planar FAS. The trade-off between covertness and transmission success is characterized in this paper. In \cite{10770248}, a CUMA architecture with FAS is studied for reliable and secure 6G communications. Closed-form results for SIR, outage, rate, and secrecy outage show improved reliability but potential secrecy loss when eavesdroppers adopt CUMA, which is mitigated via partial interference cancellation. In \cite{mai2024}, a secure beamforming strategy for FAS-assisted downlink NOMA is proposed, where beamforming and FA positions are jointly optimized via alternating optimisation with MM algorithm. The scheme maximizes the secrecy rate and achieves clear gains over fixed and random antenna designs.

\subsubsection{SWIPT}
SWIPT technology is a key enabler of massive Internet of Things (mIoT). Integrating FAS into SWIPT system can provide significant performance for both wireless information transfer (WIT) and WPT. The performance analysis and optimisation schemes of FAS adied SWIPT system are studied in \cite{10916605,10980171,linx2025,10506795,zhanglong2025}. Specifically, a unified analytical framework is developed for FAS-aided SWIPT systems by using stochastic geometry and copula theory in \cite{10916605}. The authors of \cite{10980171} analyze the outage probabilities of WIT and WPT of FAMA assisted SWIPT system. Then, the SWIPT outage probabilities and multiplexing gains of the proposed system are defined and evaluate under Rayleigh and Rician channel model.  In \cite{linx2025}, a FAMA-assisted WPCN under the SWIPT framework is studied, proposing four practical port selection strategies. Closed-form and approximate outage probabilities are derived under a block correlation model.  The authors of \cite{10506795} study the FAS assisted MU-SWIPT system. The beamforming vector and port selection are jointly optimized to maximize the weighted energy harvesting power, while satisfying the SINR constraint of each information receiver.  In \cite{zhanglong2025}, the authors study the FAS-assisted SWIPT system that jointly optimizes beamforming and port selection while accounting for the switching delay and energy consumption of FA movement. The short -term and long-term WPT efficiency are maximized by using traditional optimisation algorithm and DRL approach, respectively. 
\subsubsection{Index Modulation}
Index modulation is a promising solution to enhance spectral efficiency (SE) and energy efficiency (EE) by delivering additional information through the indices of antennas, subcarriers, time slots, channel states.In \cite{10440042}, a fluid antenna-assisted MIMO scheme with index modulation (FA-IM) is proposed to improve spectral efficiency without extra hardware cost. Information bits are mapped to both symbols and antenna position patterns using an optimized codebook, and a sparse Bayesian detector enables low-complexity signal detection.  However, the work in \cite{10440042} neglects the issues of correlated channel between multiple activated ports, which decreases the performance of BER. In \cite{guox2024}, a fluid antenna grouping-based index modulation (FAG-IM) scheme is proposed to mitigate performance loss from spatial correlation in FAs. Ports are divided into groups with one active port per group, improving spatial separation and robustness. A closed-form average bit error probability (ABEP) upper bound is derived, and a low-complexity S-AMP detector is developed. Results show that FAG-IM outperforms FA-IM and MIMO-FAS. The authors of \cite{10640106} investigate advanced channel coding strategies for FA-IM system. The authors derive a closed-form BER expression under spatially correlated fading, then enhance performance using set partition coding (SPC) and turbo-coded modulation to combat spatial correlation. The authors of \cite{10653737} propose a novel position index modulation (PIM) scheme for FAS. Analytical expressions for BER and data rate are derived by considering general fading conditions and imperfect channel estimation.

\section{Movable antenna for next-generation mobile communications}\label{s3}
In this section, we first introduce the fundamental principles of movable antenna (MA). Then, we investigate the channel modelling, channel estimation techniques, and performance analysis related to MA. Next, we present various resource allocation strategies tailored for MA-assisted systems. Finally, we discuss the integration of MA with other advanced wireless technologies.
\subsection{Basic Principle of MA}
\subsubsection{Fundamental Principle}
MA represents a novel antenna technology that enhances wireless communication performance by allowing physical adjustments in location or orientation \cite{10286328, shao20256dma}.  Unlike traditional fixed-position antenna (FPA), which is limited in their ability to adapt to dynamic propagation environments, {MA exploits a continuous reconfigurable domain by actively repositioning themselves within a predefined spatial region to better align with favorable channel conditions}. This spatial adaptability enables several key performance benefits, including improving received signal strength, more flexible and accurate beamforming, interference suppression, and enhancing spatial multiplexing capabilities \cite{10278220, 10354003}. Specifically, MA introduces an additional spatial degree of freedom (DoF) {(ranging from 1D to 6D)} into wireless systems, which augments conventional antenna processing techniques by leveraging physical displacement. This extra DoF facilitates finer beam steering and dynamic spatial alignment, thereby contributing to significant gains in both spectral efficiency and energy efficiency. {In practical deployment, such capabilities allow MAs to be integrated into diverse platforms including base stations, UAVs, vehicles, and high-speed railways\cite{10741192,11162069,9770353,yi2024performanceanalysisxlmimorotary}, where they leverage physical position and orientation adjustments to fully exploit these spatial degrees of freedom for enhanced beamforming accuracy, interference suppression, and adaptability to dynamic channel conditions.}As a result, MA offers a promising hardware-level approach to meeting the increasing performance demands of next-generation wireless networks.

\begin{figure}[!t]
	\centering
	\includegraphics[width=0.7\linewidth]{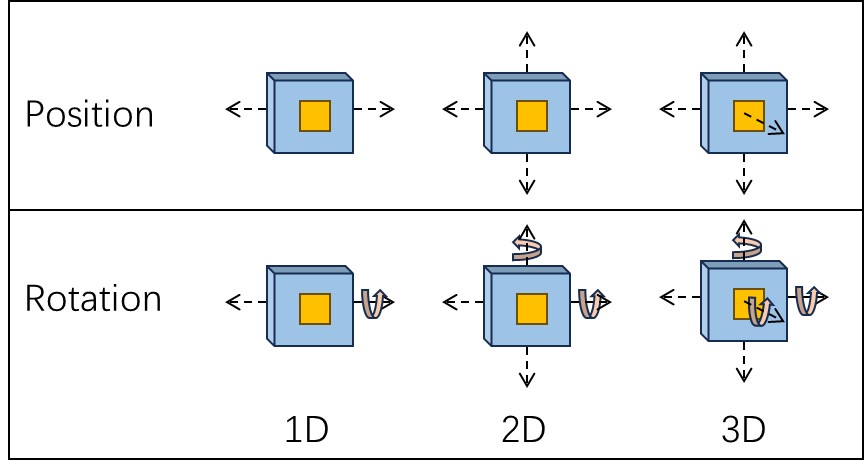}
	\caption{Antenna movement model for MAs.}
	\label{fig:MA types}
\end{figure}
As shown in Fig. \ref{fig:MA types}, based on the dimensionality of the movement space, MA can be classified into two types as follows:
\begin{itemize}
	\item \textbf{Position Adjustable MA:}
	In a one-dimensional movable antenna (1DMA) array, elements move along a linear track, allowing joint optimisation of the antenna position vector (APV) and antenna weight vector (AWV) to improve beamforming \cite{10278220}. Under proper conditions, this achieves full array gain toward the target while steering nulls toward interference, overcoming the trade-off faced by fixed-position arrays. Extending mobility to a two dimensional plane, 2DMA enables joint optimisation of APV, receive combining, and user powers, significantly
	improving the max-min user rate\cite{10741192}. Compared with 1DMA, 2DMA provides more DoF, finer beam control, and stronger interference suppression. Further extending
	to three dimensions, 3DMA allows full 3D placement of
	antennas, enhancing beamforming accuracy, spatial multiplexing, and adaptability to complex channels \cite{10354003}.
	\item \textbf{Six-Dimensional MA (6DMA):}
	{6DMA has been proposed as a reconfigurable transceiver architecture in~\cite{shao20246d,shao20256d,10737418,shao20256dma,shao2025distributed}. It provides additional spatial degrees of freedom and can enhance MIMO capacity without deploying extra antenna elements. A 6DMA-equipped transceiver integrates multiple independently adjustable antenna surfaces. Each surface supports 3D translation and 3D rotation, enabling flexible position--orientation reconfiguration. The CPU adapts the surface configurations according to user spatial distributions and the associated long-term CSI, so as to leverage spatial variability for improved propagation conditions. Each surface is connected to the CPU via an extendable and rotatable arm with flexible wiring, which supplies power and carries RF/control signals.}
\end{itemize}

\subsubsection{Hardware Design}
\begin{figure}[!t]
	\centering
	\includegraphics[width=0.7\linewidth]{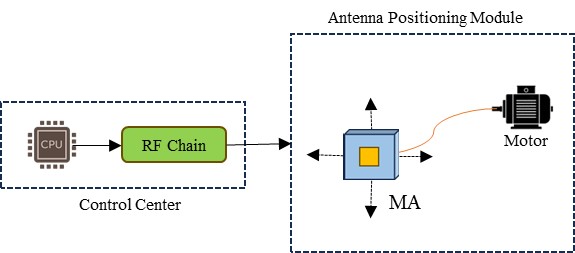}
	\caption{Structure of MA unit.}
	\label{MA structure}
\end{figure}
{In terms of hardware deployment, the practical feasibility of MA hinges on the utilization of mechanical displacement.} Fig. \ref{MA structure} illustrates the architecture of an MA unit installed at the transmitter (Tx) or the receiver (Rx) moving in 2D space, which  typically consists of four parts: a control unit, a RF link, antenna elements, and a drive structure. The control unit usually utilizes CPU for digital signal processing and controlling the position of the antenna, while the antenna is connected to the CPU via radio frequency chain to achieve information transmission. As for the drive structure, it mainly includes the drive motor and the mechanical sliding rail. After receiving the signaling from the CPU, the drive motor controls the antenna component to adjust its position on the mechanical sliding rail. 

{Although mechanically driven MAs offers cost advantages through mature fabrication processes, it faces inherent limitations in response latency and operational maintenance. In terms of speed, performance depends heavily on the driving mechanism: while compact  MEMS-based solutions\cite{6416007} can achieve rapid response times ranging from microseconds to milliseconds, conventional motor-based systems typically lag between milliseconds and seconds. As noted in \cite{11100705}, this electromechanical delay driven by stepper motors can significantly encroach upon valid data transmission windows. Physically, the integration of motors and rails expands the antenna's form factor (volume and weight) and imposes constraints on wiring and trajectory planning to mitigate coupling effects. Consequently, despite lower upfront costs, the susceptibility to mechanical wear necessitates frequent calibration, resulting in higher long-term maintenance burdens compared to static electronic counterparts.}

{Regarding 6DMA, higher dimensions entail increased deployment complexity. To balance scalability and hardware complexity,~\cite{shao20256d,ning2025movable} propose hybrid architectures that combine fixed and movable arrays. Spatial scalability can also be enhanced by deploying MAs on unmanned aerial vehicles (UAVs) to extend coverage~\cite{liu2024uav,ren20256d}. However, as the number of 6DMA units grows, the control and processing burden of a centralized CPU scales up rapidly. To mitigate this centralized bottleneck, distributed intelligence has been advocated in~\cite{shao2025distributed,hua2025hierarchically}. Accordingly, two representative coordination paradigms are commonly considered for 6DMA: centralized and distributed architectures.}
\begin{itemize}
	\begin{figure}
		\centering
		\includegraphics[width=1\linewidth]{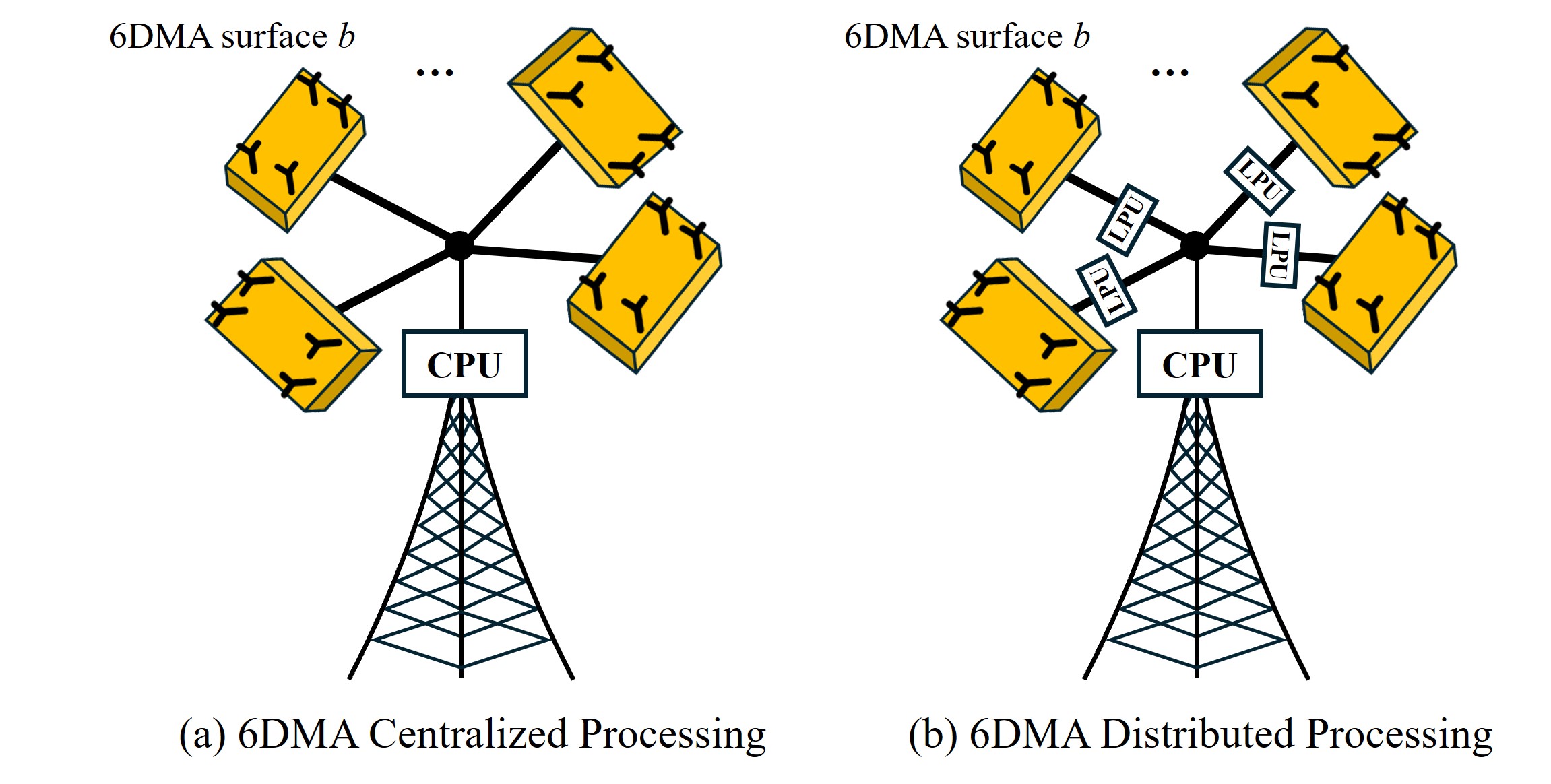}
		\caption{Illustration of Different 6DMA processing architectures.}
		\label{6DMA Architectures}
	\end{figure}
	
	\item \textbf{6DMA Centralized Architecture: }
	As illustrated in Fig. \ref{6DMA Architectures} (a), {the centralized architecture from \cite{shao20246d} uses multiple 6DMA surfaces based on uniform planar arrays (UPAs),} each connected to the BS CPU via a rotatable mechanical rod with flexible wiring for power and control. The CPU uniformly controls the 3D position and 3D rotation of each surface and jointly processes all signals, assuming statistical CSI is available.
	
	\item \textbf{6DMA Distributed Architecture: }
	As illustrated in Fig. \ref{6DMA Architectures} (b), to overcome high computational cost, delay, and CPU burden in centralized setups, \cite{shao2025distributed} proposes a distributed architecture. Each 6DMA surface has a local processing unit (LPU) for tasks like channel estimation and precoding/detection. The CPU in the BS is responsible for optimizing position and rotation, while the LPUs are connected to the CPU via flexible cables, enabling distributed and parallel processing.
\end{itemize}

The performance, precision, and scalability of MA systems heavily depend on the design of the drive mechanism. Various driving mechanisms have been proposed to physically reposition the antenna element in MA systems. The main types of feasible methods for implementing MA architecture are mechanically driven\cite{ning2025movable}.

In reconfigurable antenna systems, one common approach for enabling physical movement or rotation of the antenna is the use of external mechanical structures equipped with actuators that convert control signals and electrical energy into torque or displacement. These actuators-such as stepper motors, digital servo motors, precision gears, motor drive shafts, or MEMS-allow for accurate positioning and orientation of antenna elements, thereby supporting dynamic adaptation to varying communication environments\cite{9770353}.

Alternatively, rotational adjustment of antenna elements can be achieved using servo motors or other rotational actuators. For example, in \cite{dai2025rotatable}, dynamic adjustment of a directional antenna is implemented using a two-dimensional digital servo motor controlled by a microcontroller. The servo motor adjusts both the azimuth and elevation angles by converting pulse signals from the microcontroller into mechanical torque. In \cite{shao20246d}, each rod in a 6D movable antenna structure is equipped with two rotating motors at both ends, enabling contraction or extension to adjust the position and orientation of each 6DMA surface relative to the CPU. A simplified design is proposed in \cite{zhou2025rotatable}, where a rotatable 6DMA system supports only 3D rotational adjustment using a single rotary motor for each transceiver. Although this design sacrifices some performance compared to fully movable 6DMA systems, it significantly reduces deployment, operation, and maintenance costs by minimizing the number of motors required per antenna unit.

\subsection{Channel Modelling and Estimation}
\subsubsection{Channel Modelling}
The wireless channel, serving as the physical medium connecting the transmitter and receiver, dictates signal fading, multipath effects, and interference levels, thereby profoundly impacting system coverage, data rates, and reliability. In this section, we introduce the field-response channel model for MA systems based on the general 2D framework. For simplicity, we star with the basic field-response channel model of the MA position in narrowband single-input single-output (SISO) systems under far-field conditions, and then extends it to MIMO systems and near-field channel models.
\begin{itemize}
	\item \textbf{Field-Response Channel Model:}
	We study a MA-assisted communication system, which MAs are deployed at both the Tx and the Rx. Flexible cables link each MA to an RF chain, allowing the antennas to move within a local 2D area $\mathcal{C}_t$ or $\mathcal{C}_r$, and the coordinates of the transmit MA and the receive MA are denoted as $\mathbf{t} = [x_t, y_t]^T \in\mathcal{C}_t$ and $\mathbf{r} = [x_r, y_r]^T \in\mathcal{C}_r$ respectively. For simplicity, the 2D area $\mathcal{C}_t$ and $\mathcal{C}_t$ are both simplified as an $A \times A$ square area. With $A$ as the length of the antenna moving region per dimension at the Tx/Rx and $\lambda$ as the carrier wavelength, the Rayleigh distance is $2A^2/\lambda$. If separations between the Tx and Rx as well as that between
	the Tx/Rx and scatterers exceed this distance, the far-field condition applies, and a planar wave model can be used. Thus, the overall channel response arises from the combined coefficients of $L_t$ paths originating from the Tx and $L_r$ paths arriving at the Rx\cite{10318061}. Consequently, the field responses vectors for the Tx/Rx are given by:
	\begin{align}\label{FRV}
		\mathbf{g}(\mathbf{t})= [e^{j \frac{2\pi}{\lambda} \rho_{t,1}(\mathbf{t})}, e^{j \frac{2\pi}{\lambda} \rho_{t,2}(\mathbf{t})},\ldots, e^{j \frac{2\pi}{\lambda} \rho_{t,L_t}(\mathbf{t})}]^T,\notag\\
		\mathbf{f}(\mathbf{r})= [e^{j \frac{2\pi}{\lambda} \rho_{r,1}(\mathbf{r})}, e^{j \frac{2\pi}{\lambda} \rho_{r,2}(\mathbf{r})},\ldots, e^{j \frac{2\pi}{\lambda} \rho_{r,L_r}(\mathbf{r})}]^T,
	\end{align}
	where $\rho_{t,i}(\mathbf{t})$ and $\rho_{r,l}(\mathbf{r})$ are the propagation phase difference between the TX/RX and the reference point, $\mathbf{r}_0 = [0, 0]^T$.
	For MA-assisted MIMO systems, the Tx and Rx are equipped with $M_t$ and $M_r$ MAs, respectively. Then, the channel matrix from the Tx to Rx is denoted as
	\begin{align}\label{MIMO_channel}
		\mathbf{H}(\tilde{\mathbf{t}},\tilde{\mathbf{r}})=\mathbf{F}(\tilde{\mathbf{r}})^\mathrm{H} \mathbf{\Sigma}\mathbf{G}(\tilde{\mathbf{t}}),
	\end{align} 
	where $\tilde{\mathbf{t}} =  \left[\mathbf{t}_1,\mathbf{t}_2,\cdots, \mathbf{t}_{M_t}\right]^T$ and $\tilde{\mathbf{r}} =  \left[\mathbf{r}_1,\mathbf{r}_2,\cdots, \mathbf{r}_{M_r}\right]^T$, $\mathbf{G}(\tilde{\mathbf{t}})$ and $\mathbf{F}(\tilde{\mathbf{r}})$ are the field response of all the MAs at the TX and RX, respectively, and $\mathbf{\Sigma} \in \mathbb{C}^{L_r \times L_t}$ denotes the path response. 
	
	\item \textbf{Near-Field Channel Model:}
	In the far-field channel model, the channel response mainly depends on the AoD and AoA. However, when the antenna is deployed in a compact area or close to the transceivers, the near-field effect becomes non-negligible. In this case, the spherical characteristics of the wavefront must be accurately captured, and the channel response depends not only on the angles but also on the absolute positions and distances between the transceivers and the antennas. Therefore, we consider adopting the near-field channel model of uniform spherical waves in a stationary propagation environment to accurately characterize the position-dependent channel changes caused by antenna motion. 	
	Correspondingly, The non-line-of-sight (NLoS) paths between the Tx-MA and Rx-MA are formed through reflections and scattering caused by surrounding environmental objects. The near-field response vector at the Tx/Rx can be obtained as 
	\begin{align}\label{near_FRV}
		\mathbf{g}_\mathrm{near}(\mathbf{t})= [e^{j \frac{2\pi}{\lambda} \lVert \mathbf{t} - \mathbf{v}_{t,1} \rVert_2},\ldots, e^{j \frac{2\pi}{\lambda} \lVert \mathbf{t} - \mathbf{v}_{t,L_t} \rVert_2}]^T,\notag\\
		\mathbf{f}_\mathrm{near}(\mathbf{r})= [e^{j \frac{2\pi}{\lambda} \lVert \mathbf{r} - \mathbf{v}_{r,1} \rVert_2}, \ldots, e^{j \frac{2\pi}{\lambda} \lVert \mathbf{r} - \mathbf{v}_{r,L_r} \rVert_2}]^T,
	\end{align}
	where $\mathbf{v}_{t,i} $ for $1\leq i \leq L_t$, $\mathbf{v}_{r,l}$ for $1\leq l \leq L_r$ represent the coordinates of the $i$-th point scatterers at the Tx side and the $l$-th point scatterers at the Rx side, respectively. Then, let $\mathbf{\Sigma}_\mathrm{NLoS}$ denote the path response matrix of the NLoS paths between the Tx and Rx, so the channel vector from the Tx to Rx is denoted as
	\begin{align}\label{NLoS_channel}
		\mathbf{h}_\mathrm{NLoS}(\mathbf{t},\mathbf{r})=\mathbf{f}_{near}(\mathbf{r})^\mathrm{H} \mathbf{\Sigma}_\mathrm{NLoS}\mathbf{g}_{near}(\mathbf{t}),
	\end{align} 
	As for the LoS path, the channel vector is represented as 
	\begin{align}\label{LoS_channel}
		\mathbf{h}_\mathrm{LoS}(\mathbf{t},\mathbf{r})=\eta e^{j \frac{2\pi}{\lambda} \lVert \mathbf{r}_0 + \mathbf{T}^T \mathbf{r} - \mathbf{t} \rVert_2},
	\end{align}
	where $\eta$ is the amplitude of the LoS channel gain, $\mathbf{r}_0$ is the coordinates of the reference point for the Rx region at the Tx side, and $\mathbf{T}$ is the coordinate transform matrix between the	Tx and Rx.
	Finally,  the channel coefficient based on near-field is obtained as 
	\begin{align}\label{near_channel}
		\mathbf{h}_\mathrm{near}(\mathbf{t},\mathbf{r})= \mathbf{h}_\mathrm{NLoS}(\mathbf{t},\mathbf{r}) + \mathbf{h}_\mathrm{LoS}(\mathbf{t},\mathbf{r}).
	\end{align}
	\begin{figure}[!t]
		\centering
		\includegraphics[width=0.7\linewidth]{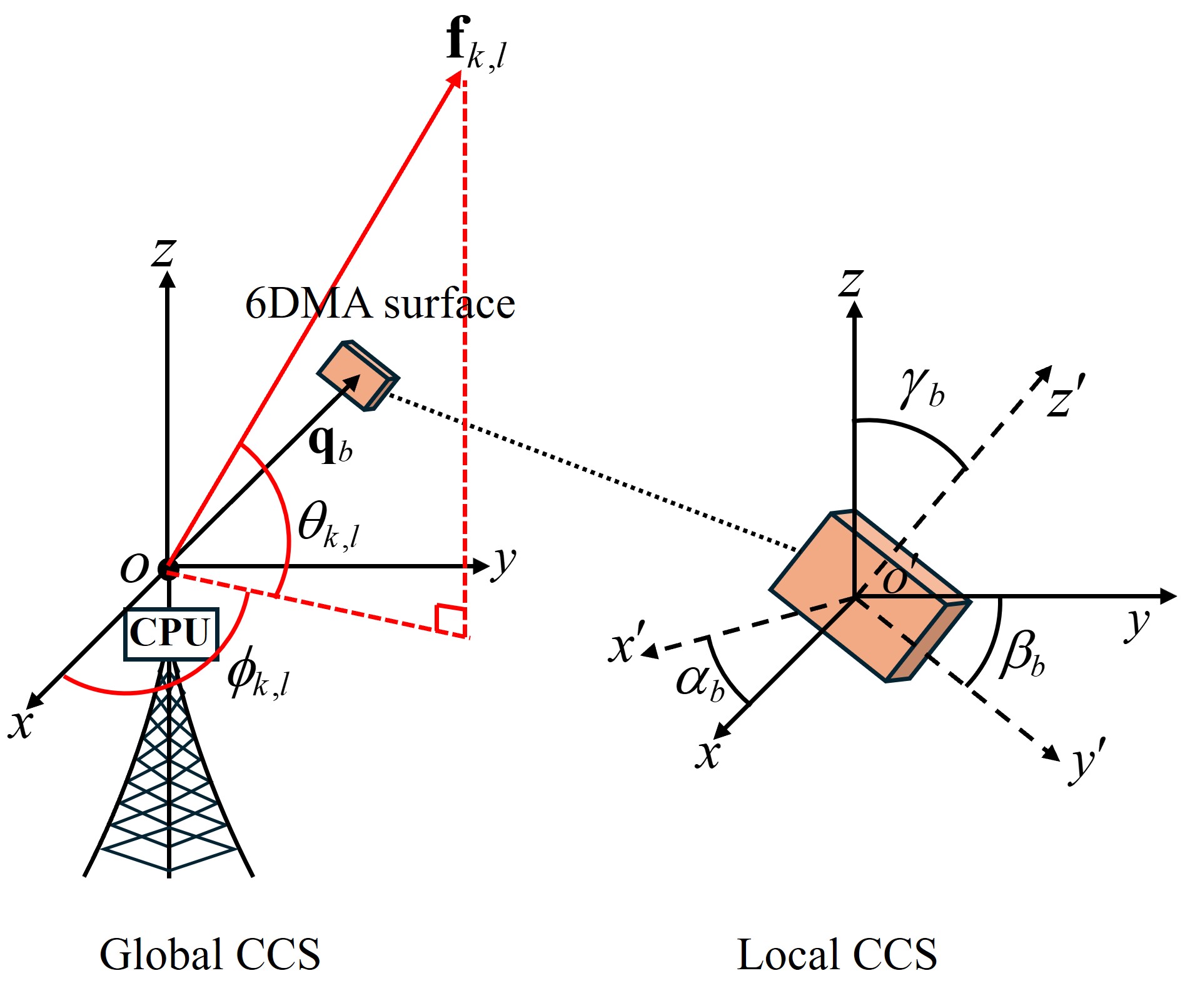}
		\caption{Illustration of the geometry of the 6DMA channel model.}
		\label{6DMA Channel Model}
	\end{figure}
	\item \textbf{Basic 6DMA Channel Model:}
	{As shown in Fig.~\ref{6DMA Channel Model}, the global Cartesian coordinate system (CCS) of the 6DMA system is denoted by $o$-$xyz$, where the origin $o$ is located at the system reference position. The 6DMA system comprises $B$ reconfigurable surfaces, and the $b$-th surface is equipped with $N\ge 1$ antenna elements. For the $b$-th surface, a local CCS $o^{\prime}$-$x^{\prime}y^{\prime}z^{\prime}$ is introduced, whose origin $o^{\prime}$ is located at the geometric center of the surface. Accordingly, the surface placement can be described in the global CCS, while the antenna-element coordinates are specified in the corresponding local CCS.
	The effective channel between the 6DMA surface and the user is characterized by three components: (i) the 6D steering vector of each 6DMA surface, (ii) the effective antenna gain, and (iii) the propagation channel between the surface and the user~\cite{shao20256dma}.
		
	The 6D steering vector of the $b$-th 6DMA surface for the $l$-th path of the $k$-th user can be expressed as}
	\begin{align}
		\mathbf{a}_{b,k,l}(\mathbf{u}_b)=[e^{-j\frac{2\pi}{\lambda}\mathbf{f}_{k,l}^T\mathbf{r}_{b,1}(\mathbf{q}_b,\mathbf{u}_b)},\cdotp\cdotp\cdotp, \notag\\
		e^{-j\frac{2\pi}{\lambda}\mathbf{f}_{k,l}^T\mathbf{r}_{b,N}(\mathbf{q}_b,\mathbf{u}_b)}]^T,
	\end{align}
	where $\lambda$ denotes the carrier wavelength,  $\mathbf{r}_{b,n}(\mathbf{q}_b,\mathbf{u}_b)$ denotes the position of the $n$-th antenna in the $b$-th 6DMA surface in the global CCS,  $\mathbf{f}_{k,l}$ denotes the pointing vector.
	
	{The effective gain of each antenna on the $b$-th 6DMA surface is defined as $A(\tilde{\theta}_{b,k,l},\tilde{\phi}_{b,k,l})$, where $(\tilde{\theta}_{b,k,l},\tilde{\phi}_{b,k,l})$ represents the corresponding arrival direction of the signal in the local CCS (the specific form is determined by the radiation pattern of the antenna used).} Then, the linear proportion of the effective antenna gain in the direction $(\tilde{\theta}_{b,k,l},\tilde{\phi}_{b,k,l})$ can be defined as
	\begin{equation}
		g_{k,l}(\mathbf{u}_b)=10^{\frac{A(\tilde{\theta}_{b,k,l},\tilde{\phi}_{b,k,l})}{10}},
	\end{equation}
	
	Assume that there are $L_k$ channels between 6DMA-BS and users. The $l$-th channel between $k$-th  user and the $b$-th 6DMA surface antenna can be expressed as
	\begin{equation}
		\mathbf{h}_{k,b}(\mathbf{q}_b,\mathbf{u}_b)=\sum_{l=1}^{L_k}\eta_{k,l}\sqrt{g_{k,l}(\mathbf{u}_b)}\mathbf{a}_{b,k,l}(\mathbf{u}_b),
	\end{equation}
	where $\eta_{k,l}$ denotes the channel coefficient that the $k$-th user reaches 6DMA-BS through the $l$-th path, $g_{k,l}(\mathbf{u}_b)$ denotes the effective antenna gain of the $k$-th user and the $l$-th channel of the $b$-th 6DMA surface, and $\mathbf{a}_{b,k,l}(\mathbf{u}_b)$ denotes the 6D steering vector of the $b$-th 6DMA surface that receives the $k$-th user signal through the $l$-th path.

\end{itemize}

{
Beyond analytical channel modeling, MA has been validated by hardware prototyping and channel measurements. In \cite{11224420}, the authors built a movable-antenna testbed with closed-loop motion control and performed over-the-air measurements at sub-6~GHz and mmWave bands. This testbed helps characterize position-dependent channel variations under practical hardware constraints. Moreover, \cite{wang2024movable} developed a broadband THz measurement platform and sampled the channel over a dense 2D grid of candidate antenna locations. The measured statistics were then used to parameterize spatially correlated channel models for MA, providing measurement-grounded modeling support.}

\subsubsection{Channel Estimation}
Conventional channel estimation techniques, designed for fixed-position antennas, often rely on stationary angular features and assume quasi-static propagation environments.  However, in MA-assisted systems, the dynamic repositioning of antenna elements introduces new challenges, such as time-varying channels, location-dependent fading, and increased estimation dimensionality. The channel response becomes highly sensitive to the absolute position of the antenna, invalidating traditional angle-domain methods. Despite these challenges, the controllable mobility of MA also enables new opportunities.  By sampling the channel across different spatial locations, richer training data can be acquired, facilitating high-resolution channel reconstruction. Recent studies have explored compressed sensing and position-assisted estimation methods to demonstrate promising performance gains in estimation accuracy.

\begin{itemize}
	\item \textbf{Compressed sensing based channel estimation:}
	In an MA-assisted communication system based on a multipath channel model, compressed sensing (CS) is used to estimate key multipath components, including the AoD, AoA, and complex gain of each path\cite{10571235}. {However, the computational complexity of this method is typically high, scaling as $\mathcal{O}(L \cdot M \cdot G)$, where $G$ denotes the grid size of the angular domain. Consequently, achieving high-resolution estimation requires a dense grid, which significantly increases the computational burden.} The accuracy of channel estimation depends heavily on the measurement positions of the Tx-MA and Rx-MA, as these positions define the measurement matrix. Furthermore, the CS-based framework mainly considers the SISO case. In more complex systems, such as multiple access MIMO, {the prohibitive complexity may render it inapplicable}, leading to poor estimation accuracy.
	
	\item \textbf{Tensor decomposition based channel estimation:}
	In\cite{10659325}, a novel channel estimation and reconstruction method is proposed for MA-enabled MIMO systems based on tensor decomposition techniques. A two-stage training procedure is designed, where the transmitter and receiver perform successive antenna movements. This results in pilot signals that can be naturally arranged into a third-order tensor. By applying canonical polyadic (CP) decomposition, the factor matrices are extracted to estimate the multipath components. {Unlike CS-based methods that suffer from grid-dependent complexity, the computational cost of this approach is determined by the efficient CP decomposition algorithm (e.g., alternating least squares).} Consequently, the full channel can be reconstructed for arbitrary Tx/Rx MA positions. As a result, the proposed tensor decomposition-based method offers significant advantages in channel estimation accuracy, pilot overhead reduction, { and computational efficiency}.
\end{itemize}
To obtain the optimal 6DMA position and rotation, CSI between all 6DMA candidate positions/rotations and all users is essential. However, 6DMA simultaneously adjusts position and rotation, significantly increasing the pilot overhead of channel estimation. Aside from the cost, different 6DMA surfaces may exhibit very different channel distributions with users due to differences in position and rotation. The existing literature mainly focuses on two methods used for 6DMA channel estimation, namely statistical 6DMA channel estimation and instantaneous 6DMA channel estimation.
\begin{itemize}
	\item \textbf{Statistical 6DMA Channel Estimation:}
	{
		In \cite{shao2025directional}, a directional sparsity property in 6DMA channels is identified, indicating that notable gains are attained only for a subset of position-rotation pairs. The work in \cite{jiang2025statistical} implements statistical channel estimation (SCI) via a three-stage protocol: (I) The BS collects channel measurements at a limited number of training position-rotation pairs using user pilots; (II) Multipath parameters (e.g., DOA, power) are estimated via orthogonal matching pursuit (OMP) algorithm, and SCI is reconstructed under a sparsity prior to reduce overhead; (III) Optimized SCI guides antenna positioning and rotation to enhance performance. \cite{shao2025hybrid} further introduced a hybrid-field THz channel model, which exploits directional sparsity and the non-stationary service nature of users.}
	
	\item \textbf{Instantaneous 6DMA Channel Estimation:}
	{
		Instantaneous channels can be fast time-varying, leading to increased computational burden. To mitigate this, \cite{11142311} uses a distributed 6DMA architecture,  where LPUs concurrently estimate instantaneous channels. After statistical optimization, the 6DMA surfaces are configured accordingly. The instantaneous estimation is achieved via the sparse model, focusing only on valid channel regions. A sparsity-constrained least-squares algorithm ignores invalid entries (e.g., zero-gain channels), reducing pilot length via about 50$\%$ and improving accuracy. The computational complexity of the proposed algorithm can be shown to be $\mathcal{O}(L^{2}KM_{\mathrm{g}}+M_{\mathrm{a}}G)$, where $M_{\mathrm{g}}$ represents the number of position-rotation pairs per group, $M_{\mathrm{a}}=\max\{M_{1},M_{2},\cdots,M_{K}\}$. Compared to the covariance-based exhaustive measurement approach, the proposed algorithm exhibits a slightly higher normalized mean square error (NMSE). Furthermore, this algorithm achieves significantly higher estimation accuracy than traditional LS algorithm. This is because the proposed  algorithm utilizes the directional sparsity from Stage I, thereby requiring fewer measurements than the traditional LS method.}
\end{itemize}

\subsubsection{Performance Analysis}
Recent studies have demonstrated that MAs can significantly enhance wireless system performance by exploiting spatial diversity and enabling dynamic channel adaptation. A number of theoretical works have established analytical frameworks to quantify these performance gains under different channel models and system configurations.

Through rigorous mathematical modelling and derivation, \cite{10318061} systematically investigates the fundamental performance limits of MA systems. The authors analyze the statistical characteristics of the channel gain, the maximum achievable gain within a given movement region, the average SNR improvement due to antenna mobility, and the corresponding outage probability. Their analysis spans multiple channel conditions, including both LoS and multipath fading scenarios, and considers the impact of different numbers of propagation paths. These results provide a solid theoretical basis for understanding how spatial mobility enhances link robustness, and offer valuable performance predictions that guide the design and optimisation of MA-assisted communication systems. In a related work, \cite{10709885} focuses on the performance analysis and optimisation of broadband orthogonal OFDM systems enhanced by movable antennas. Leveraging a multi-tap field-response channel model, the study reveals that MA localization can be effectively utilized to synthesize a desired channel impulse response CIR with both maximum gain and tunable phase characteristics. This capability allows for flexible spatial signal shaping, which is particularly beneficial in broadband and frequency-selective channels.

6DMA transceivers enhance wireless performance by dynamically allocating antennas to match user locations. This spatial matching provides superior array gain, spatial multiplexing, interference suppression, and geometric gain for sensing compared to fixed antennas.
\begin{itemize}
	{
		\item \textbf{Array Gain: }By steering antenna positions and orientations toward the desired direction, 6DMA can provide higher array gain than fixed planar arrays (FPAs). This helps compensate path loss via more directional radiation. In~\cite{yi2025performance}, the channel capacity of a bilateral XL-MIMO system with ROMA is about 1.4 times higher than that with an FPA.
		
		\item \textbf{Spatial Multiplexing Gain: }6DMA improves spatial multiplexing by positioning and rotating antennas to shape the singular-value profile of the MIMO channel matrix. This can mitigate practical limitations such as antenna coupling and limited scattering. In~\cite{shao20256dma}, the network capacity increases by 60\%, 305\%, and 656\% over FAR, CAM, and FPA baselines, respectively.
		
		\item \textbf{Interference Suppression: }In multi-user scenarios~\cite{pi20256d}, 6DMA enhances the desired-signal power while suppressing interference from other directions. Long-term, channel-aware configuration of antenna placement and orientation can further improve beamforming, together with instantaneous CSI. In~\cite{ren20256d}, the SINR gain is about 15~dB when the number of co-channel users is 12.
		
		\item \textbf{Geometric Gain: }6DMA can improve sensing accuracy through geometric reconfigurability. By optimizing antenna positions and rotations at the BS, the relative geometry between the transmitter and the sensing target is improved, which benefits localization~\cite{11142311}.
	}
\end{itemize}
\subsection{Resource Allocation}
Efficient resource allocation plays a critical role in fully unleashing the potential of MA-assisted wireless systems. Unlike conventional static antenna architectures, the dynamic and controllable nature of movable antennas introduces new degrees of freedom, such as spatial position and movement trajectory, which significantly affect channel characteristics, interference patterns, and system performance. This flexibility, while advantageous, also brings new challenges to the design of resource allocation strategies, including antenna positioning, beamforming and power control. In what follows, we explore key issues and recent developments in resource allocation tailored to MA-assisted communication systems.

\subsubsection{Beamforming Design}
Recent studies have explored intelligent optimisation strategies for MAs to improve spectral efficiency and robustness under practical constraints. 
{To reduce the overhead of frequent MA repositioning based on instantaneous CSI, \cite{11071262} proposes a two-timescale MU-MIMO design. MA positions are updated using statistical CSI at a large timescale, while MRT/ZF beamforming is updated using instantaneous CSI at a small timescale, which improves ergodic sum-rate over fixed-position antennas. }
Complementarily, \cite{10534854} develops a heterogeneous multi-agent reinforcement learning (MADDPG) framework, where one agent learns beamforming and the other learns antenna mobility. This design supports online decisions under incomplete CSI, but it depends on training stability and generalization. 
{Beyond overhead, hardware latency also matters. \cite{11082461} explicitly models the movement delay within a transmission block and optimizes MA positions accordingly, which clarifies the throughput--delay tradeoff and enables delay-aware design. 
In addition, MA beamforming has been extended to multicast and coverage enhancement. \cite{10694747} jointly optimizes Tx/Rx MA positions and transmit beamforming to maximize the minimum SINR across multicast groups. \cite{11218733} further integrates MA with IRS, showing how node-side geometry adaptation and environment-side reconfiguration can be coordinated for coverage improvement, at the cost of higher control and coordination overhead.
 }
 
For 6DMA and related 6D reconfigurable arrays (including position/orientation adjustable designs), beamforming is often optimized jointly with array pose parameters. 
In \cite{wu2024modeling}, a rotatable-antenna model is adopted and the beamforming vector and antenna deflection angles are jointly optimized. For the single-user case, the optimal receive beamforming under MRC is derived in closed form, while for multiuser cases AO combined with SCA is used to design ZF/MMSE receivers for interference suppression. 
In \cite{sun2025rotatable}, multi-objective designs are investigated, where SDR-based methods and heuristic search (e.g., PSO) are combined with AO-type iterations to handle coupled beamforming-and-parameter optimization. 
In \cite{zhang20246dma}, the authors develop a fractional-programming-aided AO framework to jointly optimize digital and unit-modulus analog beamformers, leveraging Lagrange-multiplier reformulation, manifold optimization, and gradient descent. This highlights a typical tradeoff: stronger joint optimization usually yields higher gains, but also increases computational and calibration burden. 
Related designs for spectrum sharing with rotatable antennas are also studied in \cite{peng2025rra}, where antenna orientation provides an additional DoF to manage interference alongside beamforming.

\subsubsection{Power Allocation}
In the context of MA-assisted systems, power allocation plays a critical role in optimizing communication and sensing performance under mobility-induced spatial reconfiguration. In \cite{10741192}, for given the fixed position of the movable antenna (MA), a block coordinate descent method is employed to jointly optimize the receive matrix and the users' transmit power, with the objective of maximizing the minimum achievable rate among all users.
In \cite{10638767}, a piecewise power allocation strategy based on channel state information (CSI) and constraint conditions is proposed to maximize system capacity by adjusting the power allocation coefficient among different users. The strategy is categorized into five typical cases according to the relationship between the upper and lower bounds of the power allocation. An interruption mechanism is introduced: when the minimum rate requirements of all users cannot be simultaneously satisfied, priority is given to users with better channel conditions, while the power allocation coefficient of users in outage is set to zero to enhance overall transmission efficiency.

\subsubsection{Antenna Positioning}
Various studies have explored the optimisation of MA positioning from multiple angles to enhance wireless communication system performance. These include improvements in channel capacity, system throughput, and user data rates, with strategies ranging from signal sampling theory to algorithmic optimisation and antenna mobility modelling. In \cite{10879671}, antenna positioning is optimized by interpolating received signal strength (RSS) from a minimum number of samples determined via the Nyquist theorem, reducing measurement overhead while enabling real-time deployment. For multi-user systems, \cite{10464791} proposes a two-loop PSO-based framework, where the outer loop optimizes antenna positions and the inner loop refines receive combining and user transmit powers, jointly enhancing sum rate and fairness. {In~\cite{shao20246d,shao20256d}, alternating-optimization (AO) algorithms are proposed for different 6DMA architectures and reconfiguration methods. In~\cite{shao20246d}, an AO algorithm decomposes the joint position-and-rotation design into several subproblems. These subproblems are solved by combining a conditional-gradient method with successive convex approximation (SCA). Practical implementation may be constrained by the difficulty of realizing continuous translation and rotation. Therefore,~\cite{shao20256d} restricts the surface to a finite set of discrete position–rotation configurations. Under this discrete model, the authors develop offline and online algorithms. The offline method is based on Monte Carlo evaluation, while the online method uses conditional sampling averaging (CSM).}

\subsection{Convergence with Other Wireless Technologies}
With the increasing complexity and performance demands of next-generation wireless networks, standalone communication technologies often face limitations in scalability, adaptability, or efficiency. MA systems, due to their reconfigurable spatial sampling and dynamic channel interaction capabilities, exhibit strong potential for integration with other emerging wireless technologies. Such convergence not only enhances the performance of MA systems themselves, but also complements and extends the capabilities of the technologies they are combined with. The following discusses several promising directions where MA can be effectively integrated with key wireless paradigms to achieve synergistic gains.
\subsubsection{ISAC}
MA-assisted ISAC systems have attracted increasing attention, with recent works focusing on joint optimisation of antenna positioning, beamforming, and signal design to enhance both communication and sensing performance under complex system constraints.
In \cite{10839251}, a bistatic radar-enabled ISAC system is introduced, where a movable antenna array mounted on the base station operates within a multi-user MISO framework. The objective is to maximize both communication rate and sensing mutual information by jointly optimizing antenna coefficients and positions.
Additionally, in \cite{10693833}, the authors consider a low-altitude platform (LAP)-based MA-ISAC system where a UAV equipped with movable antennas functions as an aerial base station supporting dual communication and sensing tasks. A data rate maximization problem is formulated, jointly optimizing the information and sensing beamforming along with antenna locations. {In~\cite{10737418,hua2025hierarchically}, 6DMA-assisted integrated sensing and communication (ISAC) networks are investigated. In~\cite{10737418}, the authors consider a 6DMA-enhanced sensor network. For each target, the channel vector associated with the horizontal direction angle depends jointly on the antenna position, the surface rotation, and the target angle. The complexity of the particle swarm optimization (PSO) method in~\cite{10737418} depends on the number of particles and the iterations required for convergence. It is given by $\mathcal{O}(IT_{\mathrm{PSO}})$, where $T_{\mathrm{PSO}}$ denotes the maximum number of iterations. In~\cite{hua2025hierarchically}, the authors design an HT-6DMA BS that can either serve multiple single-antenna users or sense potential drones. In the communication mode, the objective is to maximize the long-term average sum rate by optimizing the 6DMA positions and rotations. In the sensing mode, the goal is to maximize the minimum received signal power via joint optimization of the array configurations and the BS transmit covariance matrix.}

\subsubsection{Physical Layer Security}
To address the security requirements of wireless communication, researchers have increasingly focused on integrating mobile antenna  technology with physical layer security techniques \cite{10684758}. In \cite{10749968}, a system is considered where the transmitter, equipped with a linearly movable antenna, sends confidential information to legitimate users while being monitored by multiple eavesdroppers. The proposed system model employs a block coordinate descent algorithm for iterative optimisation, aiming to maximize the secrecy rate and significantly enhance both the stealth and efficiency of the communication. In \cite{10716477}, a jammer equipped with an MA is deployed to monitor the communication between the transmitter and receiver. With the aid of the jammer, the legitimate monitor's probability of successfully eavesdropping is maximized, thereby improving the wireless surveillance capability.

\subsubsection{IRS}
The integration of MA and IRS presents a promising direction for enhancing the flexibility and efficiency of next-generation wireless systems. MAs provide active spatial adaptability by physically adjusting antenna positions, while IRSs offer passive wavefront control by dynamically altering the reflection coefficients of large-scale surface elements. The joint use of MAs and IRSs enables a complementary synergy: MAs can adjust their positions to better align with IRS-reflected paths, while IRSs can dynamically tune their reflection behavior to reinforce MA-enabled beamforming\cite{10946610, zhang2024risaidedwirelesscommunicationmovable, 10746645}.
{In~\cite{10946610}, an algorithm is proposed to jointly optimize the mobile-antenna position and the IRS phase shifts, where the beamforming matrix is introduced as an intermediate variable. The considered IRS-assisted multi-user system aims to maximize the overall throughput. In~\cite{liu2024uav,wang2025passive}, 6DMA-assisted passive IRS communication and/or sensing systems are further investigated. In~\cite{liu2024uav}, a passive 6DMA-aided multicast framework is studied. For the single-user case, near-optimal performance is achieved using only the IRS 1D direction. For the multi-user case, an enhanced AO algorithm with Gibbs sampling is developed. It updates the IRS position and direction through probabilistic Markov-chain iterations. In~\cite{wang2025passive}, a distributed passive IRS-based 6DMA system is proposed. The sum-rate objective is maximized by jointly optimizing IRS positions, reflection coefficients, and BS receive beamforming. The resulting problem is solved via an AO procedure with three alternating subproblems. However, enabling individual motion control of multiple 6DMA surfaces may incur higher hardware cost and algorithmic complexity.} {Moreover, distinct from the aforementioned architectures, \cite{11212818} proposed a movable intelligent surface technology that employs a differential position shifting mechanism between two closely stacked static metasurfaces to synthesize beam patterns. This design eliminates the need for element-wise electronic tuning, and a manifold optimisation-based algorithm was developed to jointly design the phase shifts and shifting positions for worst-case SNR maximization.}

\subsubsection{UAV}
The integration of MAs with UAVs has attracted growing attention in recent research due to its potential to enable reconfigurable and adaptive wireless systems in dynamic environments.
In \cite{10654366}, the authors propose a system where UAVs equipped with MAs dynamically reconfigure their onboard antenna positions to enhance beamforming performance and spectral efficiency. By leveraging both UAV trajectory control and antenna displacement, the system demonstrates improved adaptability to channel variations and user distributions. Similarly, \cite{10892658} investigates the potential of MA-enabled UAVs for interference suppression and precise beam steering. The study shows that micro-scale antenna movement adds a new degree of freedom for fine-tuning signal propagation, which is crucial in dense urban or mission-critical scenarios.
{In~\cite{ren20256d}, the authors consider a downlink cellular-connected UAV system, where each BS serves one user per resource block. The design jointly optimizes the UAV 3D position, 3D rotation, receive beamforming, and BS selection, with the objective of maximizing the received SINR. The problem is solved via a block coordinate descent (BCD) procedure, which alternately updates the optimization variables. The resulting algorithm has low computational complexity. Its order is $\mathcal{O}\!\left(K I_{k}\!\left(T_{x}N+T_{x}T_{1}N+I_{r}N^{3}+T_{a}J+T_{a}T_{2}J\right)\right)$, where $T_{x}$, $T_{1}$, $T_{a}$, $T_{2}$, $I_{r}$, and $I_{k}$ denote the maximum numbers of iterations associated with the corresponding subproblems.}

\subsubsection{Mobile Edge Computing (MEC)}
The integration of MA with MEC introduces a new dimension of adaptability to edge intelligence and wireless resource management. MEC enables low-latency computation, storage, and caching at the network edge, while MAs offer fine-grained control over spatial resources through dynamic antenna positioning. When jointly optimized, MAs and MEC nodes can collaboratively enhance user experience, improve wireless link quality, and reduce end-to-end latency for computation-intensive or delay-sensitive applications\cite{ xiu2024delayminimizationmovableantennasenabled}. In \cite{10971386}, the BS equipped with MAs is designed to assist in computing tasks offloaded from multiple devices. To efficiently find a stationary solution to the associated optimisation problem, the authors propose an algorithm based on penalized dual decomposition (PDD). Meanwhile, by employing the block coordinate descent (BCD) method, they derive closed-form solutions for each variable type within the PDD framework, aiming to maximize the overall task completion rate across all devices. 
In \cite{10620306}, an MA-enhanced WP-MEC system is investigated, where adjustable antenna positions improve both energy transfer and task offloading. By introducing dynamic, semi-dynamic, and static positioning schemes and solving the um computational rate maximization problem via a hybrid PSO-VLS algorithm, the study demonstrates clear advantages of MAs over fixed-position arrays.
\subsubsection{ AirComp}
To enhance the efficiency and accuracy of wireless data aggregation, recent research has explored the integration of MAs into AirComp systems.
In \cite{cheng2023movableantennaempoweredaircomp}, an MA-enabled AirComp framework is proposed, where transmit power, antenna positions, and receive combining are jointly optimized to minimize the computation mean-squared error (CMSE) via an alternating optimisation algorithm. Extending this, \cite{10729877} investigates a 2DMA array at the access point (AP), jointly optimizing APV, receive combining, and user transmit coefficients. A two-loop algorithm is employed, with PSO for APV design and alternating updates for the other variables, effectively reducing CMSE and improving aggregation performance.

\section{Pinching antenna for next-generation mobile communications}\label{s4}
In this section, we focus on pinching antenna systems (PASS), including the basic principle, channel modelling and estimation as well as the resource allocation. A detailed discussion and investigation on pinching antenna is provided.
\begin{figure}[t]
	\centering
	\includegraphics[width=0.8\linewidth]{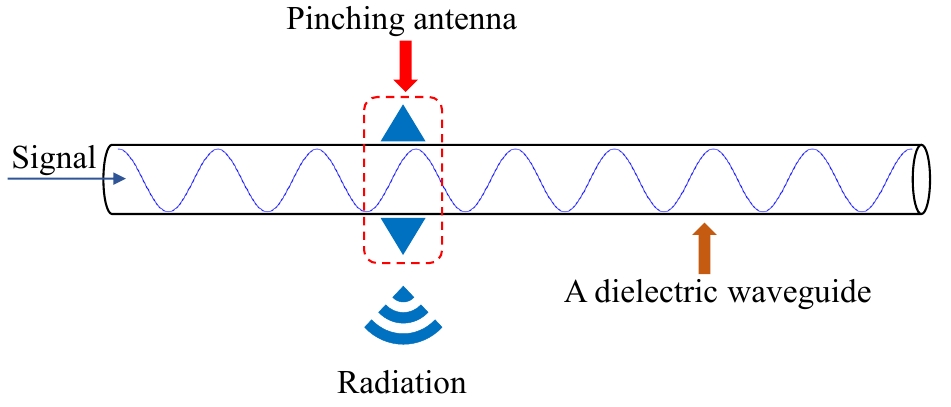} 
	\caption{An illustration of the physical principles of pinching antenna.}
	\label{fig:PA}
\end{figure}
\subsection{Basic Principle of PA}
\subsubsection{Waveguides}
A waveguide is a physical structure engineered to direct electromagnetic waves from one point to another with minimal energy loss. Waveguide is typically realized as hollow metallic tubes or dielectric-filled channels, which confines wave propagation to specific modes—such as transverse electric (TE) and transverse magnetic (TM)—by leveraging internal surface reflections. This confinement is particularly effective at microwave and millimeter-wave frequencies, making waveguides fundamental components in high-frequency communication systems, radar installations, and satellite links. Compared to conventional transmission lines, waveguides offer enhanced power-handling capabilities and significantly lower propagation losses, particularly over long distances. The shape and dimensions of a waveguide determine its cutoff frequencies and supported propagation modes, ensuring efficient and reliable signal transmission in advanced electromagnetic applications.

When electromagnetic waves propagate through a waveguide, their wavelength is altered due to the waveguide’s geometric and material properties. In dielectric waveguides, the guided wavelength is typically shorter than the free-space wavelength because the wave is confined within the structure and propagates more slowly. This relationship highlights the role of dielectric loading in modulating wave propagation characteristics.
\subsubsection{Fundamental Principle}
Pinching antennas represent an innovative approach in antenna design, where small dielectric particles (e.g., plastic inserts) are strategically placed along dielectric waveguides to dynamically manipulate electromagnetic wave propagation. This concept has been experimentally validated by DOCOMO in 2021 \cite{docomo2021waveguide}. Compared to conventional antenna systems, the PASS exhibits two significant advantages. First, PASS facilitates line-of-sight (LoS) communication through its highly reconfigurable antenna positioning capability. The antennas can be flexibly relocated over distances much larger than the wavelength, allowing deployment in close proximity to target receivers to establish new LoS links or enhance existing ones. In contrast, movable or fluid antennas \cite{9264694} are typically limited to displacements within a few wavelengths and are primarily intended to mitigate non-line-of-sight (NLoS) effects. Therefore, PASS is inherently more effective in combating large-scale fading. Second, PASS offers exceptional reconfigurability in antenna structure. By simply adding or releasing dielectric “pinching” elements, the physical dimensions of the system can be dynamically adjusted. This mechanism enables multiple pinching antennas to be flexibly and cost-effectively integrated into one or more waveguides, opening up new possibilities for adaptive and scalable antenna system design.
\subsubsection{Hardware Design}
\begin{figure}[t]
	\centering
	\includegraphics[width=\linewidth]{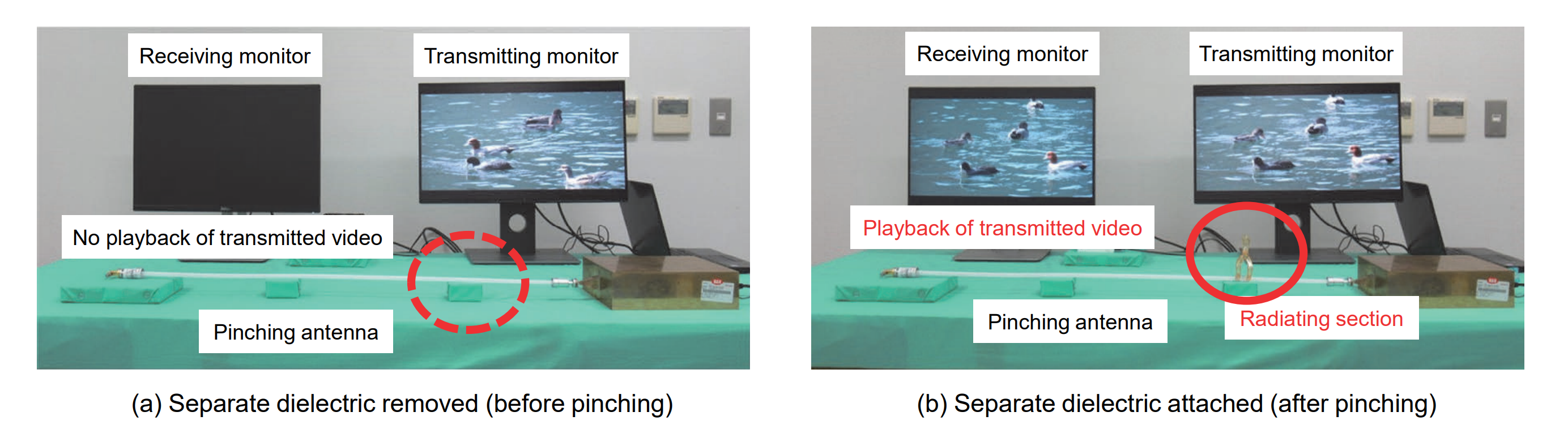} 
	\caption{Video transmission experiment through PA in a dielectric waveguide \cite{docomo2021waveguide}.}
	\label{fig:NTT}
\end{figure}
{
	A typical pinch antenna system consists of three main components, namely a BS, a dielectric waveguide, and a set of pinch antennas implemented as discrete dielectric units. The BS serves as the central processing unit (CPU) for baseband signal processing and is connected to the dielectric waveguide via a wired interface, as illustrated in Fig. \ref{fig:NTT}. The dielectric waveguide acts as the transmission medium, enabling efficient distribution of radio-frequency signals along its length. Specifically, a dielectric waveguide is composed of a rod-shaped dielectric core surrounded by a medium with a lower permittivity. When the permittivity of the inner dielectric exceeds that of the surrounding medium, high-frequency electromagnetic waves are primarily confined within the core and guided along the waveguide. In the implementation shown in Fig. \ref{fig:NTT}, polytetrafluoroethylene (PTFE) with a relative permittivity of 2.1 is adopted as the guiding core, while air, whose relative permittivity is close to unity, serves as the cladding. Owing to this refractive-index contrast, the dielectric waveguide exhibits low propagation loss, particularly in the millimeter-wave frequency band. Overall, pinch antennas can be realized by mechanically attaching dielectric units, referred to as pinches, onto a pre-installed dielectric waveguide. These pinches introduce localized perturbations to the electromagnetic field within the waveguide, thereby inducing a controlled leakage effect that enables electromagnetic radiation to be emitted into free space at the clamping locations \cite{kogelnik19752}. Each pinch unit thus functions as a localized radiating element and is capable of both signal transmission and reception. The resulting radiation characteristics, including radiation strength and directivity, are determined by the geometric configuration, spacing, and dielectric properties of the pinch units relative to the waveguide \cite{xu2025generalizedpinchingantennasystemstutorial}.}
\subsection{Channel Modelling and Channel Estimation}
\subsubsection{ Channel Modelling}

To accurately characterize the signal radiated by a pinching antenna, the work in \cite{wang2025modeling,11169486} models it as an open-ended directional waveguide coupler. This physics-based approach enables adjustable radiation characteristics and simplifies signal modelling. For analytical tractability, it is assumed that electromagnetic (EM) waves radiate from one end of the pinching antenna with negligible reflection, ideally achieving full radiation from the open end.

\begin{itemize}
	\item \textbf{Signal from a Single Pinching Antenna:}
	Consider a single pinching antenna attached to a waveguide at position $x_p$, fed with a baseband communication signal $c_0$. Based on coupled-mode theory and assuming matched effective refractive indices between the waveguide and the pinching antenna (i.e., $\beta_g = \beta_p$), the radiated signal $s_{\text{rad}}$ is given by
	\begin{equation}
		s_{\text{rad}} = \sin(\kappa L)e^{-j\beta_g x_p}c_0,
	\end{equation}
	where $\kappa$ is the mode coupling coefficient, $L$ is the coupling length of the pinching antenna, and $\beta_g$ is the propagation constant of the waveguide.
	
	Taking free-space path loss into account, the signal $y$ received by a user located at distance $r$ from the pinching antenna is
	\begin{equation}
		y = \frac{\eta e^{-j\beta_0 r}}{r}s_{\text{rad}} + n,
	\end{equation}
	where $\eta$ represents the combined channel gain and radiation pattern, $\beta_0=\frac{2\pi}{\lambda}$ is the free-space propagation constant, $r$ is the distance between the pinching antenna and the user, and $n$ is additive white Gaussian noise (AWGN). In this derivation, in-waveguide propagation loss is assumed negligible.
	\begin{figure}[t]
		\centering
		\includegraphics[width=0.8\linewidth]{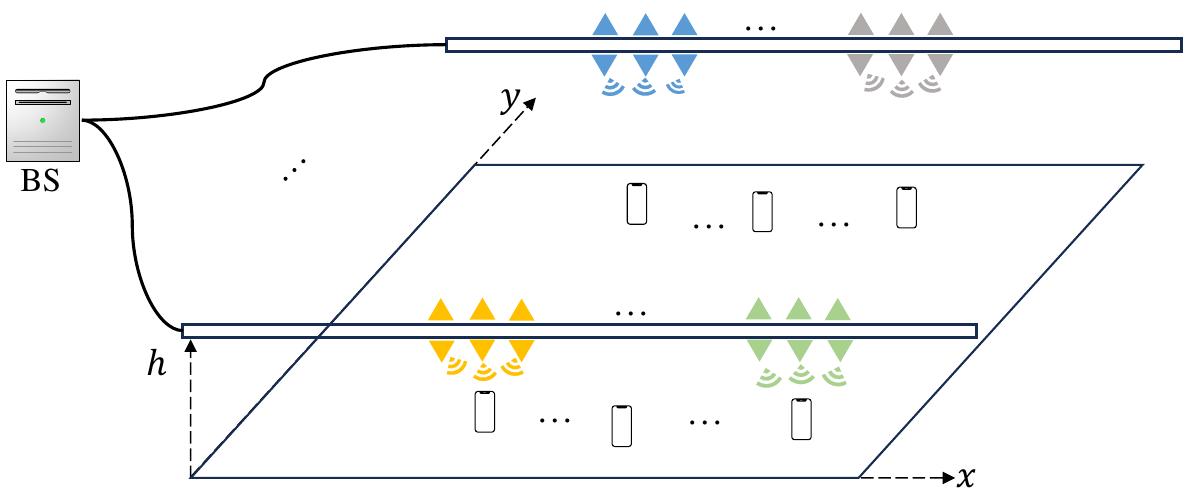} 
		\caption{PASS-enabled multi-user communication mechanism.}
		\label{fig:modelforMPASU}
	\end{figure}
	\item \textbf{Signal from Multiple Pinching Antennas:}
	As shown in Fig. \ref{fig:modelforMPASU}, consider $M$ pinching antennas are sequentially placed on the same waveguide, the power radiated by the $m$-th antenna is influenced by the power extracted by the preceding $m-1$ antennas. The signal radiated from the $m$-th antenna is expressed as
	\begin{equation}
		s_{\text{rad},m} = \sin(\kappa L_m) \prod_{i=1}^{m-1} \cos(\kappa L_i) e^{-j\beta_g x_{p,m}} c_0,
	\end{equation}
	where $L_i$ denotes the length of the $i$-th pinching antenna. 
	This formulation captures the inherent coupling between the radiation powers of pinching antennas on the same waveguide. Two practical power allocation models are then considered:
	\begin{itemize}
		\item \textbf{Equal Power Model}:  Each antenna radiates the same fraction of total power, denoted by $\delta_{\text{eq}}$ (with $0 < \delta_{\text{eq}} \le 1/M$). 
		\item \textbf{Proportional Power Model}: All antennas share the same length $L$, so that each antenna radiates the same proportion of the residual power in the waveguide. 
	\end{itemize}
\end{itemize}

Then, the total received signal from all $M$ pinching antennas is the superposition of their individual contributions as
\begin{equation}
	y = \sum_{m=1}^{M} \frac{\eta e^{-j\beta_0r_m}}{r_m} s_{\text{rad},m} + n
\end{equation}
where $r_m$ is the distance from the $m$-th antenna to the user.

Overall, Equal Power Model is insightful for performance analysis, though it may increase hardware complexity due to differing antenna lengths, while the Proportional Power Model simplifies hardware design by using identical antennas, which reduces hardware costs as antennas are identical.

{
	Since PA is a newly emerging RA paradigm, and its channel modeling studies are currently dominated by analytical and simulation-based evaluations. To the best of our knowledge, publicly available channel measurement campaigns or open datasets that directly validate PA channel models are still missing in the open literature. This is largely because building a repeatable measurement platform requires careful calibration of the waveguide pinching mechanism and controllable activation of radiating points. Therefore, existing PA channel models are currently mainly evaluated via analytical studies and simulations. We view measurement-driven modeling and open PA datasets as an important future research direction.}

\subsubsection{Channel Estimation}

Accurate and efficient channel estimation is essential for unleashing the full potential of PASS. Unlike traditional fixed antenna arrays, where channel state information (CSI) for each antenna element is typically acquired independently, PASS allows for the activation of antenna elements at arbitrary positions—potentially continuous—along a waveguide. This spatial flexibility opens opportunities for performance gains. However, it also poses new challenges for efficient CSI acquisition, especially under constraints on  hardware switching and pilot overhead.

To address this, recent works have focused on reducing pilot overhead and computational complexity by leveraging either physical priors of the wireless channel or data-driven inference techniques. Broadly, existing channel estimation strategies for PASS can be categorized into two types: 1) Model-based parametric methods that exploit the sparse multipath structure of the channel—particularly effective in millimeter-wave (mmWave) scenarios; 2) Learning-based methods that utilize deep neural networks to directly infer the channel state from limited observations.

We elaborate on both classes of methods with representative techniques and discuss their respective strengths, limitations, and applicable scenarios.
\begin{itemize}
	\item \textbf{PAMoE: Mixture of Experts-Based Channel Estimation:}To address the challenge of high-dimensional and dynamic channel estimation from inherently low-dimensional received pilot signals in PASS, Xiao \textit{et al.} \cite{11018390} proposed a deep learning-based estimator named PAMoE. This model leverages a mixture of experts (MoE) architecture, integrating several key modules to effectively process PA position information and pilot signal. PAMoE offers high estimation accuracy through expert specialization but is constrained by the predefined $N_{\text{max}}$, which may limit scalability to unseen PA counts during deployment.
	\item \textbf{PAformer: Transformer-Based Channel Estimation:}
	To overcome the scalability limitation of the PAMoE estimator, which is constrained by a predefined maximum number of PAs ($N_{\text{max}}$) during training, Xiao \textit{et al.} \cite{11018390} further introduced PAformer. This estimator employs a Transformer-style architecture, leveraging self-attention mechanisms to dynamically accommodate an arbitrary number of PAs and predict channel coefficients on a per-antenna basis. This design offers enhanced flexibility, particularly when encountering PA configurations with more antennas than seen during training. A significant advantage of PAformer is its inherent ability to handle varying numbers of PAs without requiring retraining. The self-attention mechanism allows the network to adapt to different input dimensionalities (\textit{i.e.}, different numbers of PAs) during the test phase, even if those dimensionalities were not encountered during training, showcasing a zero-shot learning capability.
	\item \textbf{Parametric Channel Estimation with Sparse Subarrays:}
	Although learning-based channel estimation can achieve high accuracy, it typically requires a substantial number of training samples (\textit{e.g.}, $10^5$) for dataset construction, leading to significant pilot overhead. These challenges highlight the pressing need for low-complexity and scalable channel estimation methods that are specifically tailored to the unique characteristics of PASS. In millimeter-wave (mmWave) communication environments, the wireless channel often exhibits a sparse multipath structure, which motivates the adoption of model-based channel estimation frameworks.
\end{itemize}

If conventional LS estimation is applied, it requires activating all candidate antennas, leading to high pilot overhead. PASS operate in the near-field, requiring distance-dependent channel models. Recently, sparse array assisted parametric estimation has emerged as a promising technique for CSI acquisition in PASS. The core idea is to activate a limited number of PAs to emulate a large aperture array while minimizing switching and pilot overhead. A representative method is presented in \cite{zhou2025channel}, where the full channel is reconstructed by sequentially estimating the angle of arrival (AoA), path distances, and complex gains of dominant multipath components. The estimation method for mmWave PASS activates two small subarrays, \textit{i.e.}, the near-end subarray (close to the waveguide feed point) and the far-end subarray (far from the feed point).

\subsubsection{Performance Analysis}
A growing body of work has analyzed PASS from both theoretical and practical perspectives, confirming its superiority over conventional antenna systems.

Early work by Ding \textit{et al.} \cite{10945421} showed that flexible antenna repositioning in PASS mitigates large-scale path loss and significantly improves ergodic sum rate, with further gains when combined with NOMA. Tyrovolas \textit{et al.} \cite{tyrovolas2025performance} incorporated waveguide losses, deriving outage probability and rate expressions, and highlighted the critical role of waveguide length and optimal antenna placement. Ouyang \textit{et al.} \cite{ouyang2025array} examined array-level behavior, revealing that array gain does not grow monotonically with antenna number due to power dispersion and mutual coupling, and proposed position optimisation and spacing strategies.

In uplink scenarios, Hou \textit{et al.} \cite{hou2025performance} compared different PASS configurations (MPSU, SPSU, SPMU), showing that MPSU yields the highest rates, while also identifying distinct spatial distribution patterns in near- and far-field conditions. Robustness against impairments was studied by Ding and Poor \cite{11036558}, which showed that LoS blockage can paradoxically improve performance by suppressing interference, enabling unbounded rate growth under certain conditions.

Overall, PASS delivers notable gains in rate, reliability, and flexibility, while maintaining resilience under practical impairments such as blockage, waveguide loss, and mutual coupling, making it a promising candidate for next-generation reconfigurable wireless systems.

\subsection{Resource Allocation}
Resource allocation is critical for fully exploiting the reconfigurability and spatial flexibility enabled by PASS. In this section, we survey the recent advancements in beamforming design, antenna placement optimisation, and power allocation, with emphasis on the unique characteristics of PASS and their impact on different system configurations and transmission directions.
\subsubsection{Beamforming Design}
In PASS, beamforming must account for the joint effect of digital precoding and the spatial configuration of pinching antennas. Early work by Bereyhi \textit{et al.} \cite{bereyhi2025downlink} proposed a hybrid beamforming strategy that jointly optimizes digital precoders and antenna positions via fractional programming and Gauss-Seidel iterations to maximize the weighted sum-rate in multi-user MIMO downlink systems. This was later extended to a bidirectional MIMO setup \cite{bereyhi2025mimo}, incorporating joint user detection and pinching optimisation. To reduce beam training overhead, Lv \textit{et al.} \cite{lv2025beam} developed a three-stage beam training (3SBT) scheme utilizing hierarchical and scalable codebooks for waveguide-based PASS architectures. For real-time optimisation, Xu \textit{et al.} \cite{xu2025joint} compared a model-based algorithm (MM-PDD) and a deep learning approach (KDL-Transformer) for jointly optimizing transmit beamforming and antenna positioning. In terms of energy efficiency, Xu \textit{et al.} \cite{xu2025pinching} proposed a joint beamforming design to minimize transmit power under SINR constraints by leveraging the spatial controllability of pinching antennas.

To improve interference management and multi-user separation, Zhao \textit{et al.} \cite{zhao2025waveguide} introduced a Waveguide Division Multiple Access (WDMA) scheme for PASS, where each user is served by an independent waveguide. A two-step alternating optimisation combining successive convex approximation (SCA) and matching theory was proposed. Zhang \textit{et al.} \cite{11048566} proposed a two-timescale hybrid beamforming architecture that separates long-term antenna placement optimisation (using stochastic SCA) from short-term digital precoding (using KDL learning based on instantaneous CSI). To support real-time and scalable deployment, Guo \textit{et al.} \cite{guo2025gpass} introduced GPASS, a GNN-based joint beamforming framework with two specialized sub-networks for pinching position optimisation and digital precoding. Leveraging permutation invariance and advanced GNN modules (e.g., edge updates, nested equivariance), GPASS achieves state-of-the-art spectral efficiency with inference latency under 6 ms for an 8-user scenario.
\subsubsection{Placement Design and Power Allocation}
A major advantage of PASS lies in its ability to dynamically reposition antennas along waveguides, enabling fine-grained spatial control and intelligent energy usage. We now discuss placement and power allocation strategies, grouped by system architecture.

\textbf{Single-waveguide multi-antenna with single-user:}
For LoS-dominant scenarios, optimal antenna placement is crucial. In \cite{10896748}, the authors formulated a rate-maximization problem balancing path loss and phase alignment, solved via relaxed optimisation to efficiently search spatial configurations with near-optimal performance.

\textbf{Single-waveguide multi-antenna with two users:}
For downlink multicast, Mu \textit{et al.} applied PSO to optimize antenna deployment. Xu \textit{et al.} \cite{xu2025qos} proposed a double-loop algorithm for joint power and placement optimisation with QoS guarantees in NOMA-assisted systems. Zhou \textit{et al.} \cite{11029492} developed a bisection-based method for constrained sum-rate maximization with per-user minimum rate requirements.

\textbf{Single-waveguide multi-antenna with multiple users:}
In \cite{10912473}, discrete antenna positions were optimized for sum-rate via a one-sided matching model. Xie \textit{et al.} \cite{11016750} extended this to OMA and NOMA, providing low-complexity deployment algorithms. To improve energy efficiency, Xie \textit{et al.} \cite{11063467} introduced a Bipartite Graph Attention Network (BGAT) that jointly optimizes antenna positions and power allocation, outperforming MLP and GAT baselines.

\textbf{Multi-waveguide multiple users}
Fu \textit{et al.} \cite{fu2025power} proposed an iterative power allocation algorithm based on standard interference functions, achieving up to 76\% power reduction and revealing oscillatory behaviors linked to waveguide separation. Wang \textit{et al.} \cite{wang2025antenna} studied joint optimisation of waveguide assignment, antenna activation, SIC order, and power allocation in NOMA-assisted PASS via coalition games and monotonic/convex optimisation.

\textbf{Uplink communication system:}
Zeng \textit{et al.} \cite{zeng2025sum} applied PSO for uplink antenna placement in SISO systems to maximize throughput. Zhang \textit{et al.} \cite{zhang2025uplink} analyzed MISO PASS with MMSE receivers, jointly optimizing placement and transmit power. For energy efficiency, Zeng \textit{et al.} \cite{11131179} proposed a two-block alternating optimisation, while Tegos \textit{et al.} \cite{tegos2025minimum} addressed max–min uplink rates via convex power allocation and SCA-based placement optimisation.
\subsection{Convergence with Other Wireless Technologies}
\subsubsection{ISAC}
PASS have shown promising potential in enhancing ISAC by exploiting their flexible and reconfigurable structures. In \cite{Ouyang2025Rate}, the rate region of PASS in both communication and sensing is thoroughly investigated. Specifically, in the single-antenna scenario, three design strategies—communication-centric (CC), sensing-centric (SC), and Pareto-optimal—are proposed to characterize the achievable communication-sensing tradeoff. It is demonstrated that PASS achieves a strictly larger rate region compared to traditional fixed-antenna systems. In the multi-antenna case, inner and outer bounds of the rate region are derived using element-wise alternating optimisation and time-sharing approaches. The application of PASS in ISAC from the perspective of Cramér-Rao bounds (CRB) was studied in \cite{ding2025pinching} and \cite{bozanis2025cram}, which highlight the  potential of PASS to enhance sensing accuracy and flexibility in 6G ISAC systems through analytical and simulation results.  Specifically, \cite{ding2025pinching} focuses on the CRLB analysis for PASS-assisted ISAC, demonstrating that PASS can provide uniform positioning accuracy across users, unlike conventional antennas, and leverage reconfigurability for user-centric positioning. While \cite{bozanis2025cram} derived closed-form CRB for joint range and angle estimation in a bistatic PA-ISAC system, where a PA acts as the transmitter and a uniform linear array (ULA) serves as the receiver. It reveals that PASS can achieve significantly lower CRBs than conventional ULAs due to their non-uniform antenna placement and in-waveguide phase diversity, enabling centimeter-level ranging and sub-degree angular resolution with fewer hardware resources. Further advancing ISAC implementation, \cite{zhang2025integrated} proposes a dual-waveguide PASS architecture, where one waveguide transmits joint communication-sensing signals, and the other receives reflected echoes. To maximize target illumination under communication QoS constraints, a penalty-based AO algorithm is developed. By integrating SDR, element-wise search, and semi-continuous activation beamforming structures, the system achieves significant gains in near-field sensing resolution and communication reliability.
\subsubsection{SWIPT and WPN}
PASS has also demonstrated notable potential in enhancing SWIPT systems. In \cite{11106459}, a PASS-assisted SWIPT architecture is developed, where multiple pinching antennas are dynamically activated along a single waveguide to concurrently serve information receivers (IRs) and energy receivers (ERs). To maximize total received energy, a joint optimisation framework is proposed for power allocation and antenna positioning. The original non-convex problem is decomposed into two subproblems: a convex optimisation for IR power allocation and a hybrid algorithm for PA positioning, involving both a high-accuracy iterative method and a low-complexity linear decreasing weight particle swarm optimisation (LDW-PSO) scheme.  Besides, \cite{11096622} proposed a wireless powered pinching-antenna network (WPPAN) to address the ``double near-far" problem in conventional wireless powered networks. The system utilizes a single waveguide with selectively activatable pinching antennas to provide users with closer energy harvesting points in the downlink and uses this harvested energy for uplink data transmission. Three resource allocation approaches of varying complexity (Search-based, Greedy, and Naive WPPAN) were proposed, transforming the original problem into solvable convex optimisation problems.  
\subsubsection{Physical Layer Security}
\begin{itemize}
	\item Secrecy communication: PASS offers significant advantages in physical layer security due to its reconfigurable structure. In \cite{sun2025physical}, the secrecy capacity of a PASS with a single pinching antenna on a dielectric waveguide is analyzed. Closed-form expressions for secrecy metrics—average secrecy capacity, strictly positive secrecy capacity, and secrecy outage probability—are derived, revealing the impact of antenna height and eavesdropper location. In \cite{zhu2025pinching}, a secure communication framework is proposed by dynamically adjusting PA positions over single and multiple waveguides. A PA-wise successive tuning (PAST) algorithm enhances legitimate channels while suppressing eavesdroppers, supplemented by AN injection with novel WD and WM architectures.
	\item Covert communication: PASS also supports effective covert communication through flexible beam control and spatial diversity. In \cite{jiang2025pinching}, both single-waveguide single-antenna (SWSP) and multi-waveguide multi-antenna (MWMP) scenarios are considered. For SWSP, a closed-form optimal antenna position is derived to maximize the covert rate, and a one-dimensional power search method is applied for efficient optimisation. For MWMP, a Twin Particle Swarm Optimisation (TwinPSO) algorithm is proposed to jointly optimize antenna positions and beamforming vectors in a multimodal optimisation landscape.
	\item Artificial noise injection: Artificial noise (AN)-based beamforming can further boost the security of PASS-enabled systems. \cite{papanikolaou2025secrecy} investigates AN beamforming strategies in both single- and multi-waveguide scenarios. In the single-waveguide case, a closed-form solution is achieved via alternating optimisation between fixed beamforming and AN covariance matrix, leveraging PA repositioning to increase secrecy rate. For the multi-waveguide case, an AO framework is developed, combining beamforming optimisation with grid-based PA position search. 
\end{itemize}

\subsubsection{AirComp}
PASS has also been explored in the context of AirComp, where it enables improved aggregation accuracy by dynamically optimizing spatial antenna configurations. In \cite{11133436}, a PASS-assisted AirComp framework is proposed to jointly optimize antenna position, transmit power, and decoding vectors in order to minimize the mean squared error (MSE) of the aggregated result. Due to the highly non-convex nature of the problem, an alternating optimisation scheme is introduced, with PA positions refined through a Gauss-Seidel strategy.

\section{Reconfigurable Holographic Antenna for Next-Generation Mobile Communications}\label{s5}
In this section, we firstly introduce the basic principle of the reconfigurable holographic antenna (RHA). Secondly, the channel modeling and estimation are presented. Thirdly, we discuss the resource allocation strategies of the RHA-based. Finally, we explore the integration of the RHA with other emerging wireless technologies.

\subsection{Basic Principle of RHA}

\subsubsection{ Fundamental Principle} 

\begin{figure}
	\centering
	\includegraphics[width=1\linewidth]{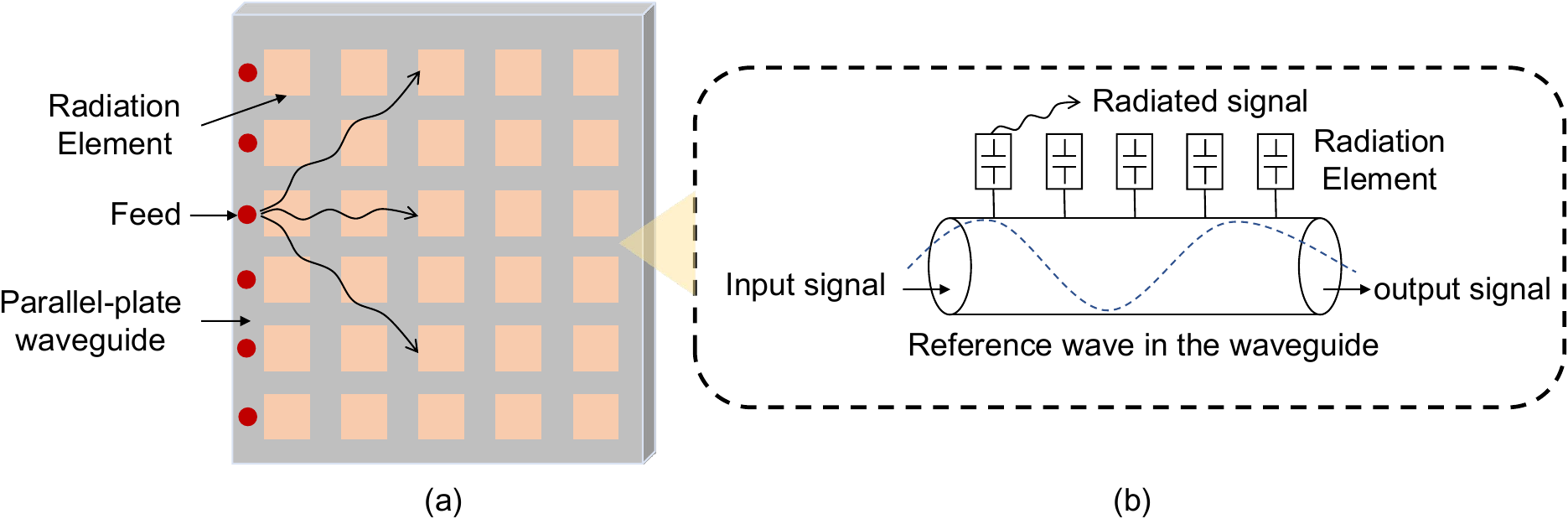}
	\setlength{\abovecaptionskip}{0pt}
	\setlength{\belowcaptionskip}{0pt} \caption{ Illustration of the RHA \cite{deng_reconfigurable_2022}: (a) Schematic structure of the RHA; (b) The work principle of the RHA.} \label{fig1}
\end{figure}

The RHA is a type of leaky-wave antenna \cite{deng_reconfigurable_2022}. Owing to its unique structure, the spacing between adjacent element is  half-wavelength, allowing a higher element density within a given area. Unlike traditional phased array, the RHA generates the desired beamforming by tuning resonance circuits of the elements rather than relying on the phase shifters, thereby resulting a compact structure. By leveraging the holographic principles, the RHA is capable of generating the desired analog beamforming efficiently.
As depicted in Fig. \ref{fig1} (a), the RHA consists of a parallel-plate waveguide, $Q$ feeds and $M$ radiation elements. The RF signals are generated by RF chains and subsequently delivered to the feeds in the form of high-frequency currents. These currents are then converted into the electromagnetic waves by the feeds, which are also referred to as the reference wave. The signal propagation model is illustrated in Fig. \ref{fig1} (b). The reference waves first propagate through the waveguide.
As they reach each radiating element, a portion of the energy leaks into the free space. The amplitude and phase of the resulting leaky waves are controlled by adjusting the individual elements. By carefully tuning these parameters across the RHA, the desired holographic beamforming is synthesized.

Next, the transmit signal at the $q$-th element of the RHA is expressed as \cite{smith_analysis_2017,deng_hdma_2022,che_reconfigurable_2020}
\begin{align}
	x_m = \sum_q \Phi_m \Psi_{ q \rightarrow m } s_q,
\end{align}
where $\Psi_{q\rightarrow m}$ represents the electromagnetic response from the $q$-th feed to the $m$-th element, $\Phi_m$ is the beamforming coefficient at the $m$-th element. Specifically, three typical RHA beamforming models are considered:
\begin{itemize}
	\item \textbf{Amplitude-only beamforming \cite{deng_reconfigurable_2023} :} $\Phi_m \in [a,b]$;
	\item \textbf{binary-amplitude beamforming \cite{deng_reconfigurable_2023_JSAC1} :} $\Phi_m \in \{a,b\}$;
	\item \textbf{Lorentzian-constrained phase beamforming \cite{9847609} :} $\Phi_m \in \{ \frac{e^{j\phi}+j}{2} | 0 \leq \phi \leq 2\pi  \}$,
\end{itemize}
where all the three beamformers are implemented by tuning the resonance circuit of each element \cite{smith_analysis_2017}. By carefully designing the beamforming coefficient $\Phi_m$, the desired RHA beamforming can be achieved to support wireless communication.

\begin{figure}
	\centering
	\includegraphics[width=0.8\linewidth]{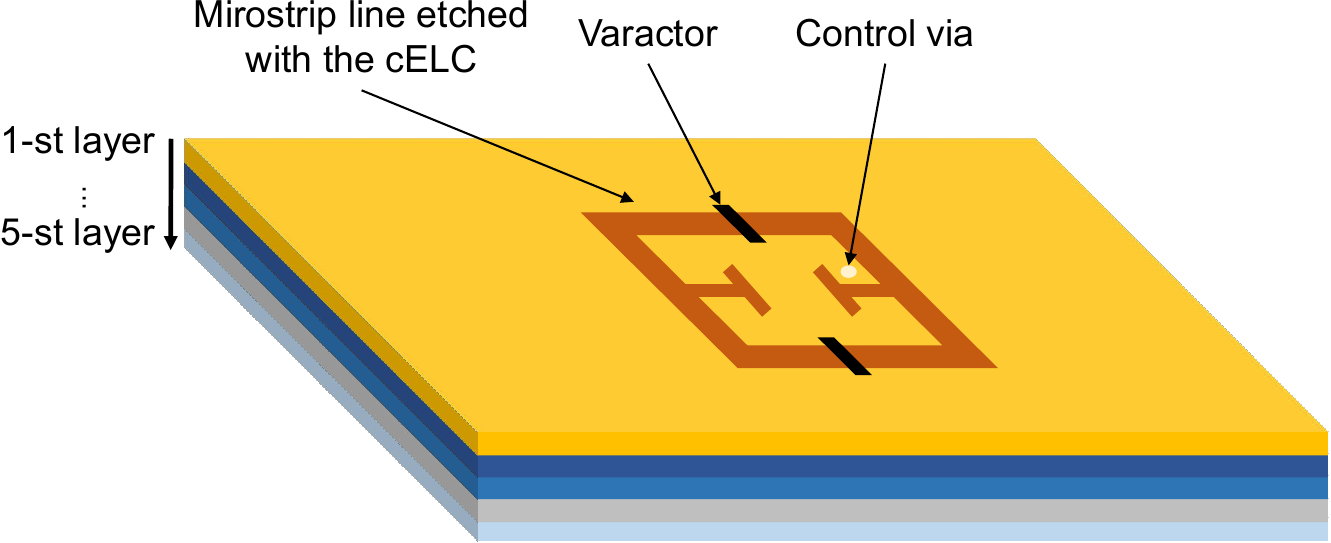}
	\setlength{\abovecaptionskip}{0pt}
	\setlength{\belowcaptionskip}{0pt} \caption{ Illustration of the hardware of the RHS element.} \label{fig2}
\end{figure}

\subsubsection{ Hardware Design}

{The fabrication of RHA systems predominantly leverages mature printed circuit board (PCB) technologies, which ensures low manufacturing costs and facilitates mass production \cite{deng_reconfigurable_2023}. The core tuning elements typically consist of PIN diodes or varactor diodes embedded within complementary electric-LC (cELC) resonators. By electronically controlling the bias voltage of these diodes, the radiation amplitude of each metamaterial element can be dynamically modulated. Unlike movable antennas that rely on mechanical motors for position adjustment, RHAs operate electronically with a solid-state nature. This absence of moving parts grants RHAs superior mechanical durability and vibration resistance, making them highly suitable for dynamic environments.}

As depicted in Fig. \ref{fig2}, the RHA is consists of 5 layers, which are detailed as follows \cite{deng_reconfigurable_2023_MAG1}:
\begin{itemize}
	\item \textbf{1-st layer:} The cELC is positioned within this layer, while two varactors are placed across the capacitive gaps between the metamaterial structure and the upper conductor of the surrounding waveguide. By adjusting the varactors, the resonance of the cELC circuit can be tuned, thereby enabling control over both the amplitude and phase of the transmitted signal.
	\item \textbf{2-st layer:} This layer serves as the RF substrate, within which the waveguide is embedded and the reference wave propagates.
	\item \textbf{3-st layer:} This layer acts as the ground plane in the RHA, offering electromagnetic shielding and serving as a reference for both RF and DC circuitry.
	\item \textbf{4-st layer:} This layer functions as a spacer dielectric layer, providing electrical insulation and structural separation between adjacent layers.
	\item \textbf{5-st layer:} This layer serves as the DC bias feed line layer. A control via extends from the 1-st layer to the 5-st layer, enabling vertical signal delivery. By applying a bias voltage through the control via, the state of the varactors can be adjusted, thereby tuning the resonance of the cELC circuit.
\end{itemize} 

{However, the unique series feeding architecture of the RHA introduces a critical physical constraint known as the leakage power constraint. As the reference wave propagates along the waveguide, its energy gradually decays due to radiation leakage at each element. Consequently, the beamforming algorithms must satisfy the law of energy conservation, ensuring that the total radiated power does not exceed the input power \cite{di_holographic_2025}.}


\subsection{Channel Modelling and Estimation}

\subsubsection{ Channel Modelling}

To analyze the theoretical performance of the RHA system, an accurate channel model is essential and has been extensively researched \cite{an_tutorial_2023, 9110848}. Due to the dense arrangement of elements in the RHA, the spatial correlation among elements is significantly higher than that in conventional half-wavelength phased arrays \cite{9110848}. As a result, the channel models of the RHA differ from those of traditional phased array systems. The main approaches are introduced below.

\textbf{Parametric physical channel model:} Ghermezcheshmeh \textit{et al.} \cite{ghermezcheshmeh_parametric_2022} proposed a parametric physical channel model that accurately characterizes LoS-dominated scenarios and facilitates simplified channel estimation. However, this model is primarily suited for environments with strong LoS components and may degrade in performance when non-line-of-sight (NLoS) components become dominant.

\textbf{Electromagnetic-based channel model:} Wei \textit{et al.} \cite{wei_electromagnetic_2025} proposed a stochastic Green's function-based channel model, where the LoS component is characterized using the dyadic Green's function, and NLoS components are modeled via an expansion of plane waves. This model is directly derived from Maxwell's equations and incorporates practical considerations such as the RHA's physical structure, current excitation, and radiated fields. However, the model is challenging to apply in practice, as some parameters are difficult to obtain and the model lacks a closed-form expression, leading to high computational complexity.
In a related effort, Liu \textit{et al.} \cite{10500399} proposed a physically constrained channel that incorporates polarization leakage and mutual coupling effects. This model reveals fundamental trade-offs among array density, radiation efficiency, and channel capacity, offering guidance for practical system deployment. Its applicability is limited, however, by the difficulty of obtaining the required mutual-coupling-aware radiation patterns.

\textbf{ Wavenumber-domain channel model:} Pizzo \textit{et al.} \cite{9110848} proposed a wavenumber-domain channel model characterized by its simple expression and ease of use for performance analysis. 
The channel model is based on the spatial autocorrelation function, which, for a 3D isotropic scattering environment, simplifies to the sinc function \cite{an_tutorial_2023, molisch_wireless_2011}
\begin{equation}
	C(\mathbf{\Delta}) = \frac{ \sin( \beta_0 \left\| \mathbf{\Delta} \right\|_2 ) }{ \beta_0 \left\| \mathbf{\Delta} \right\|_2 },
\end{equation}
where $\mathbf{\Delta}$ is the displacement vector between two points and $\beta_0= 2\pi/\lambda$. This function implies that the channel responses are uncorrelated at integer multiples of a half-wavelength.
Based on this principle, the wavenumber-domain channel matrix $\mathbf{H} \in \mathbb{C}^{N_r \times N_t}$ is modeled as \cite{an_tutorial_2023, 9110848}
\begin{align}
	\mathbf{H} ={}& \sqrt{N_r N_t} \sum_{\substack{(\bar{l}_x,\bar{l}_y) \in \mathcal{E}_r, \\ (\tilde{l}_x,\tilde{l}_y) \in \mathcal{E}_t}} H_a(\bar{l}_x,\bar{l}_y,\tilde{l}_x,\tilde{l}_y) \nonumber \\
	& \times \mathbf{a}_\text{r}(\bar{l}_x,\bar{l}_y) \mathbf{a}_\text{t}^H(\tilde{l}_x,\tilde{l}_y),
\end{align}
where $\mathcal{E}_\text{r}$ and $\mathcal{E}_\text{t}$ are the sets of receiver and transmitter wavenumber grids. The term $H_a(\cdot) \sim \mathcal{CN}(0, \sigma_{(\cdot)}^2)$ is the complex channel gain at a specific grid. The steering vectors $\mathbf{a}_\text{r}(\cdot)$ and $\mathbf{a}_\text{t}(\cdot)$ are defined by their $m$-th element $[\mathbf{a}(l_x, l_y)]_m = \frac{1}{\sqrt{N}} e^{j \phi_m(l_x, l_y)}$, with the phase $\phi_m(\cdot)$ given by
\begin{align} \label{eq:steering_phase_condensed}
	\phi_m(l_x, l_y) ={}& \beta_0 \left( \frac{p_{x,m} l_x}{L_x} + \frac{p_{y,m} l_y}{L_y} \right) \nonumber \\
	& + \beta_0 p_z \sqrt{1 - \left(\frac{l_x}{L_x}\right)^2 - \left(\frac{l_y}{L_y}\right)^2}.
\end{align}
The parameters in the phase term are substituted accordingly for the receiver ($\mathbf{p}_m = [r_{x,m}, r_{y,m}, r_z]^T$) and the transmitter ($\mathbf{p}_m = [t_{x,m}, t_{y,m}, t_z]^T$).

{ While theoretical models provide valuable insights, physical validation is crucial for assessing the real-world performance of RHA systems. Wang \textit{et al.} \cite{10279384} reported a 2.6~GHz channel measurement campaign using an electrically-controlled virtual dense array (EC-VDA). The setup employs a conventional horn antenna as the transmitter and a fabricated EC-VDA prototype as the receiver in a 10~m line-of-sight link. The reported measurements provide empirical evidence for the benefits of spatial oversampling and offer useful guidance on the applicable range of correlation models (e.g., sinc-type behavior under moderate element spacing). Meanwhile, the study also indicates that practical hardware effects, such as reduced antenna efficiency under extremely dense sampling, can become a key limiting factor. These observations motivate calibration- and efficiency-aware RHA modeling and highlight the importance of measurement-driven validation. }
 
\subsubsection{Channel Estimation}

To enhance system performance through beamforming design for RHA, accurate channel state information (CSI) is essential. Consequently, high-quality CSI estimation is essential and has been extensively studied \cite{guo_wavenumber_2024,chen_unified_2025,guo_statistical_2025,cui_near-optimal_2024,yu_learning_2024,yue_hybrid_2024,chen_holographic_2024,zeng_csi-gpt_2025,yu_bayes-optimal_2024,chen_angular-distance_2024}. In the following, we provide an overview of existing channel estimation techniques for acquiring instantaneous CSI.

\textbf{Classical linear channel estimation methods:} Similar to conventional linear estimation methods used in phased arrays, linear techniques can also be applied to estimate the CSI of the RHA. The authors in \cite{demir_channel_2022} proposed three linear channel estimation methods. Given the uplink received pilot signal $\mathbf{Y}_L$,  The corresponding estimated CSI using the three linear methods is given by
\begin{align}
	\hat{\mathbf{H}}=
	\begin{cases}
		\frac{1}{\sqrt{\rho}} \mathbf{Y}_\text{L},\ \text{LS estimation}, \\
		\sqrt{\rho} \mathbf{J}( \rho \mathbf{J} + \mathbf{I}  )^{-1} \mathbf{Y}_\text{L},\ \text{MMSE estimation}, \\
		\frac{1}{\sqrt{\rho}} \mathbf{U}_r \mathbf{U}_r^H \mathbf{Y}_\text{L},\ \text{RS-LS},
	\end{cases}
\end{align}
where $\mathbf{J}$ is the spatial correlation matrix, $\mathbf{I}$ is the identity matrix and $\mathbf{U}_r$ is the low-rank approximation of $\mathbf{J}$. These estimation methods do not require prior statistical knowledge of the receiver and are relatively easy to implement. However, their performance degrades in large-scale RHA systems due to the following limitations: (1) A large number of RF chains are needed to sample the uplink pilot signals; (2) The computational complexity is high owing to matrix inversion and eigenvector decomposition operations.

\textbf{Sparse reconstructive channel estimation methods:} 
Considering the sparse nature of wireless channels, especially in the mmWave and THz bands, compressed sensing techniques are widely adopted for RHA channel estimation. Various sparse reconstruction methods have been proposed to address challenges such as near-field effects and energy diffusion in the angular domain. However, conventional angular-domain sparse models suffer from power leakage. To overcome this, wavenumber-domain sparse estimation frameworks have been introduced, which exploit clustered sparsity for more accurate channel recovery \cite{guo_wavenumber_2024}. While these methods focus on instantaneous CSI, recent work has also addressed the estimation of statistical CSI using similar wavenumber-domain models \cite{guo_statistical_2025}.

\textbf{AI-based channel estimation methods:} 
AI-based methods offer strong generalization and efficiency by learning directly from pilot observations without relying on explicit channel priors, making them suitable for large-scale and dynamic environments. For instance, self-supervised deep learning frameworks and generative pre-trained models like Transformers have been developed to achieve near-optimal performance in complex electromagnetic environments without requiring prior knowledge or labeled data \cite{yu_bayes-optimal_2024, zeng_csi-gpt_2025}.

\textbf{Other channel estimation methods:} 
Other notable approaches include applying Electromagnetic Information Theory to Gaussian process regression for physically consistent channel modeling \cite{zhu_benefits_2024}, and tensor-based methods that model the channel as a sparse 3-mode tensor to improve recovery accuracy in high-dimensional settings \cite{du_tensor_2024}.

\subsubsection{ Performance Analysis}

To guide the design of the RHA, extensive efforts have been devoted to its performance analysis, particularly concerning channel capacity.

\textbf{Capacity performance analysis:} 
Ouyang \textit{et al.} \cite{ouyang_performance_2024} established an electromagnetic field-based framework to analyze continuous-aperture RHA-based systems\footnote{The continuous-aperture RHA is an idealized RHA configuration in which the spacing between adjacent elements is zero. Studying its performance allows us to characterize the upper bound of RHA performance.}, deriving closed-form expressions for channel capacity. Their results reveal that a continuous-aperture RHA can be viewed as a theoretical upper bound for discrete arrays. Subsequent analyses have extended this framework to account for factors such as Rician fading channels and inter-receiver interference. 
Moreover, the performance comparison between continuous and discrete aperture RHA has been investigated, showing that with sufficient antenna density, the latter can approach the performance limit of the former, though mutual coupling can degrade performance in dense deployments \cite{wan_performance_2023}. 
In addition, researchers have analyzed the capacity of RHA by considering practical constraints, such as finite blocklength information theory, realistic array apertures, hardware impairments, and specific propagation scenarios.

\textbf{Degrees of freedoms (DoFs):} 
Degrees of freedom (DoF) quantify the number of independent spatial channels available for information transmission in RHA systems. Studying DoF reveals the fundamental capacity limits imposed by physical aperture size, not merely the number of antennas. Research in this area has investigated the impact of mutual coupling on DoF and proposed novel 3D antenna topologies to overcome the limitations of conventional 2D RHA systems, thereby enhancing spatial multiplexing capabilities \cite{yuan_breaking_2024}.

{\textbf{Ultra-Low Latency:} RHA systems exhibit exceptional responsiveness due to their electronic tuning nature. Experimental prototypes have achieved circuit-level beam switching within 2 $\mu$s and system-level switching within 50 $\mu$s \cite{di_holographic_2025}. Furthermore, advanced FPGA-based implementations have demonstrated switching times as low as 3 $\mu$s \cite{hu_reconfigurable_2025}, making RHA highly suitable for fast-fading environments.}

\textbf{Energy efficient performance analysis:} 
To enhance the energy efficiency of RHA systems, research efforts have focused on developing energy-efficient beamforming frameworks that account for hardware impairments and on formulating optimization problems based on closed-form achievable rate expressions \cite{li_energy-efficient_20241}. {Quantitative studies demonstrate the superior energy efficiency of RHA. For instance, a 256-element RHS consumes approximately 10 W, offering a 56.52\% power saving compared to a 23 W phased array with equivalent gain \cite{di_holographic_2025}. This efficiency is partly attributed to the minimal power consumption of individual PIN diodes, which is around 0.01 W \cite{deng_reconfigurable_2023}.}

{\textbf{Computational Complexity:} To address the challenge of high-dimensional channel estimation, hierarchical codebook designs utilizing the "scale-changeable" aperture feature have been proposed. These designs effectively reduce the beam training overhead to a logarithmic scale ($O(\log N)$), significantly lowering the computational complexity compared to exhaustive search methods \cite{zhang_hierarchical_2024}.}

\textbf{Other Aspects of Performance Analysis:} 
Several other key performance aspects have also been investigated, such as coupling-aware beamforming design to improve gain \cite{wang_beamforming_2024} and the development of comprehensive electromagnetic-based communication models that integrate physical layer effects into a tractable framework \cite{williams_electromagnetic_2022}.

\subsection{Resource Allocation}
Resource allocation is crucial for the RHA to achieve the optimal performance.

\subsubsection{ Beamforming Design} Beamforming design plays a vital role in RHA systems by focusing spatial power toward a desired direction or location, thereby enhancing transmission efficiency. 
Deng \textit{et al.} \cite{deng_reconfigurable_2022} proposed a hybrid beamforming scheme for RHA-enabled multi-user communication systems, combining digital beamforming at the base station with holographic beamforming at the surface. A joint optimization algorithm was developed to maximize the sum-rate, revealing that the RHA can outperform same-sized phased arrays with lower cost and higher beamforming flexibility.
Wang \textit{et al.} \cite{wang_beamforming_2024} proposed a coupling-aware beamforming design and jointly optimized the characteristic and load impedances to maximize realized gain.
Xu \textit{et al.} \cite{xu_near-field_2024} proposed a beam combining framework for near-field wideband RHA systems. By modeling spherical-wave propagation and accounting for dual-wideband effects, they jointly optimize analog and digital combiners under practical hardware constraints.
Designing holographic beamforming for RHA systems typically entails high computational complexity due to the large number of elements. To address this, various simplified beamforming design schemes have been proposed. Deng \textit{et al.} \cite{deng_hdma_2022} proposed a novel holographic-pattern division multiple access (HDMA) scheme that enables the multi-user beamforming by superimposing the known patterns on the RHA. The key advantage of HDMA lies in its low complexity. Similarly, Zhang \textit{et al.} \cite{zhang_pattern-division_2023} proposed a pattern-division multiplexing technique for multi-user continuous aperture RHA, projecting pattern functions onto orthogonal bases to enable tractable coefficient optimization. A BCD-based algorithm maximizes sum-rate with higher efficiency than conventional schemes, and holographic beamforming can be simplified in specific scenarios.

AI-powered holographic beamforming enables low-complexity, data-driven solutions that can adapt to hardware imperfections, environmental dynamics and multi-objective tasks, offering significant advantages over traditional optimization-based methods in RHA systems. Zhang \textit{et al.} \cite{zhang_superdirective_2025} proposed an AI-driven approach for superdirective beamforming using a novel MultiTransUNet-GAN model. By learning the mapping from far-field electric fields to excitation coefficients, the method enables high-precision beamforming for the RHA systems without requiring explicit mutual coupling modeling. It significantly reduces computational cost and achieves near-theoretical directivity in both linear and planar arrays. 
Adhikary \textit{et al.} \cite{adhikary_integrated_2024} proposed an AI-enabled framework for integrated sensing, localization, and communication (ISLC) in the RHA systems. Using a variational autoencoder for user position sensing and a GRU for power allocation, the system dynamically activates minimal holographic grids to generate energy-efficient, direction-aware beamforming. 

\subsubsection{Hardware Selection} In addition to beamforming design, hardware selection is another effective approach to enhance the performance of RHA systems.
Ouyang \textit{et al.} \cite{ouyang_aperture_2024} proposed an aperture selection strategy for continuous aperture RHA to improve uplink performance with reduced complexity. By analyzing both LoS and NLoS channels, they showed that activating only a portion of the aperture—based on nearest-neighbor or segmented selection—can effectively enhance SNR and achieve selection diversity.
Huang \textit{et al.} \cite{huang_joint_2023} proposed a joint microstrip selection and beamforming scheme for RHA systems. By modeling the unique propagation characteristics and Lorentzian constraints of RHA, they developed a low-complexity alternating algorithm that significantly reduces the number of RF chains while maintaining high spectral efficiency.

\subsection{Convergence with Other Wireless Technologies}

As a promising antenna technology, RHA can be integrated with other advanced wireless technologies.

\subsubsection{Satellite Communications}
As a lightweight and energy-efficient transmit antenna, the RHA is a promising candidate for satellite deployment to enable communication services. 
Hu \textit{et al.} \cite{hu_holographic_2023} investigated holographic beamforming for LEO satellite communications using RHA. They proposed a sum-rate-maximizing beamforming algorithm under real-valued amplitude constraints and derived the minimum number of RHA elements required to outperform phased arrays. Deng \textit{et al.} \cite{deng_holographic_2022}  demonstrated that RHA offers a promising alternative to phased arrays in LEO satellite communications, providing high directive gain, low power consumption, and cost efficiency. By integrating a satellite tracking mechanism and holographic beamforming, the proposed scheme supports robust and high-throughput multi-satellite connectivity with compact user terminals.

\subsubsection{Wireless Sensing and ISAC}
RHA enables high-fidelity and energy-efficient sensing through near-continuous-aperture beam control and programmable radiation patterns. Unlike dense phased arrays, RHA improves sensing performance by enhancing  spatial gain, refining beam shaping, suppressing sidelobes and supporting flexible integration of communication and sensing functionalities.
Yang \textit{et al.} \cite{yang_near-field_2023} proposed a near-field localization framework using RHA with reduced RF chains. By leveraging curvature-of-arrival (CoA) features in spherical wavefronts, the authors developed a direct localization method and an alternating optimization algorithm for joint RHA tuning and position estimation. Simulation results show that RHA-based localization approaches the accuracy of fully digital arrays with significantly lower hardware cost.
Zhang \textit{et al.} \cite{zhang_parameter_2022} proposed a RHA-based radar system for low-power, high-precision parameter estimation. They developed a beamforming algorithm tailored to amplitude-controlled architectures and derived the Cramér-Rao bound (CRB) for range and reflection coefficient estimation. 
Zhang \textit{et al.} \cite{zhang_reconfigurable_2023} proposed a collaborative multi-vehicle SLAM system for autonomous driving, powered by RHA and federated learning. By replacing phased arrays with cost-efficient RHA-based radars and optimizing beamforming and localization models collaboratively, the system achieves high-accuracy, privacy-preserving localization with lower hardware and communication overhead compared to conventional approaches.
Hu \textit{et al.} \cite{hu_multi-band_2023} proposed a multi-band RHA-based ISAC system to overcome the bandwidth limitations of single-band RHA. By employing a position-then-transmit protocol and developing an efficient alternating analog-digital beamforming algorithm, the system significantly improves positioning accuracy and reduces communication capacity loss. 
Zhang \textit{et al.} \cite{zhang_holographic_2022} proposed an RHA-enabled ISAC system that jointly optimizes digital and holographic beamforming to support multi-user communication and multi-target sensing. An iterative algorithm is developed to maximize radar utility while satisfying SINR and power constraints. 

\subsubsection{WPT and SWIPT}
Due to its high spatial gain, the RHA is a promising candidate for WPT and SWIPT. 
Zhang \textit{et al.} \cite{zhang_near-field_2022} proposed a near-field wireless power transfer framework using RHA. By jointly optimizing holographic beamforming and digital beamforming, the system forms focused energy beams for efficient multi-user charging. 
Azabahram \textit{et al.} \cite{azarbahram_waveform_2025} proposed a joint waveform optimization and beam focusing framework for near-field WPT systems using RHA and non-linear energy harvesters. By considering the non-linear characteristics of the rectifiers, the authors designed an alternating SCA-based algorithm to minimize transmit power while meeting energy harvesting demands. 
Huang \textit{et al.} \cite{huang_metaswipt_2024} proposed MetaSWIPT, a RHA-assisted multi-user MISO downlink SWIPT system that jointly optimizes digital precoding and holographic beamforming to maximize the sum rate under energy harvesting constraints. Leveraging a near-field model and a WMMSE-based algorithm, the system outperforms conventional phased arrays in both spectral efficiency and power transfer efficiency.

\subsubsection{Cell Free}  RHA features lower cost, more compact form factor and simpler control circuitry, making it well-suited for deployment in cell-free large-scale wireless networks.
Li \textit{et al.} \cite{li_performance_2024} proposed a hybrid beamforming architecture for RHA-based cell-free networks and analyzed their performance under near-field conditions and hardware impairments. Their study demonstrates that RHA is well-suited for large-scale distributed deployment due to its low-cost design, local CSI-based beamforming, and robustness to phase errors and RF imperfections. The results show that increasing BS density can effectively mitigate hardware limitations and enhance spectral efficiency, especially under near-field models.

\section{Comparison Amongst Different RAs and Open Challenges}\label{s6}
{
\subsection{Overall Comparison of Different RAs}
To provide a structured and insightful comparison of RA technologies for next-generation wireless systems, this section provides a unified comparison of four representative RA types. Here we compare FA, MA, PA, and RHA along five aspects: {fundamentals}, {hardware}, {flexibility}, {deployment and applications}, and {channel estimation}. The discussion highlights the underlying trade-offs, while Table~\ref{tab:ra_comparison} provides a concise summary.

\textbf{Fundamentals:}
The most fundamental difference of RA is what is being reconfigured. FA reconfigures the radiation position by selecting among discrete or quasi-continuous ports/positions, commonly realized via RF switches, micro-pumps, or MEMS mechanisms \cite{10753482}. Its primary purpose is often NLoS enhancement, since port/position diversity can mitigate deep fades and improve link reliability; correspondingly, the spectral/energy efficiency gain tends to become more pronounced as the number of candidate ports increases due to stronger diversity and selection opportunities \cite{10208068}. MA directly reconfigures physical geometry by translating and rotating antennas in 1D--6D space using stepper/servo actuation with position feedback \cite{11142311}. Similar to FA, MA typically provides NLoS enhancement by exploring spatial regions with improved channel conditions; its efficiency gains generally grow with a larger moving region (and richer pose DoF), albeit subject to motion and control overhead. PA changes the effective radiation point along a waveguide through passive pinching or dielectric insertion, with optional motorized actuation \cite{10945421}. PA is typically designed for LoS enhancement, as waveguide-guided reconfiguration can strengthen dominant paths and mitigate blockage in directional links; its efficiency improvements tend to increase with more pinch/activation points or more waveguides that expand the set of selectable radiation states \cite{11159498}. RHA departs from macroscopic motion and reshapes the sub-wavelength aperture field through tunable resonance and biasing circuits \cite{deng_reconfigurable_2022}. Its main purpose is beamforming-focused enhancement, since high-dimensional aperture DoF enables refined wavefront shaping; accordingly, spectral/energy efficiency gains often become more pronounced as the number of radiating elements increases, providing larger aperture DoF and finer controllability \cite{di_holographic_2025}. These distinct mechanisms explain why the same term ``reconfigurability'' leads to different design constraints, calibration needs, and controllable degrees of freedom across paradigms.

\textbf{Hardware:}
Hardware complexity and lifecycle cost differ substantially across paradigms. FA exhibits a split profile: liquid/actuated FA is typically high in hardware complexity and cost due to fluidic actuation, packaging, and reliability considerations, while solid-state pixel/meta FA is generally low in complexity and more moderate in cost \cite{tong2025designs}. This split also carries over to replacement: liquid/actuated FA tends to be harder to replace, whereas solid-state pixel/meta FA is usually easier to swap since it behaves more like an electronic module. MA has moderate hardware complexity because it relies on precision mechanics together with sensing and calibration; correspondingly, its replacement complexity is also moderate because mechanical modules often require re-alignment and re-calibration after replacement, and the cost remains moderate due to precision components and maintenance overhead \cite{tong2025designs}. PA is typically low in hardware complexity and cost thanks to its simple waveguide structure and modular dielectric ports; its replacement is also low-complexity in many implementations, although repeatability and alignment can affect how “plug-and-play” the swap is in practice \cite{10945421}. RHA is low in hardware complexity at the mechanical level because it is commonly PCB-architecture based, yet its overall cost is moderate since fabrication and calibration/control infrastructure can be nontrivial; replacement is generally low-complexity as a solid-state module, but re-calibration is typically required to restore EM-consistent operation and maintain performance \cite{hu_reconfigurable_2025,deng_reconfigurable_2023}.

\textbf{Flexibility:}
Flexibility is primarily reflected by switching latency, energy consumption, and  switching complexity. FA switching speed is implementation-dependent: solid-state pixel switching can reach the microsecond regime \cite{10740058}, reported meta-based prototypes can be even faster (sub-microsecond) \cite{liu2025metafass}, while liquid FA is fundamentally bounded by fluid motion and is often characterized by physical movement speed (e.g., mm/s) \cite{shen2024d}. MA is constrained by mechanical actuation and typically operates at ms--s timescales \cite{11142311}. PA is often ms--s and commonly slower than purely electronic tuning \cite{10945421}. RHA can support microsecond-to-millisecond updates through electronic tuning \cite{hu_reconfigurable_2025}. In energy terms, mechanical actuation usually dominates for MA, liquid actuation can increase FA energy consumption, PA is mostly passive with moderate operational overhead, and RHA typically requires relatively low bias/control power but may incur system-level overhead due to control and calibration.

\textbf{Deployment and applications:}
The deployment suitability of RA paradigms depends on how their reconfiguration mechanisms impact system integration, scalability, and mobility support. Table~\ref{tab:ra_comparison} summarizes these practical outcomes. FA is broadly deployable at both BS and UE by integrating into conventional RF chains with moderate complexity for port control and CSI selection. Mobility support is generally high due to fast port switching, though tracking can be limited by scanning and CSI overhead. Scalability is moderate; adding ports improves diversity but increases training load, whereas meta-based dense arrays scale more favorably. These features suit wearables, IoT, and WPT applications. MA is best matched to platforms accommodating motion hardware, such as BSs, UAVs, and vehicles. It offers rich 1D–6D spatial exploration but demands high integration for actuation, sensing, safety, and re-calibration. Scalability is limited by mechanical complexity. Mobility support is low due to mechanical latency, favoring slow-varying links or two-timescale designs. Key applications include UAV tracking, ISAC, and dynamic MIMO. PA offers a modular add-on for BS and UE with simple waveguide structures, enabling high hardware scalability by adding pinches \cite{11169486}. Integration complexity is low, though performance depends on alignment and site planning. Configurations are typically quasi-static, resulting in low mobility support. PA is ideal for mmWave/THz LoS enhancement and blockage mitigation. RHA aligns with infrastructure deployment (BS, RIS nodes). Scalability is high regarding aperture size, but calibration and control workloads grow significantly. Integration is demanding due to dense bias networks and EM-aware CSI needs. While electronic tuning is fast, high-dimensional CSI and calibration often limit practical tracking. RHA suits smart environments, ISAC, and secure communications.

\textbf{Channel estimation:}
Channel estimation is the key enabler for RA gains, since configuration selection and beamforming are CSI-driven. Table~\ref{tab:RA_CE_consolidated} summarizes representative estimation methods across RA paradigms. For FA, the challenge is estimating channels {and} selecting ports under inter-port correlation, which often requires scanning/training across candidate ports. For MA, CSI depends on position/pose, so estimation is coupled with the movement trajectory and is often structure-exploiting. For PA, estimation includes activation-point selection along the waveguide and may be affected by waveguide-induced correlation, while open measurement-based benchmark is still limited. For RHA, CSI acquisition can be high-dimensional and EM-constrained, and calibration errors may directly degrade the synthesized aperture field \cite{demir_channel_2022}. From a system-level viewpoint, the dominant complexity is typically the end-to-end CSI acquisition burden (training/pilots, scanning/selection, and calibration/feedback), rather than arithmetic operations alone. Moreover, the scalability is largely governed by how this burden grows with the reconfiguration DoF: FA overhead increases with the number of candidate ports due to port scanning, MA incurs trajectory/pose-dependent sampling overhead, PA scales with the number of activation points along the waveguide, and RHA scales with aperture size due to high-dimensional CSI and calibration requirements. This structured view clarifies the relative efficiency, scalability, and implementation challenges across RA families.
}

\begin{table*}[!t]
	\centering
	\caption{Comparison of Four Reconfigurable Antenna Types.}
	\label{tab:ra_comparison}
	\renewcommand{\arraystretch}{1.35}
	\setlength{\tabcolsep}{4pt}
	\begin{tabular}{|>{\centering\arraybackslash}m{2.1cm}
			|>{\centering\arraybackslash}m{3.1cm}
			|>{\centering\arraybackslash}m{3.1cm}
			|>{\centering\arraybackslash}m{4.1cm}
			|>{\centering\arraybackslash}m{4.1cm}|}
		\hline
		\textbf{Category} & \textbf{FA} & \textbf{MA} & \textbf{PA} & \textbf{RHA} \\
		\hline
		
		\multicolumn{5}{|c|}{\textbf{Fundamentals}} \\
		\hline
		\textbf{Reconfiguration domain} 
		& Discrete/continuous port/position 
		& 1D--6D position and orientation 
		& Dielectric location along waveguide 
		& Sub-wavelength aperture field (phase/amplitude) \\
		\hline
		\textbf{Reconfiguration mechanism} 
		& RF switches / micro-pump / MEMS \cite{10753482} 
		& Stepper/servo motor with position feedback \cite{11142311} 
		& Passive pinching / insertion (optional motorized actuation) \cite{10945421} 
		& Tunable resonance/biasing circuits \cite{deng_reconfigurable_2022} \\
		\hline
		\textbf{Spectral/energy efficiency trend} 
		& More pronounced with more ports (diversity gain) \cite{10208068} 
		& More pronounced with larger moving region 
		& More pronounced with more pinches/waveguides \cite{11159498} 
		& More pronounced with more radiating elements (aperture DoF) \cite{di_holographic_2025} \\
		\hline
		\textbf{Purpose} 
		& NLoS enhancement 
		& NLoS enhancement
		& LoS  enhancement 
		& Beamforming focus enhancement\\
		\hline
	
		\multicolumn{5}{|c|}{\textbf{Hardware}} \\
		\hline
		\textbf{Hardware complexity} 
		& High (liquid/actuated); low (solid-state pixel/meta) \cite{tong2025designs} 
		& Moderate (precision mechanics + sensing/calibration) \cite{tong2025designs} 
		& Low (simple waveguide structure) \cite{10945421} 
		& Low (PCB architecture based) \cite{hu_reconfigurable_2025} \\
		\hline
		\textbf{Antenna replacement complexity} 
		& High (liquid/actuated); low (solid-state pixel/meta) 
		& Moderate (mechanical modules + re-alignment/re-calibration) 
		& Low (modular dielectric ports; repeatability dependent) 
		& Low (solid-state module, but re-calibration is typically required) \cite{deng_reconfigurable_2023}\\
		\hline		
	
		\textbf{Hardware Cost} 
		& High (liquid/actuated); moderate (solid-state pixel/meta) 
		& Moderate (precision mechanics and maintenance) 
		& Low (waveguide + dielectric ports; modular) 
		& Moderate (fabrication + calibration/control) \\
		\hline
		\multicolumn{5}{|c|}{\textbf{Flexibility}} \\
		\hline
		\textbf{Switching speed / latency} 
		& $\mu$s--ms (impl.-dependent): $\mu$s pixel switching \cite{10740058}; $\sim$mm/s liquid motion \cite{shen2024d}; sub-$\mu$s meta switching \cite{liu2025metafass} 
		& ms--s (mechanical actuation limited) \cite{11142311} 
		& ms--s (impl.-dependent; often slower than electronic tuning) \cite{10945421} 
		& $\mu$s--ms (electronic tuning) \cite{hu_reconfigurable_2025} \\
		\hline
		\textbf{Switching energy consumption} 
		& Low (liquid actuation higher) 
		& High (actuation energy) 
		& Moderate (mostly passive) 
		& Low (bias/control power) \\
		\hline
		\textbf{Switching complexity} 
		& Moderate (liquid/actuated); low (solid-state pixel/meta) 
		& High (motion control + feedback/calibration) 
		& Moderate (activation along waveguide; alignment dependent) 
		& Low (bias/control network) \\
		\hline

		\multicolumn{5}{|c|}{\textbf{Deployment and applications}} \\
		\hline
		\textbf{Deployment target} 
		& BS / UE 
		& BS / UAV / vehicle 
		& BS / UE \cite{11169486}
		& BS, RIS-enabled nodes \\
		\hline
		\textbf{Deployment scalability} 
		& Moderate (ports increase overhead); high for meta-based dense arrays 
		& Low (mechanical complexity scales) 
		& High (add/remove pinches along waveguide) 
		& High (scales with aperture size but calibration/control grows) \cite{li_performance_2024} \\
		\hline
		\textbf{Integration complexity} 
		& Moderate (fits conventional RF chain; needs port control/CSI interface) 
		& High (actuation control, sensing, safety, and re-calibration in the loop) 
		& Low (modular add-on; alignment and site planning dependent) 
		& High (dense bias/control, calibration, and EM-aware CSI/pattern synthesis) \\
		\hline
		\textbf{Mobility support} 
		& High (fast port switching; limited by CSI/scan overhead) 
		& Low (mechanical latency; better for slow-varying links / two-timescale control) 
		& Low (often quasi-static; suited to slow updates) 
		& High (fast tuning, but high-dimensional CSI/calibration may limit tracking) \\
		\hline
		\textbf{Application scenarios} 
		& Wearables devices, IoT, WPT  
		& UAV beam tracking, ISAC, dynamic MIMO 
		& mmWave LoS enhancement, blockage mitigation 
		& Smart environments, ISAC, secure and semantic communication \\
		\hline

	\end{tabular}
\end{table*}

\begin{table*}[t]
	\centering
	\caption{Consolidated Comparison of Channel Estimation Methods Across RA Paradigms}
	\label{tab:RA_CE_consolidated}
	\small
	\renewcommand{\arraystretch}{1.25} 
	\setlength{\tabcolsep}{4pt}      
	\begin{tabularx}{\textwidth}{|c|c|X|X|X|X|}
		\hline
		\textbf{RA Type} & \textbf{Ref.} & \textbf{System / Scenario} & \textbf{Setup / Configuration} & \textbf{Estimation Method} & \textbf{Complexity} \\
		\hline
		
		\multirow{11}{*}{\textbf{FA}} 
		& \cite{9992289} & Point-to-point system & Tx-FPA + Rx-FAS & Skipped-LMMSE based & \multirow{11}{=}{\textbf{High} \par \footnotesize (Port correlation + port selection/scan overhead)} \\
		\cline{2-5}
		& \cite{10233765} & Point-to-point mmWave & Tx-FPA + Rx-FAS & LS based & \\
		\cline{2-5}
		& \cite{10375559} & Multiuser uplink mmWave & Tx-FAS + Rx-FPA & LS \& Compressed sensing & \\
		\cline{2-5}
		& \cite{10751774} & Point-to-point system & Tx-FPA + Rx-2DFAS & Nyquist sampling \& MLE & \\
		\cline{2-5}
		& \cite{10807122} & Point-to-point uplink & Tx-FPA + Rx-FAS & S-BAR based estimation & \\
		\cline{2-5}
		& \cite{10495003} & Multiuser downlink & Tx-FPA + Rx-2DFAS & AGMAE based estimation & \\
		\cline{2-5}
		& \cite{tang2025} & Point-to-point system & Tx-FPA + Rx-2DFAS & Diffusion model based & \\
		\hline
		
		\multirow{8}{*}{\makecell{\textbf{MA}}}
		& \cite{10571235,10236898,10497534} & Point-to-point system & Tx-MA + Rx-MA & Compressed sensing & \multirow{8}{=}{\textbf{Moderate} \par \footnotesize (Position/pose-dependent CSI; trajectory overhead)} \\
		\cline{2-5}
		& \cite{10978584} & Point-to-point system & Tx-MA + Rx-MA & Tensor decomposition & \\
		\cline{2-5}
		& \cite{shao2025directional,11142311} & Multiuser uplink & Tx-6DMA + Rx-FPA & Directional sparsity based & \\
		\cline{2-5}
		& \cite{jiang2025statistical} & Multiuser uplink & Tx-6DMA + Rx-FPA & Orthogonal Matching Pursuit & \\
		\cline{2-5}
		& \cite{shao2025hybrid} & Multiuser hybrid-field THz & Tx-6DMA + Rx-FPA & Directional-sparsity-driven & \\
		\hline
		
		\multirow{3}{*}{\textbf{PA}}
		& \cite{11018390} & Multiuser uplink & Tx-FPA + Rx-PA & Mixture of experts-based & \multirow{3}{=}{\textbf{Moderate}  \footnotesize ( Waveguide correlation)} \\
		\cline{2-5}
		& \cite{zhou2025channel} & Point-to-point mmWave & Tx-PA + Rx-FPA & Sparse array parametric & \\	
		\hline
		
		\multirow{10}{*}{\textbf{RHA}}
		& \cite{guo_wavenumber_2024} & Point-to-point uplink & Tx-FPA + Rx-RHA & Graph-cut-based SWAP & \multirow{10}{=}{\textbf{High}~\cite{demir_channel_2022} \par \footnotesize (Large number of antennas)} \\
		\cline{2-5}
		& \cite{guo_statistical_2025} & Point-to-point system & Tx-RHA + Rx-FPA & WD-EM with variational inference & \\
		\cline{2-5}
		& \cite{cui_near-optimal_2024} & Multiuser uplink & Tx-FPA + Rx-RHA & Bayesian regression method & \\
		\cline{2-5}
		& \cite{yue_hybrid_2024} & Point-to-point uplink & Tx-FPA + Rx-RHA & PD-aware OMP & \\
	    \cline{2-5}
		& \cite{yu_bayes-optimal_2024} & Multiuser uplink & Tx-FPA + Rx-RHA & Score-based OAMP (unsupervised) & \\
		\cline{2-5}
		& \cite{chen_angular-distance_2024} & Multiuser uplink & Tx-FPA + Rx-RHA & Decomposition + CS-based & \\
		\hline
	\end{tabularx}
\end{table*}

\subsection{Complement and Coexistence}
{Although FA, MA, PA, and RHA are often introduced as distinct paradigms, they are better understood as complementary design options that realize reconfigurability through different physical mechanisms and therefore exhibit different trade-offs in spatial DoF, reconfiguration latency, hardware complexity, and control/CSI overhead. In some scenarios, an RA paradigm can replace fixed antennas when additional spatial flexibility is essential; however, in many practical deployments it coexists with conventional antenna arrays by providing an additional degree of freedom on top of standard beamforming and scheduling. For example, FA and PA typically offer fast, discrete configuration selection that can be aligned with scheduling timescales and are therefore more often used as a complement to conventional beamforming to improve diversity or beam alignment with moderate overhead. MA provides geometry adaptation over a larger range and can be viewed as a complement when reconfiguration latency is acceptable; in infrastructure-supported settings, it may partially substitute dense arrays by achieving comparable gains through position/pose adaptation. RHA can enable high-dimensional field shaping and potentially strong near-field focusing, but its gains come with increased requirements on dense control, calibration, and EM-compliant modeling and is therefore most naturally deployed as a complement to MIMO beamforming.}

{From a system viewpoint, coexistence is straightforward when technologies act on different parts of the link or provide orthogonal control knobs (node-side aperture reconfiguration versus environment-side wave control), which enables joint optimization rather than direct replacement. These paradigms can also be combined in hybrid architectures, such as conventional MIMO beamforming assisted by RA configuration selection, or BS-side RA jointly optimized with multiuser scheduling, thereby exploiting both digital beamforming and physical aperture reconfiguration. Moreover, RA technologies naturally coexist with other existing solutions that act on different parts of the link. For instance, RIS manipulates the propagation environment by controlling reflection/scattering, whereas RAs reshape the transceiver-side effective aperture; thus, the two approaches are not competing but can be jointly deployed to combine node-side aperture reconfiguration with environment-side wave control. Similarly, RA-enabled designs can coexist with ISAC, SWIPT, AirComp, and physical-layer security by tailoring the aperture response to meet communication, sensing, or energy-transfer objectives under the same overarching overhead and hardware constraints.}

{
	From a practical 6G deployment perspective, whether an RA paradigm complements or substitutes conventional antennas is largely determined by mobility support, hardware cost, and integration complexity. Regarding mobility, FA and PA are often more attractive for fast adaptation because their configuration can be selected at (or close to) scheduling timescales, making them suitable for mobility and blockage-prone scenarios where channel coherence is limited. MA offers larger geometry-induced gains but is more appropriate when reconfiguration latency is acceptable (e.g., slower-varying large-scale channel changes), or when the infrastructure can support reliable actuation and calibration. In terms of hardware cost and power, FA typically feature simpler front-end structures (e.g., switching or waveguide-based mechanisms), whereas MA requires actuators and sensing, and RHA usually demands dense control, EM-compliant calibration, and tighter manufacturing tolerances. Finally, integration complexity differs substantially: RHA and MA may incur higher control and calibration overhead, while FA and PA can often be integrated as a relatively lightweight configuration dimension on top of existing MIMO beamforming. Consequently, in many realistic systems, these paradigms are deployed in a complementary manner (RA-assisted beamforming) rather than as a full replacement, with the preferred choice governed by the latency--overhead--DoF trade-off and the targeted deployment role (BS-side infrastructure versus UE-side constraints).}

\subsection{Outage Probability Performance of Different RAs}
\begin{figure}[!t]
	\centering
	\includegraphics[width=0.8\linewidth]{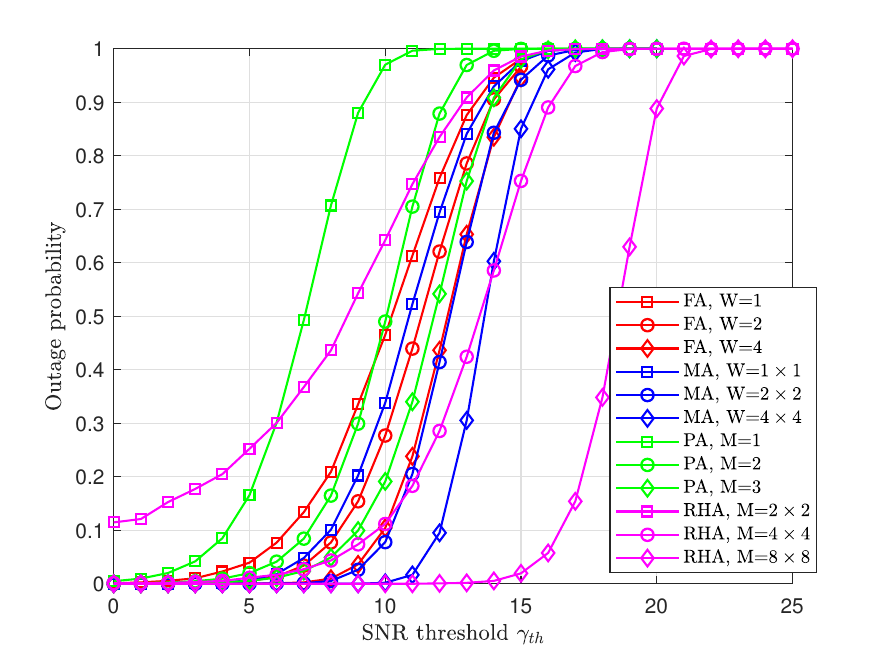}
	\caption{Outage probability of different reconfigurable antennas versus SNR threshold .}
	\label{fig:op_ra}
\end{figure}

\begin{figure}[!t]
	\centering
	\includegraphics[width=0.8\linewidth]{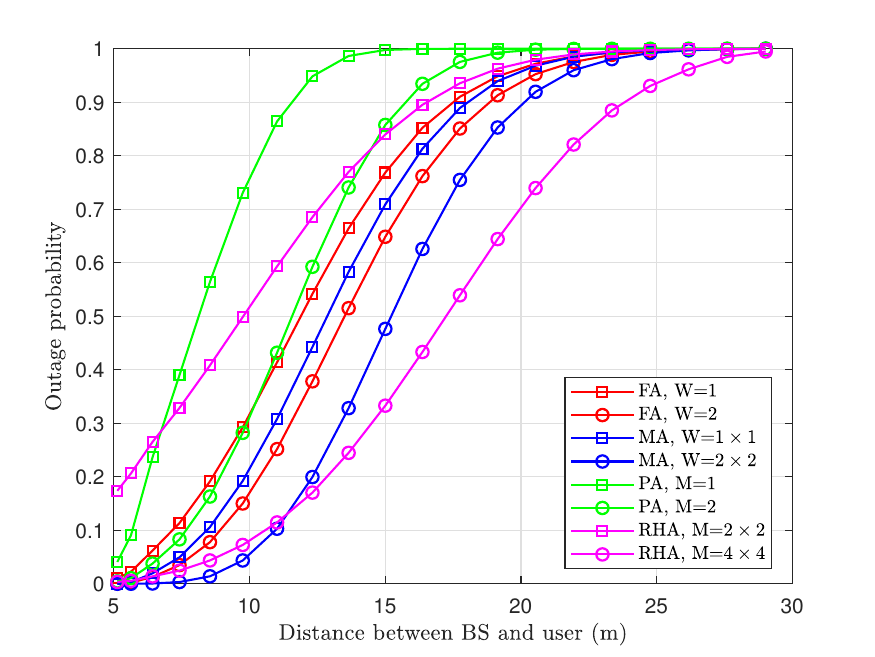}
	\caption{Outage probability of different reconfigurable antennas versus distance.}
	\label{fig:op_ra_distance}
\end{figure}

Outage probability is an important indicator for evaluating the performance of communication systems. Fig.~\ref{fig:op_ra} presents the outage probability versus SNR threshold $\gamma_{th}$ for four representative types of RAs, namely FA, MA, PA, and RHA. In the considered scenario, the RA is deployed at the transmitter side, while the receiver is equipped with a FPA. For the FA and MA configurations, only one active antenna element is utilized within a spatially constrained region. In the FA case, the antenna is fixed along a 1D linear structure of length $\lambda W$, where 
$W$ indicates the normalized size. As illustrated in the figure, increasing $W$ leads to significant improvements in outage performance, as it provides higher spatial diversity. In contrast, the MA configuration leverages a 2D movement area of size $\lambda W \times\lambda W$, allowing the antenna to dynamically search for the best transmission position \cite{10464791}. Again, a larger motion range $W$ enhances the performance by expanding the search space for favorable propagation directions, thus reducing outage probability. The PA configuration involves a waveguide-embedded structure where the antenna can dynamically slide along predefined paths to optimize link quality for a specific user. The user is randomly located in the $x\!-\!y$ plane, and the PA can dynamically adjust the positions of $M$ antennas along the waveguide \cite{tyrovolas2025performance}. Observe from Fig.~\ref{fig:op_ra} that increasing the number of PA $M$ significantly improves the outage performance due to the unique degree of freedom provided by the dynamic placement of the PAs along the waveguide. Finally, the RHA employs a 2D surface populated with $M\times M$ evenly spaced radiating elements. The RHA can electronically reconfigure the activation pattern of these elements to steer beams toward desired directions. It is evident from the figure that increasing the array size significantly improves outage probability performance, thanks to its large beamforming and spatial resolution capability.
{Fig.~\ref{fig:op_ra_distance} further depicts the outage probability as a function of the distance between the BS and the user. As expected, for all RA paradigms, the outage probability monotonically increases with distance due to path-loss attenuation and the reduced average received SNR. Moreover, the relative performance trends remain consistent with Fig.~\ref{fig:op_ra}: enlarging the FA region (larger $W$) or expanding the MA movement area (larger $W$) yields lower outage across the entire distance range, since a larger reconfiguration space provides more opportunities to exploit favorable channel realizations. For PA, increasing the number of sliding antennas $M$ also reduces outage, particularly at medium-to-long distances where additional placement DoF becomes more beneficial for compensating propagation loss. For RHA, a larger aperture (e.g., from $2\times2$ to $4\times4$) achieves the lowest outage among the compared settings at the same distance, indicating that stronger beamforming gain and higher spatial resolution are effective in extending coverage. }

{
	\subsection{Security Vulnerabilities of Different RAs}
	Reconfigurable antennas improve link adaptability, but they also introduce new attack surfaces because their performance depends on configuration signaling, CSI acquisition, and (in some designs) cyber--physical actuation. In what follows, we summarize key vulnerabilities for FA, MA, PA, and RHA under a unified view.
	
	For {FA}, the main vulnerabilities are related to configuration integrity and CSI-driven state selection. First, FA relies on switching among ports or positions, so the control interface becomes a direct attack surface. If an adversary tampers with the switch-control logic or its firmware, it may force the antenna to stay in suboptimal ports, or trigger frequent state toggling to waste energy and disrupt scheduling. Second, many FA protocols select ports based on pilots and CSI. Pilot contamination, spoofed feedback, or training-phase jamming can bias the selection toward poor ports, so the system loses diversity/beamforming gains even when the data channel is not heavily jammed. Third, the reconfiguration process can create observable side-channel cues, such as switching timing or EM leakage, which may help an adversary infer the active port/position and improve targeted interception or jamming. These issues highlight the need for authenticated configuration signaling, robust CSI validation, and simple fallback states when abnormal switching behavior is detected.
	
	For {MA}, the main risk comes from its cyber--physical reconfiguration loop. First, 6DMA relies on motors or robotic actuators to adjust 3D positions and 3D rotations. An attacker may induce frequent reconfiguration by sending high-rate interference or by manipulating CSI feedback, which leads to over-actuation and accelerates mechanical wear \cite{zhang2025dependability}. In hard-to-maintain BS deployments, such accelerated degradation may even evolve into persistent or permanent denial-of-service (PDoS) effects \cite{abaimov2024understanding}. Second, many 6DMA designs exploit directional sparsity in the wireless channel \cite{shao2025distributed}. Pilot contamination or crafted interference can weaken this sparsity, which degrades sparse recovery and results in incorrect position/pose decisions \cite{akgun2017pilot}. Third, the position/rotation optimization is often non-convex and may use DRL or heuristic control \cite{tang2025deep}. In this case, adversarial perturbations can mislead the controller and push it toward inefficient or unsafe configurations \cite{sun2020stealthy}.
	
	For {PA}, the flexibility of pinching-based radiating points brings both security and privacy concerns. Strong LoS links can be beneficial, but they may also create more predictable propagation paths for interception in the coverage region. Moreover, PA often relies on CSI to determine activation patterns and beamforming. This makes it vulnerable to pilot contamination, jamming, and CSI poisoning, which can cause activation errors and beam misalignment. In addition, PA optimization may require user location information. This can leak privacy if control links are compromised or if activation patterns can be inferred from observations. Finally, the activation process itself may leave side-channel signatures (e.g., switching-related or EM leakage), which could expose antenna states or beam directions in practice.
	
	For {RHA}, the key vulnerability is the controllable excitation interface. RHA is an active radiating aperture controlled by FPGA or microcontroller-based circuits. If this interface is hijacked, an attacker can directly distort the radiated wavefront and disrupt the link. For example, the attacker may force a non-radiative state and effectively silence transmission, which resembles information-absorption manipulation attacks discussed for RIS \cite{9941040}. The attacker may also steer energy toward an eavesdropper by manipulating the holographic pattern, which increases information leakage \cite{9941040}. In addition, RHA remains vulnerable to training-phase threats such as pilot contamination. A corrupted channel estimate can produce an erroneous holographic pattern and degrade service to legitimate users.
	
	Overall, these vulnerabilities suggest that RA-enabled systems should protect configuration signaling and CSI acquisition, and should monitor abnormal reconfiguration behaviors. This is especially important for designs with dense control interfaces or cyber--physical actuation.
	}

\subsection{Open Challenges}

\subsubsection{Integrated AI for Intelligent Reconfiguration} 
Current RA control strategies are predominantly fixed or rule-based, which limits adaptability in dynamic and uncertain environments. Although the inherent reconfigurability of RAs offers substantial optimisation potential, static control logic struggles to cope with mobility, interference, and rapidly varying propagation conditions. The integration of AI, particularly machine learning (ML) and reinforcement learning (RL), can enable adaptive control mechanisms capable of learning from environmental feedback and optimizing antenna states in real time. For instance, RL agents could dynamically determine the optimal antenna port, mechanical position, or metasurface configuration, while supervised learning models could leverage historical data and contextual features for predictive configuration. Nevertheless, practical deployment is challenged by constraints on computational resources, inference latency (often below 1 ms), and on-device learning capability, especially for size- and power-limited platforms such as wearables or UAVs.
{From a system-level perspective, AI-enabled reconfiguration should be co-designed with the full control loop, including CSI acquisition, feedback signaling, and reconfiguration actuation. In particular, the learning objective should explicitly incorporate training/selection overhead, switching delay, and calibration cost, rather than optimizing instantaneous channel gain only. Moreover, unresolved gaps differ across paradigms: (i) for FA, learning must reduce port scanning and handle inter-port correlation; (ii) for MA, learning should jointly plan motion trajectories and communication actions under safety and energy constraints; (iii) for PA, learning must account for activation-point repeatability and waveguide-induced correlation; and (iv) for RHA, learning must operate under high-dimensional EM-constrained control with calibration drift. Developing lightweight, reliable, and interpretable policies that can be deployed within realistic 6G control planes remains an open challenge.}

\subsubsection{Cross-Layer Optimisation}
RA optimisation efforts are predominantly confined to the physical layer, often overlooking interactions with higher-layer protocols and network-level decisions. A cross-layer design approach can significantly enhance system efficiency by coupling antenna reconfiguration decisions with medium access control (MAC) scheduling, QoS priorities, and user mobility predictions. For example, the orientation and beam direction of a mechanically movable antenna on a UAV should be informed not only by instantaneous channel gain but also by MAC-layer scheduling and application-level latency constraints. Similarly, RHA-based beamforming could adapt to traffic patterns or application-specific requirements, such as minimizing latency for immersive AR/VR streams. Achieving such integration necessitates low-latency feedback mechanisms, efficient signaling protocols, and reconfiguration-aware scheduling and routing algorithms.
{More importantly, cross-layer co-design should be formulated for large-scale multi-user deployments, where RA control interacts with beam management, handover, interference coordination, and multi-AP cooperation (e.g., cell-free architectures). A key unresolved issue is timescale mismatch: MA and some PA operations may only update at ms--s scales, whereas scheduling and link adaptation operate at much faster scales. This motivates two-timescale architectures and mobility-aware policies that decide {when} to reconfigure, not only {how}. In addition, cross-layer designs should expose standardized interfaces for reporting reconfiguration state/latency and for allocating pilot resources, enabling the network to budget training overhead and maintain QoS guarantees.}

\subsubsection{Standardization and Interoperability}
The absence of unified control protocols, calibration methodologies, and performance metrics remains a major barrier to the widespread adoption of RA technologies. Current implementations are typically vendor-specific, lacking common interfaces for configuration, feedback, and state reporting. For network-integrated RAs, such as those in reconfigurable intelligent surface (RIS)-assisted communications, standardization is vital to ensure cross-vendor compatibility and seamless network orchestration. This includes defining standard control commands for port switching, metasurface tuning, or mechanical actuation; specifying state reporting formats for configuration and reconfiguration latency; and establishing reference calibration procedures for alignment and phase compensation. Collaboration with standardization bodies such as 3GPP, ETSI, and IEEE 802.11 will be essential to align RA capabilities with existing PHY/MAC specifications and future 6G.
{In addition to interfaces, the community lacks unified benchmarking and testing frameworks that allow fair comparison across FA, MA, PA, and RHA. Future standardization efforts should define (i) reference scenarios (indoor NLoS, urban micro, vehicular/UAV, smart-environment), (ii) agreed-upon metrics that explicitly include overhead (pilot/training, scanning, switching delay, calibration time) alongside rate/reliability/energy, and (iii) reporting templates for hardware constraints and control-loop latency budgets. Such a framework would bridge the gap between algorithmic demonstrations and deployable 6G system requirements, and enable interoperability at the level of both control signaling and performance evaluation.}

\subsubsection{Experimental Prototyping and Measurement Campaigns}
Most RA research to date has been conducted through simulation studies, with limited experimental prototyping and empirical validation across diverse channel conditions. Real-world testing is essential to evaluate performance under practical constraints, as mechanical tolerances, nonlinearities, and environmental effects can significantly deviate from theoretical assumptions. Examples include fluid dynamics latency in FAs, hysteresis in metasurface tuning for RHAs, and positioning drift in mechanically MAs. Developing modular, scalable prototypes for multiple form factors (e.g., handheld devices, drones, infrastructure) and operating bands (sub-6 GHz, mmWave, THz) is critical. Comprehensive over-the-air measurement campaigns in both anechoic and realistic multipath environments, along with standardized benchmarking protocols, will be pivotal for technology validation and fair cross-technology comparison.
{A major open gap is the scarcity of cross-technology measurement datasets and hardware-in-the-loop testbeds that jointly capture channel dynamics, reconfiguration latency, and calibration drift. Future campaigns should report not only channel statistics but also repeatability metrics (e.g., configuration-to-configuration variance), control-loop timing (sensing--decision--actuation), and long-term stability under temperature and mechanical stress. Such system-level measurements are particularly important for assessing large-scale deployment feasibility, where accumulated calibration overhead and maintenance cycles may dominate the net throughput/energy gains.}

\subsubsection{Environmental and Sustainability Impact}
Sustainability considerations for RA technologies, including material selection, manufacturing processes, and operational energy consumption, have received limited attention compared with performance metrics. Mechanical actuators in MAs may require rare earth materials with poor recyclability, fluid-based FAs might employ potentially hazardous liquids, and metasurface-based RHAs often rely on multi-layer PCBs with high-energy manufacturing processes. In addition, frequent reconfiguration cycles in dynamic environments can raise power consumption significantly, especially for actively tuned RAs. Future research should focus on eco-friendly materials such as biodegradable substrates and low-toxicity dielectrics, as well as energy-aware control algorithms that balance reconfiguration benefits with energy costs. Conducting full lifecycle assessments (from fabrication to end-of-life) can reduce total cost of ownership and enable deployment in energy-constrained or environmentally sensitive scenarios.
{To connect sustainability with deployability, future work should quantify the system-level energy and operational footprint introduced by RA control, including pilot overhead, signaling, computation, and calibration/maintenance cycles. This is especially relevant for dense 6G infrastructures and smart environments, where large numbers of RHA/RIS-enabled nodes may amplify control and calibration costs. Establishing lifecycle-aware benchmarks and reporting standards would help evaluate whether the net energy savings from improved link efficiency outweigh the additional manufacturing and operational overhead.}

\section{Conclusion}

Reconfigurable antennas, encompassing fluid antennas, movable antennas, pinching antennas, and reconfigurable holographic antennas, have emerged as key enablers for next-generation mobile networks by introducing the additional spatial flexibility. This article has systematically reviewed their fundamental principles, hardware architectures, channel modelling and estimation techniques, performance analysis, and resource allocation strategies, as well as their integration with other emerging wireless technologies. Comparative analysis reveals that each RA type offers distinct advantages in terms of reconfigurability, deployment complexity, and energy efficiency, making them suitable for diverse application scenarios. Despite remarkable progress, several open challenges remain, including adaptive RA control under dynamic environments, joint optimisation with network-level functions, and hardware design trade-offs between flexibility and complexity. Addressing these challenges will require interdisciplinary efforts spanning antenna engineering, signal processing, and intelligent resource management. With continuous advancements in materials, hardware architectures, and AI-driven optimisation, RAs are expected to play an increasingly pivotal role in enabling highly adaptable, efficient, and intelligent wireless communication systems in the 6G era and beyond.

\bibliography{Reference}

	\begin{IEEEbiography}[{\includegraphics[width=1in,height=1.25in,clip,keepaspectratio]{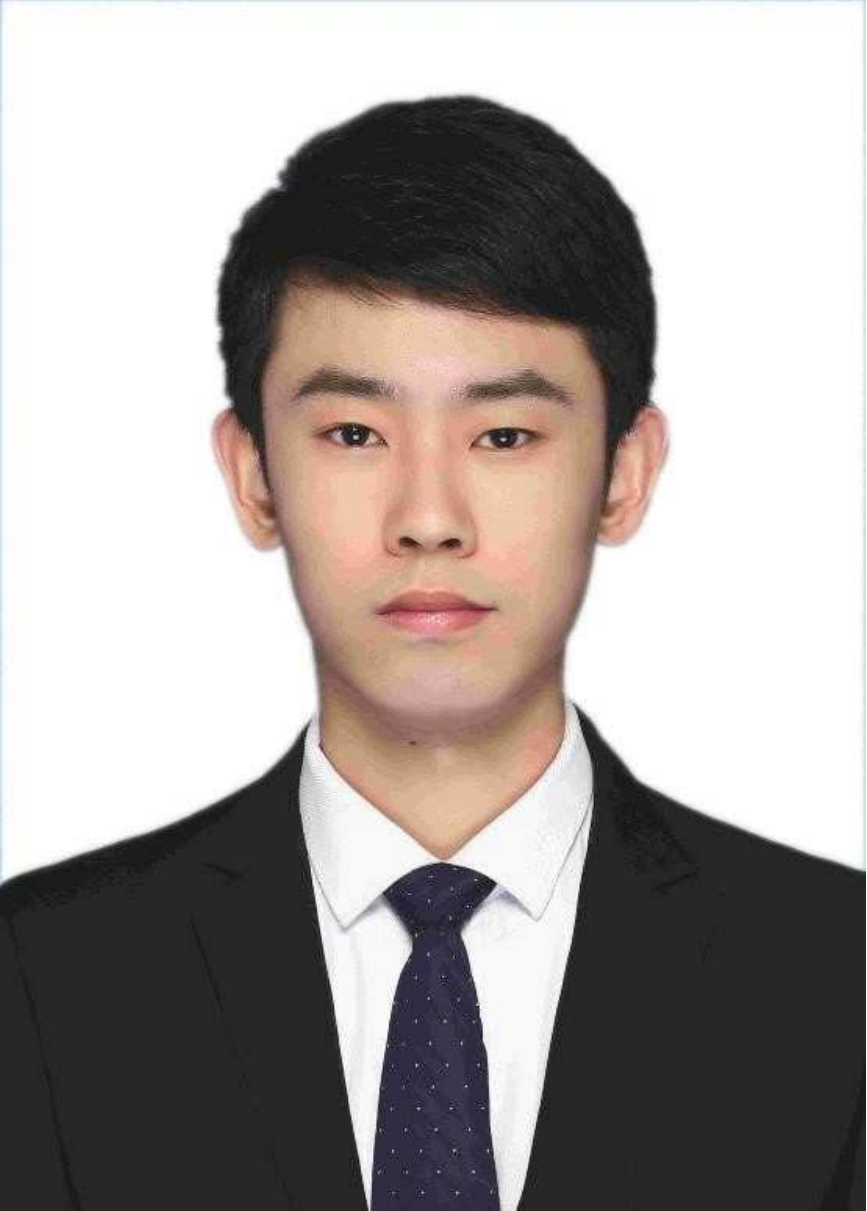}}]{Yizhe Zhao}(S'16-M'21) received the PhD in 2021 in School of Information and Communication Engineering from University of Electronic Science and Technology of China (UESTC), where he is currently an associate professor. He has been a visiting researcher with the Department of Electrical and Computer Engineering, University of California, Davis, USA. He is a member of IEEE and a senior member of China Institute of Communications. He is selected in Young Elite Scientists Sponsorship Program by China Association for Science and Technology (CAST). He serves for China Communications and Journal of Communications and Information Networks (JCIN) as the Guest Editor, and is also a TPC member of several prestigious IEEE conferences, such as IEEE ICC, Globecom. He was the recipient of IEEE CSE Best Paper Award in 2023. His research interests include reconfigurable antenna systems, integrated data and energy transfer.
	\end{IEEEbiography}
\vspace{-10mm}
	\begin{IEEEbiography}[{\includegraphics[width=1in,height=1.25in,clip,keepaspectratio]{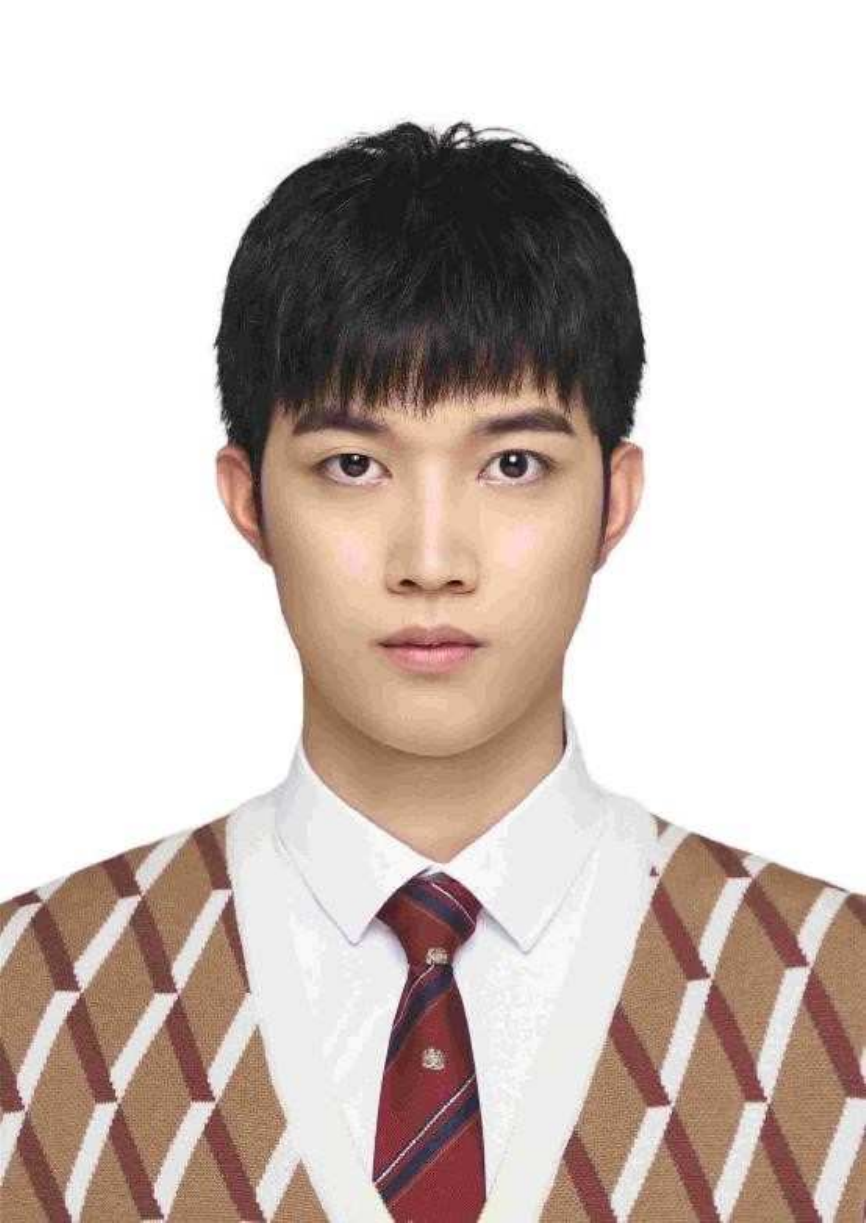}}]{Long Zhang }received the B.S. degree from Chongqing University in 2022. He is currently pursuing PHD in the School of Information and Communication Engineering, University of Electronic Science and Technology of China, Chengdu, China. His research interests include wireless communications, integrated data and energy transfer, reconfigurable antenna systems.
	\end{IEEEbiography}
\vspace{-10mm}
	\begin{IEEEbiography}[{\includegraphics[width=1in,height=1.25in,clip,keepaspectratio]{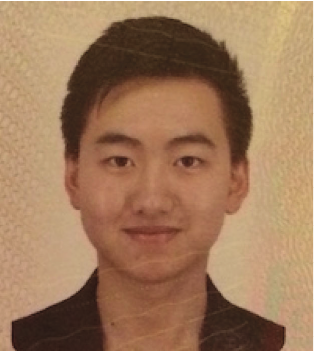}}]{Halvin Yang} is a Research Associate in the Department of Electrical and Electronic Engineering at Imperial College London, working on artificial intelligence for future communication systems. His broader research vision is to enable AI-native, energy-efficient 6G networks through the fusion of intelligent algorithms and reconfigurable communication architectures. He did his Bachelor’s degree in Electronic and Electrical Engineering at Imperial College London and received his Ph.D. from University College London (UCL), where his research focused on the theoretical and practical development of Fluid Antenna Systems (FAS). After completing his Ph.D., Halvin extended his research to Non-Terrestrial Networks (NTNs) and Orthogonal Time Frequency Space (OTFS) modulation, investigating their integration and performance under high-mobility satellite conditions. He has also explored the use of AI techniques—including large language models and learning-based channel prediction—to enhance the efficiency and adaptability of these technologies. Halvin is an active IEEE member and reviewer, and his work has appeared in leading IEEE journals such as the IEEE Journal on Selected Areas in Communications, the IEEE Transactions on Wireless Communications, and the IEEE Communications Magazine, and has earned multiple Best Paper Awards at international conferences. 
	\end{IEEEbiography}
\vspace{-10mm}
	\begin{IEEEbiography}[{\includegraphics[width=1in,height=1.25in,clip,keepaspectratio]{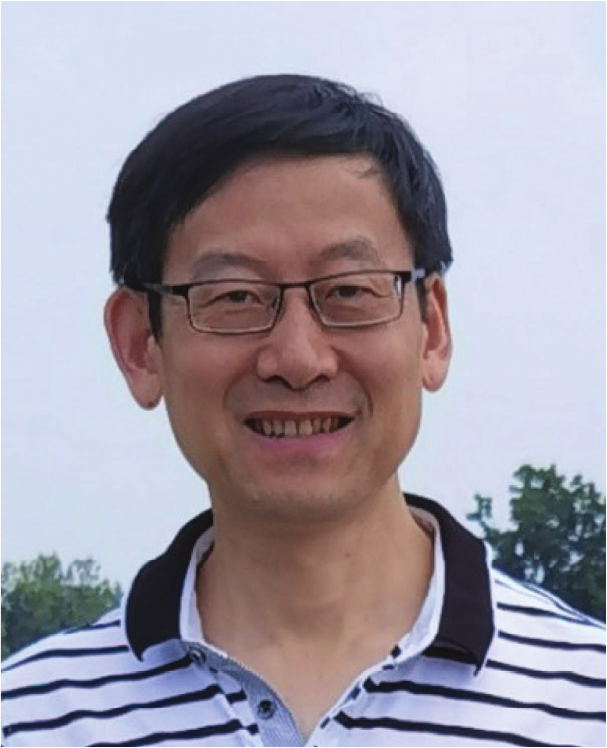}}]{Kun Yang}
	received his PhD from the Department of Electronic \& Electrical Engineering of University College London (UCL), UK. He is currently a Chair Professor of Nanjing University and an affiliated professor at the University of Essex. His main research interests include wireless networks and communications, communication-computing cooperation, and AI (artificial intelligence) for wireless. He has published 600+ papers and filed 50 patents. He serves on the editorial boards of a few IEEE journals (e.g., IEEE WCM, TVT, TNB). He is a Deputy Editor-in-Chief of IET Smart Cities Journal. He has been a Judge of GSMA GLOMO Award at World Mobile Congress – Barcelona since 2019. He was a Distinguished Lecturer of IEEE ComSoc, a Recipient of the 2024 IET Achievement Medals and the Recipient of 2024 IEEE CommSoft TC’s Technical Achievement Award. He is a Member of Academia Europaea (MAE), a Fellow of IEEE, a Fellow of IET and a Distinguished Member of ACM. 
	\end{IEEEbiography}
\vspace{-10mm}
	\begin{IEEEbiography}[{\includegraphics[width=1in,height=1.25in,clip,keepaspectratio]{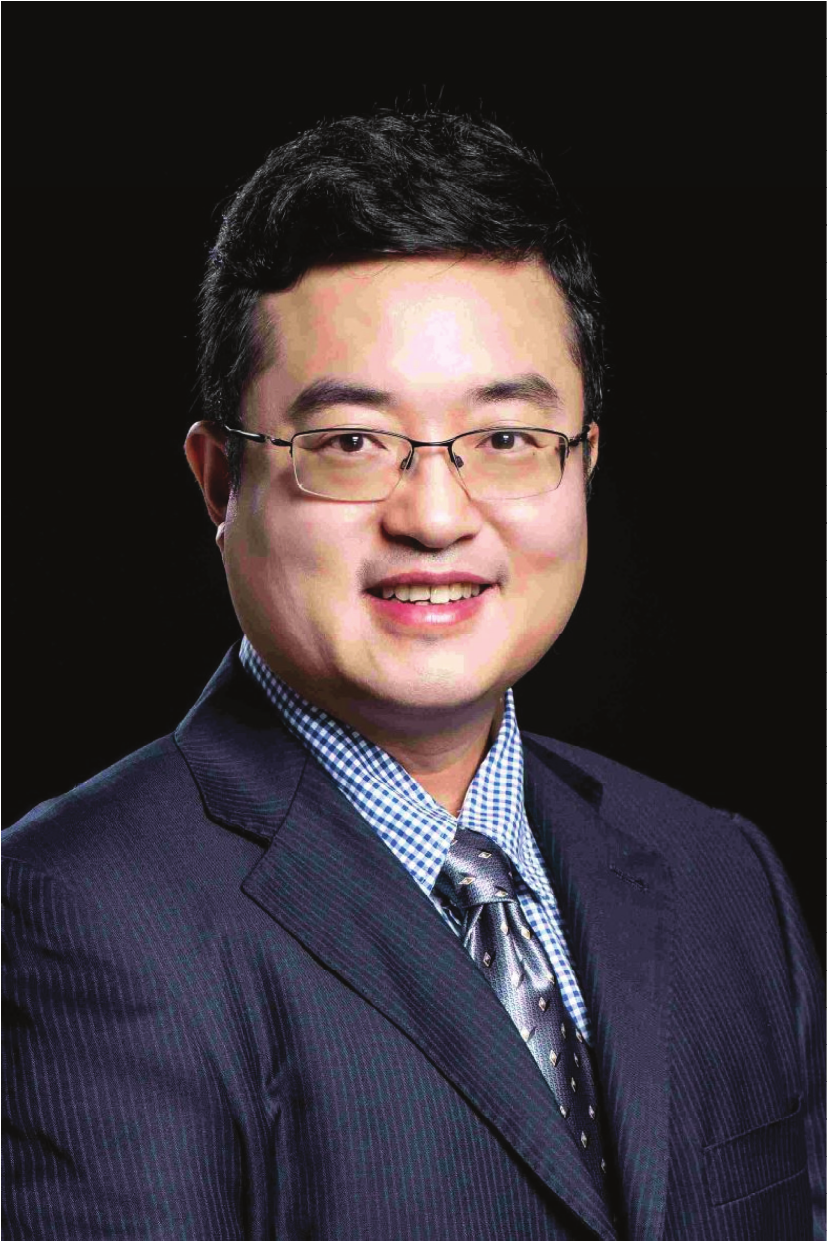}}] {Rui Zhang} (S'00-M'07-SM'15-F'17) received the B.Eng. (first-class Hons.) and M.Eng. degrees from the National University of Singapore, Singapore, and the Ph.D. degree from the Stanford University, Stanford, CA, USA, all in electrical engineering. From 2007 to 2009, he worked as a research scientist at the Institute for Infocomm Research, ASTAR, Singapore. In 2010, he joined the Department of Electrical and Computer Engineering of National University of Singapore, where he is now a Provost’s Chair Professor. He is also an Adjunct Professor with the School of Science and Engineering, The Chinese University of Hong Kong, Shenzhen, China. He has published over 600 papers, all in the field of wireless communications and networks. He has been listed as a Highly Cited Researcher by Thomson Reuters/Clarivate Analytics since 2015. His current research interests include intelligent surfaces, reconfigurable antennas, radio mapping, non-terrestrial communications, wireless power transfer, AI and optimization methods.
	\end{IEEEbiography}
	\vspace{-10mm}
	\begin{IEEEbiography}[{\includegraphics[width=1in,height=1.25in,clip,keepaspectratio]{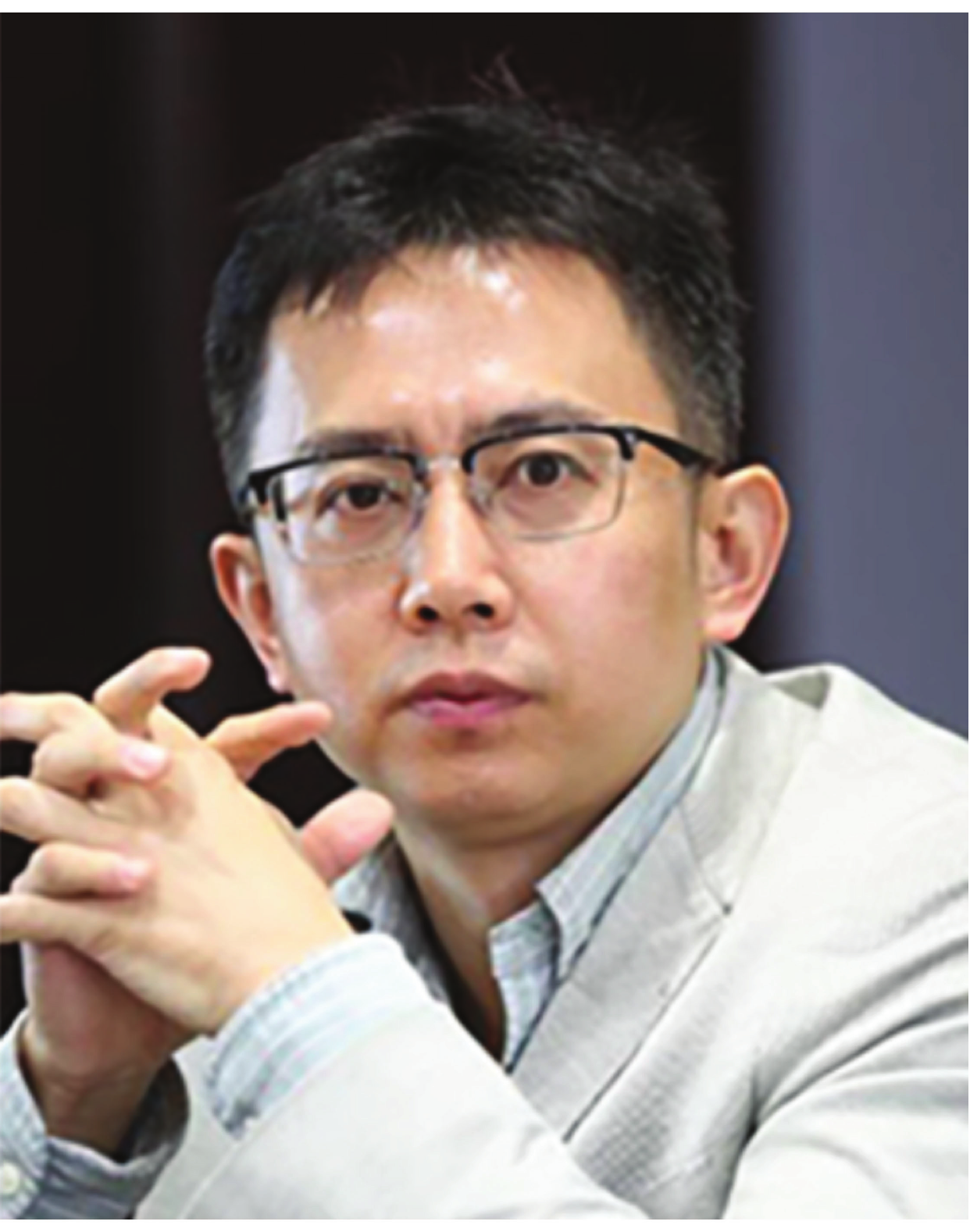}}]{Lingyang Song} (S'03-M'06-SM'12-F'19) is a Boya Distinguished Professor at the School of Electronics, Peking University, China. He received his PhD from the University of York, UK, in 2007. His research interests include wireless communication and networks, signal processing, and machine learning. Dr. Song is the coauthor of many awards, including 2025 IEEE ComSoc TCCN Publication Award and 2025 IEEE Neal Shepherd Memorial Best Propagation Paper Award. He has served as an IEEE ComSoc Distinguished Lecturer (2015-2018), an Area Editor of IEEE Transactions on Vehicular Technology (2019-), Director of IEEE Communications Society Asia Pacific Board (2024-2025). He is a Clarivate Analytics Highly Cited Researcher and an IEEE Fellow.
	\end{IEEEbiography}
	\vspace{-10mm}
	\begin{IEEEbiography}[{\includegraphics[width=1in,height=1.25in,clip,keepaspectratio]{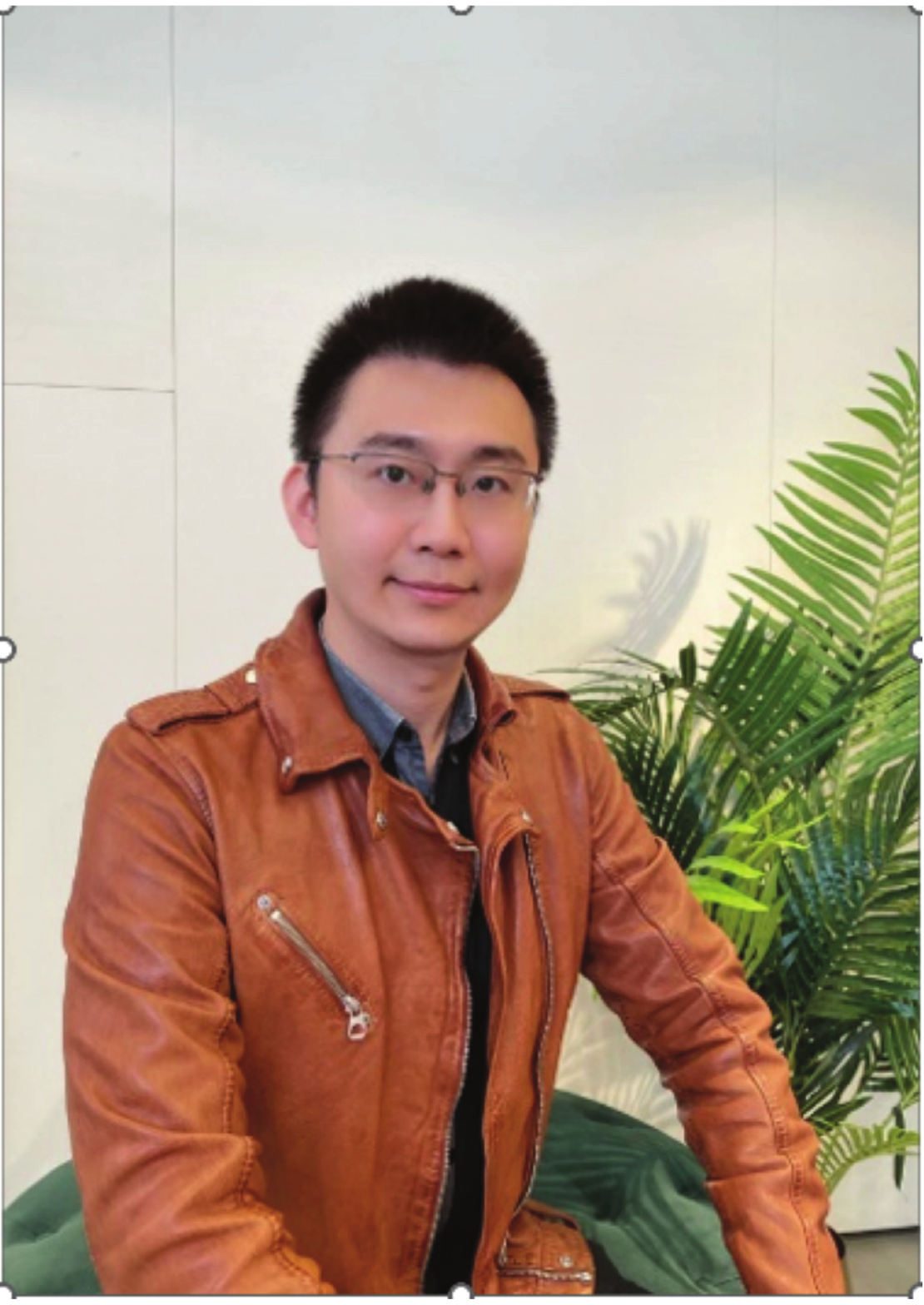}}]{Yuanwei Liu}(S'13-M'16-SM'19-F’24) is a tenured full Professor in Department of Electrical and Computer Engineering (ECE) at The University of Hong Kong (HKU). Prior to that, he was a Senior Lecturer (Associate Professor) (2021-2024) and a Lecturer (Assistant Professor) (2017- 2021) at Queen Mary University of London (QMUL), London, U.K, and a Postdoctoral Research Fellow (2016-2017) at King's College London (KCL), London, U.K. He received the Ph.D. degree from QMUL in 2016.  His research interests include generative AI/LLM for communications, low altitude networks large model, mobile edge generation, wireless digital twins, and integrated sensing and communications. 		
	\end{IEEEbiography}
\end{document}